\documentclass[prc,aps,float,twocolumn,byrevtex,superscriptaddress,showpacs,nofootinbib]{revtex4}

\usepackage{amsmath}
\usepackage{amssymb}
\usepackage{amsfonts}
\usepackage{epsfig}
\usepackage{color}

\newcommand{\etal}{{\emph{et al.}\ }}

\renewcommand{\vec}[1]{\mathbf{#1}}
\newcommand{\nuc}[2]{{$^{#1}$}{#2}}
\newcommand{\divfac}{\mathcal{V}}

\begin{document}

\title{Particle-Number Restoration within the Energy Density Functional Formalism}

\author{M. Bender}
\email{bender@cenbg.in2p3.fr}
\affiliation{National Superconducting Cyclotron Laboratory,
             1 Cyclotron Laboratory,
             East-Lansing, MI 48824, USA}
\affiliation{CEA, Centre de Saclay, IRFU/Service de Physique Nucléaire, F-91191 Gif-sur-Yvette, France}
\affiliation{Universit{\'e} Bordeaux,
             Centre d'Etudes Nucl{\'e}aires de Bordeaux Gradignan, UMR5797,
             F-33175 Gradignan, France}
\affiliation{CNRS/IN2P3,
             Centre d'Etudes Nucl{\'e}aires de Bordeaux Gradignan, UMR5797,
             F-33175 Gradignan, France}

\author{T. Duguet}
\email{thomas.duguet@cea.fr}
\affiliation{National Superconducting Cyclotron Laboratory,
             1 Cyclotron Laboratory,
             East-Lansing, MI 48824, USA}
\affiliation{CEA, Centre de Saclay, IRFU/Service de Physique Nucléaire, F-91191 Gif-sur-Yvette, France}
\affiliation{Department of Physics and Astronomy,
             Michigan State University,
             East Lansing, MI 48824, USA}

\author{D. Lacroix}
\email{lacroix@ganil.fr}
\affiliation{National Superconducting Cyclotron Laboratory,
             1 Cyclotron Laboratory,
             East-Lansing, MI 48824, USA}
\affiliation{GANIL, CEA and IN2P3, Bo\^ite Postale 5027, 14076 Caen Cedex, France}

\date{\today}



\begin{abstract}
We give a detailed analysis  of the origin of spurious divergences and finite steps that have been recently
identified in particle-number restoration calculations within the nuclear energy density functional framework. We
isolate two distinct levels of spurious contributions to the energy. The first one is encoded in the definition
of the basic energy density functional itself whereas the second one relates to the canonical procedure followed
to extend the use of the energy density functional to multi-reference calculations. The first level of spuriosity
relates to the long-known self-interaction problem and to the newly discussed self-pairing interaction process
which might appear when describing paired systems with energy functional methods using auxiliary reference
states of Bogoliubov or BCS type. A minimal correction to the second level of spuriosity to the multi-reference
nuclear energy density functional proposed in [D.~Lacroix, T.~Duguet, M.~Bender, arXiv:0809.2041]
is shown to remove completely the anomalies encountered in particle-number restored calculations. In particular, it restores
sum-rules over (positive) particle numbers that are to be fulfilled by the particle-number-restored formalism. The correction
is found to be on the order of several hundreds of keVs up to about 1 MeV in realistic calculations, which is
small compared to the total binding energy, but often accounts for a substantial percentage of the energy gain
from particle-number restoration and is on the same energy scale as the excitations one addresses with
multi-reference energy density functional methods.
\end{abstract}

\maketitle

%
%
\section{Introduction}
\label{intro}

Methods based on the use of energy density functionals (EDF)~\cite{bender03b} currently provide the only set of
theoretical tools that can be applied to all nuclei but the lightest ones in a systematic manner irrespective of
their mass and isospin. Nuclear EDF methods coexist on two distinct levels. On the first level, that is
traditionally and inappropriately called "self-consistent mean-field theory" or Hartree-Fock (HF) or
Hartree-Fock-Bogoliubov (HFB), a single product state provides the normal and anomalous density matrices the energy
is a functional of. We will call this type of method a single-reference (SR) EDF approach. On the
second level, traditionally and inappropriately called "beyond-mean-field methods", i.e.~symmetry restoration and
configuration mixing in the spirit of the Generator Coordinate Method (GCM), the set of transition density
matrices defined from an appropriate ensemble of product states enter
the EDF. We will call such a method a  multi-reference (MR) EDF approach. Although SR EDF calculations have
many similarities with Density Functional Theory (DFT) which is widely used in atomic, molecular and condensed matter
physics~\cite{hohenberg64,dreizlerBook,parrBook,lecturenotesFNM,koch01,Nagy98DFT,kohn98}, they also present key
differences, which prohibit the straightforward mapping of the concepts of electronic DFT to the nuclear case
\cite{engel06b,Giraud07a,Giraud07b}.

The reference state entering a SR EDF calculation usually breaks several symmetries of the exact eigenstates of the nuclear
Hamiltonian. This is done on purpose, as it allows one to incorporate so-called \emph{static} correlations associated
with collective modes~\cite{Rei84a,Abe90a,Naz92a,Naz93a} at moderate computational cost. One of the most important
categories of correlations which can be grasped this way are those associated with the formation of neutron and
proton Cooper pairs in the medium.

In a SR EDF approach, pairing correlations are incorporated by making the energy a functional of the anomalous density
matrix in addition to the normal one. This amounts to using an independent quasi-particle state (which will be called
a quasi-particle vacuum in what follows) of BCS or Bogoliubov type as a reference state instead of a Slater
determinant. The price to pay is breaking the $U(1)$ symmetry in gauge space that is a feature of eigenstates of
the particle-number operator. As a result the SR state is spread in particle-number space,
and one cannot associate the computed energy, even implicitly, to a state belonging to a
specific irreducible representation of $U(1)$. In condensed matter physics,
for which the BCS method was originally designed~\cite{deGennes66}, this is usually not much of a problem.
Nuclei, however, are small finite quantum many-body systems for which two problems arise in this context:
(i) the SR approach does not grasp the so-called \emph{dynamical} pairing correlations associated with
the fluctuations of both the magnitude and the phase of the order parameter of the broken $U(1)$ symmetry.
Correlations associated with this zero-energy mode may affect any observable that probes the occupation
of levels around the Fermi surface in a significant way; (ii) when the density of single-particle levels
around the Fermi energy is below a critical value, pairing correlations are entirely \emph{dynamical}
and cannot be described by the SR method, in most cases in contradiction with experiment.

All of these limitations can be overcome by performing multi-reference EDF calculations. Those allow in particular
the restoration of particle number~\cite{heenen93a,anguiano01b,samyn04,niksic06b,Robledo07a}. It has been noticed
for some time, however, that particle-number restored energies might exhibit
divergences~\cite{tajima92a,donau98,anguiano01b} and finite steps~\cite{almehed01a,doba05a} whenever a
single-particle level crosses the Fermi energy as a function of a collective coordinate. This problem is
particular to energy density functionals, but absent in approaches based on the use of a genuine Hamiltonian and
a correlated wave function. As pointed out by Anguiano \etal\ in Ref.~\cite{anguiano01b}, some of the common
assumptions and approximations made in the construction of nuclear EDFs unavoidably lead to such anomalies, and
these authors, as done earlier in Refs.~\cite{donau98,almehed01a} in a different context, advocate to use strict
antisymmetric two-body vertices and to keep all exchange terms when computing the energy. However, and contrary
to what is stated in Ref.~\cite{anguiano01b,Rodriguez07a}, using antisymmetric but density-dependent two-body
vertices is not free from pathologies, even when the divergence introduced by the density-dependent
terms is integrable. There is an additional problem that arises particularly when such a dependence is taken under
the form of a non-integer power of the density (matrix) \cite{doba05a,paperIII}.

Practitioners of EDF methods, however, recognize that it is desirable to use more general energy functionals. For
those, Particle-Number Restoration (PNR), and the MR formalism in general, still need to be formulated in a
consistent and unambiguous manner that is free from pathologies. As a first step into that direction, a thorough
analysis has been recently given by Dobaczewski \etal\ regarding the poles and steps contained in a
particle-number restored energy density functional~\cite{doba05a}. In the first of our companion
papers~\cite{lacroix06a}, hereafter referred to as Paper~I, we could connect those pathologies to an underlying level
of spuriosity that is encoded in the SR energy functional. The associated spurious terms turn out to relate to
\emph{self-interaction} processes well-known in DFT for condensed
matter~\cite{perdew81a}, a problem which was actually studied beforehand in the nuclear context~\cite{stringari78a}
but was soon forgotten, as well as to spurious \emph{self-pairing} processes, whose notion is introduced in the
present paper. The common source of both pathologies is the use of different and non-antisymmetric vertices at
different places in the EDF violating in this way the exchange symmetry of Fermi statistics. The existence of spurious
self-interaction and self-pairing in the SR energy functional is indeed a prerequisite for the appearance of
divergences and steps at the MR level, but it is not its origin as such. The pathologies that are particular
to the MR level, e.g.,
particle-number restoration, turn out to be caused by an unphysical contribution to the \emph{weight} of the
self-interaction and self-pairing contributions in multi-reference energy kernels. This is an unforeseen
consequence of the common practice of constructing the multi-reference energy functional kernel by replacing the
density matrices entering a given SR energy functional by transition density matrices~\cite{Bon90a,Rei94a}
in \emph{analogy} to the
application of the Generalized Wick Theorem (GWT)~\cite{onishi66,balian69a} within a Hamiltonian- and wave-function-based
approach. Making reference to a Wick theorem in an energy density functional without having a genuine operator to relate
to is necessarily outside the scope of that Wick theorem and might produce unexpected results. And indeed, using the
standard~\cite{wick50a} and generalized~\cite{onishi66,balian69a} Wick theorems yields different weights to self-interaction
and self-pairing contributions
to the MR energy kernel as demonstrated in Paper~I. Only the GWT-motivated procedure produces the poles that are at
the origin of the divergences and steps, thus introducing a second level of spuriosity. Using a Hamiltonian-
and wave-function-based approach, no problem arises; the vertices at play are either zero or recombine in a
particular way that cancels out dangerous poles. Our analysis in Paper~I was made without reference to a
particular MR application and aimed at the introduction of a proper framework to identify and separate both levels
of spuriosity within any MR EDF calculation. It is the aim of the present paper to apply the procedure proposed in
Paper~I to correct for the unphysical weights in the special case of particle-number restoration using a particular
energy functional the correction can be applied to. In a third paper~\cite{paperIII}, hereafter called Paper~III,
we analyze in detail in the context of PNR the reasons why the pathologies associated with more commonly used
functionals containing non-integer powers of the density (matrix)~\cite{doba05a} are very likely to be
uncorrectable. Together with Ref.~\cite{doba05a}, Paper~III demonstrates
that the density-dependent two-body forces that are advertised by some authors to be free of pathologies
\cite{anguiano01b,Robledo07a,Rodriguez07a} also have their share of problems when used in MR calculations.

The paper is organized as follows: In Section~\ref{meanfield}, we introduce single-reference EDF calculations,
paying particular attention to resemblances and key differences with the HFB method based on the use of a Hamilton operator.
In Section~\ref{projectedmeanfield}, we introduce multi-reference EDF calculations appropriate to restoring
particle number, paying particular attention to resemblances and key differences with the strict particle-number
projected HFB (PNP-HFB) method based on the use of a Hamilton operator. In Section~\ref{selfenergyselfpairing1},
we discuss the occurrence of spurious
self-interaction and self-pairing processes in SR and MR calculations. Section~\ref{complexplane} analyzes the
occurrence of spurious self-interaction and self-pairing contributions to the particle-number restored EDF using a
complex plane analysis and specifies the correction designed in Paper~I to that particular case. Section~\ref{applications}
applies the correction scheme to realistic calculations of finite nuclei. Finally, conclusions are drawn in
Section~\ref{conclusions}. Several appendices complement the paper with derivations and formulas useful for
practical applications.

\section{Single-Reference EDF approach}
\label{meanfield}

Let us first present the basic elements of the single-reference EDF method which will be needed for our
discussion. The HFB implementation of the single-reference EDF approach relies on the use of a quasi-particle
vacuum $| \Phi_{\varphi} \rangle$ as a reference state from which the normal and anomalous one-body density
matrices entering the energy density functional are calculated. In the canonical basis $\{a_{\mu}, a^+_{\mu}\}$
that diagonalizes its one-body normal density matrix, the reference state reads
\begin{equation}
\label{eq:transfo_phi}
| \Phi_{\varphi} \rangle
= \prod_{\mu>0} \big( u_{\mu} + v_{\mu} \, e^{2i\varphi} \, a^+_{\mu} \, a^+_{\bar \mu} \big) \, | 0 \rangle
\, ,
\end{equation}
where $| 0 \rangle$ is the particle vacuum. Throughout this paper we limit ourselves to time-reversal invariant
quasi-particle vacua $| 0 \rangle$ with even-number parity and thus only discuss explicitly the ground-state of
even-even systems. In addition,
we do not mix protons and neutrons when constructing quasi-particle operators. In particular, this limits the
pairing interaction to particles of the same isospin. Identical assumptions are made in most, if not all,
published work performed using particle-number projected energy density functionals
so far, and are sufficient for the purpose of the present paper.

The single-particle wave functions associated with the pair-conjugated
canonical states $(\mu,\bar{\mu})$ is denoted as $\phi_{\mu}$ and $\phi_{\bar \mu}$. A quantum number
$\eta_{\mu}$ can always be chosen to separate the single-particle basis into two halves, the "positive" and the
"negative" ones, with each partner of a given conjugated pair associated to a different half. The normalization of $|
\Phi_{\varphi} \rangle$ gives $|u^{2}_{\mu}| + |v^{2}_{\mu} \, e^{2i\varphi}| = 1$. We use phase conventions where
the $u_{\mu}$ and $v_{\mu}$ are real numbers; hence, $u^{2}_{\mu} + v^{2}_{\mu} = 1$, which also fixes the global
phase of $| \Phi_{\varphi} \rangle$. The angle $\varphi$ in the
remaining phase factor denotes the orientation of the state in the $U(1)$ gauge space.

The exact eigenstates of the nuclear many-body problem belong to a specific irreducible representation of the
$U(1)$ group. By contrast, the product state $| \Phi_{\varphi} \rangle$ behaves as a wave packet in gauge space as
it mixes states belonging to different irreducible representations. The use of such Bogoliubov product states is
at the heart of the symmetry-breaking description of static pairing correlations based on a single reference state. In spite of
the broken symmetry of the product state, all observables that are scalars in gauge space still have to be
independent on its orientation in gauge space. This allows one to choose a convenient angle on the level of SR
calculations that simplifies the calculations, a procedure similar to choosing a major axis system for quadrupole
deformed product states. In the case of gauge symmetry, a convenient orientation is provided by $\varphi = 0$.
States at different angles are obtained from this state applying the rotation operator $e^{i \varphi \hat{N}}$ in
gauge space
\begin{equation}
\label{eq:transfo_phi2}
| \Phi_{\varphi} \rangle
= e^{i \varphi \hat{N}} \, | \Phi_0 \rangle
= e^{i \varphi \hat{N}}
  \prod_{\mu>0} \big( u_{\mu} + v_{\mu} \, a^+_{\mu} \, a^+_{\bar \mu} \big) \, | 0 \rangle
\, .
\end{equation}

\subsection{Energy in the strict HFB approach}

As a strict HFB approach, we denote the method that determines the energetically most favored quasi-particle
vacuum $| \Phi_{\varphi} \rangle$ through the minimization of the expectation value of a given Hamiltonian
$\hat{H}$ in that product state, without any approximations or generalizations. For the sake of transparency, the
Hamiltonian
\begin{eqnarray}
\label{eq:hamil}
\hat{H}
= \sum_{ij} t_{ij} \, c^+_i \, c_j
  + \tfrac{1}{4} \sum_{ijkl} \bar{v}_{ijkl} \, c^+_i \, c^+_j \, c_l \, c_k
\end{eqnarray}
is assumed to be given by the sum of kinetic energy term and a two-body interaction. In Eq.~(\ref{eq:hamil})
$\{c^+_i\}$ defines a complete set of single-particle states whereas $\bar{v}_{ijkl}$ denotes antisymmetric
matrix elements (or vertices) of the two-body interaction in that basis. The discussion below can be extended
without difficulty to a Hamiltonian containing three-body or higher-body forces, but this becomes cumbersome and
is not necessary for the purpose of this paper.

An important point is that in the context of the strict HFB approach, we assume that the vertex $\bar{v}_{ijkl}$
does \emph{not} depend on density. So-called density-dependent vertices of Skyrme and Gogny type are widely used
in the literature. However, as pointed out in Ref.~\cite{doba05a}, discussed in the present paper and insisted
on further in Paper~III, \emph{any} density-dependent effective vertices do provide MR energies with (at least)
spurious finite contributions, even though the vertex is antisymmetric with respect to the remaining single-particle
degrees of freedom and all associated exchange terms are exactly accounted for in the MR energy kernels.

Using the Standard Wick theorem (SWT)~\cite{wick50a,ring80a,blaizot86}, the expectation value of $\hat{H}$ in the
product state $| \Phi_{\varphi} \rangle$ can be evaluated as
\begin{eqnarray}
\label{EHFB0}
\lefteqn{
E \left[\rho^{\varphi\varphi}, \kappa^{\varphi\varphi}, \kappa^{\varphi\varphi \, \ast} \right]
\equiv \frac{\langle \Phi_{\varphi} | \, \hat{H} \, | \Phi_{\varphi} \rangle}
            {\langle \Phi_{\varphi} | \Phi_{\varphi} \rangle}
}
      \nonumber \\
& = & \sum_{\mu} t_{\mu\mu} \, \rho^{\varphi\varphi}_{\mu\mu}
      + \sum_{\mu\nu}
        \Big[ \tfrac{1}{2} \bar{v}_{\mu\nu\mu\nu} \rho^{\varphi\varphi}_{\mu\mu}  \rho^{\varphi\varphi}_{\nu\nu}
            + \tfrac{1}{4} \bar{v}_{\mu\bar{\mu}\nu\bar{\nu}}  \kappa^{\varphi\varphi \, \ast}_{\mu\bar{\mu}}
                                                               \kappa^{\varphi\varphi}_{\nu\bar{\nu}}
        \Big]
      \nonumber \\
& = & \sum_{\mu} t_{\mu\mu} v^{2}_{\mu}
      + \sum_{\mu\nu}
        \Big[ \tfrac{1}{2} \bar{v}_{\mu\nu\mu\nu} \, v^{2}_{\mu} v^{2}_{\nu}
            + \tfrac{1}{4} \bar{v}_{\mu\bar{\mu}\nu\bar{\nu}} \, u_{\mu} v_{\mu} \, u_{\nu} v_{\nu}
        \Big]
      \nonumber \\
\end{eqnarray}
where $\rho^{\varphi\varphi}$ and $\kappa^{\varphi\varphi}$ are the normal density matrix and anomalous density
matrix (pairing tensor) constructed from $| \Phi_{\varphi} \rangle$, respectively. In the canonical basis of the Bogoliubov transformation defining $| \Phi_{\varphi} \rangle$, these
take the simple form
\begin{eqnarray}
\rho^{\varphi\varphi}_{\mu \nu}
& \equiv & \frac{\langle \Phi_{\varphi} | a^{\dagger}_{\nu} a_{\mu}
            | \Phi_{\varphi} \rangle}
           {\langle \Phi_{\varphi} | \Phi_{\varphi} \rangle}
  =   v^{2}_{\mu} \,
      \delta_{\mu \nu} \, ,
      \label{intrdens1} \\
\kappa^{\varphi\varphi}_{\mu \nu} & \equiv & \frac{\langle
\Phi_{\varphi} | a_{\nu} a_{\mu}
            | \Phi_{\varphi} \rangle}
           {\langle \Phi_{\varphi} | \Phi_{\varphi} \rangle}
  =   u_{\mu} v_{\mu} \, e^{2i\varphi} \,
\delta_{\nu\bar{\mu}} \, ,
      \label{intrdens2}  \\
\kappa^{\varphi\varphi \, \ast}_{\mu \nu} & \equiv & \frac{\langle \Phi_{\varphi} | a^{\dagger}_{\mu} a^{\dagger}_{\nu}
            | \Phi_{\varphi} \rangle}
           {\langle \Phi_{\varphi} | \Phi_{\varphi} \rangle}
  =    u_{\mu} v_{\mu} \, e^{-2i\varphi} \, \delta_{\nu\bar{\mu}} \, .
\label{intrdens3}
\end{eqnarray}
The expectation value given in Eq.~(\ref{EHFB0}) can be seen as a particular functional of
$\rho^{\varphi\varphi}$, $\kappa^{\varphi\varphi}$ and $\kappa^{\varphi\varphi \, \ast}$. The symmetries of the
Hamiltonian lead of course to a number of specific properties of this functional. In particular, since the
Hamiltonian commutes with the particle-number operator, one finds that
\begin{eqnarray}
E \left[\rho^{\varphi\varphi}, \kappa^{\varphi\varphi},
\kappa^{\varphi\varphi \, \ast} \right] =
E \left[\rho^{00}, \kappa^{00}, \kappa^{00 \, \ast} \right]
\, ,
\label{eq:invare}
\end{eqnarray}
which underlines that all states that differ only by a rotation in gauge space are degenerate. In other words, the
energy functional behaves as a scalar in gauge space as expected.

%
%
\subsection{Energy in the SR energy functional approach}

In nuclear physics, strict HFB-type approaches are frequently applied in a restricted shell-model space using
parametrized single-particle energies and an
effective Hamiltonian as a residual interaction~\cite{mang75a,ring70a,Goo79a}. For a multitude of reasons outlined
in Paper~I and references given therein, methods using the full model space of occupied particles had to resume so
far to the use of (phenomenological) density-dependent effective interactions~\cite{vautherin72a,decharge80a},
which sets the stage for what is nowadays recognized as an approximation to a more general single-reference EDF
formalism. This framework shares many features with the Density Functional Theory (DFT) widely used for
description of electronic many-body systems
\cite{hohenberg64,dreizlerBook,parrBook,lecturenotesFNM,koch01,Nagy98DFT,kohn98}, but also displays key differences,
which prohibit the straightforward mapping of all concepts of electronic DFT to the nuclear case
\cite{engel06b,Giraud07a,Giraud07b}.

In the DFT for many-electron systems, constructive schemes have been established to design the
energy functional, see for instance Ref.~\cite{lecturenotesFNM} and references given therein. In nuclear physics, such
a procedure that would suggest the structure of the functional is still missing, already on a qualitative level.
The reasons are the complexity of the nucleon-nucleon interaction on the one hand, and that in-medium correlations
are never small corrections on the other hand. In the absence of a constructive scheme, all widely used nuclear energy functionals were set up keeping an underlying two-body and sometimes three-body interaction as
guiding principle, making generalizations suggested by phenomenology and approximating or even omitting terms
that are small, but difficult to evaluate. As a consequence, the structure of these functionals resembles that of
Eq.~(\ref{EHFB0}), except that the expectation value $E \left[\rho, \kappa, \kappa^{\ast} \right]$ is replaced by a
functional $\mathcal{E} \left[\rho, \kappa, \kappa^{\ast} \right]$. Considering the simple case of a
bilinear functional for simplicity and comparison purposes, such a functional can be written as
\begin{eqnarray}
\label{eq:e00}
\mathcal{E}\left[\rho, \kappa, \kappa^{\ast} \right]
& \equiv & \mathcal{E}^{\rho} + \mathcal{E}^{\rho\rho} + \mathcal{E}^{\kappa\kappa}
      \nonumber \\
& = & \sum_{\mu} t_{\mu\mu} v^{2}_{\mu}
      + \tfrac{1}{2} \sum_{\mu\nu} \, \bar{v}^{\rho\rho}_{\mu\nu\mu\nu} \,
        v^{2}_{\mu} \, v^{2}_{\nu}
      \nonumber \\
&   & + \tfrac{1}{4} \sum_{\mu\nu} \, \bar{v}^{\kappa\kappa}_{\mu\bar{\mu}\nu\bar{\nu}} \,
      u_{\mu} \, v_{\mu} \, u_{\nu} \, v_{\nu}
\, .
\end{eqnarray}
This might appear as an unsusual way to write standard energy functional, but will turn out to be very useful below.
The corresponding explicit expressions for a Skyrme energy functional are given in Appendix~\ref{app:functional}. The crucial
point for our discussion is that the matrix elements of the effective vertex $\bar{v}^{\rho\rho}$ are in general
not necessarily antisymmetric for these energy functionals. Also, for Skyrme functionals, one almost always
chooses different vertices in the particle-hole ($\bar{v}^{\rho\rho}_{\mu\nu\mu\nu}$) and particle-particle
($\bar{v}^{\kappa\kappa}_{\mu\bar{\mu}\nu\bar{\nu}}$) channels, and exploits broken antisymmetry of
$\bar{v}^{\rho\rho}_{\mu\nu\mu\nu}$ to obtain a more versatile effective interaction, for example in the
spin-orbit~\cite{reinhard95a,chabanat98} or spin-spin parts~\cite{bender02a}. The situation is similar for the
functionals by Fayans \etal\ \cite{fayans}. By contrast, the philosophy of the Gogny force is to use the same
antisymmetrized density-dependent vertex anywhere, although in actual calculations terms that are very small in SR
calculations and at the same time very time-consuming to evaluate are often omitted~\cite{anguiano01a}. As all
standard parameterizations of the Skyrme and Gogny interactions use density-dependent vertices, they cannot be
mapped on a functional that is the strict expectation value of a many-body Hamiltonian~(\ref{EHFB0}). Almost all
relativistic mean-field models that are widely used in the literature are explicitly set up as Hartree approaches
\cite{bender03b,vretenar05a} without any explicit exchange terms at all, using phenomenological density
dependencies and non-relativistic pairing energy functionals.

Note that \emph{any} local or non-local energy functional that contains only terms proportional to integer powers of the
density matrices can be put into the form of Eq.~(\ref{eq:e00}) plus similar higher-order terms. For the rest of
this paper, however, we will assume idealized energy functionals that are linear and bilinear in the density
matrix of a given isospin projection, and possibly trilinear with the two isospin projections necessarily involved.
We postpone the discussion of functionals with non-integer powers of the density matrices to Paper~III.

We will not assume antisymmetry of $\bar{v}^{\rho\rho}$ in the formal manipulations throughout the paper. Owing to
the intrinsic antisymmetry of $\kappa$, however, only the antisymmetric part of the vertex is probed in the last term of
Eq.~(\ref{eq:e00}) and one can always take $\bar{v}^{\kappa\kappa}$ to be antisymmetric, which we do here.
The results based on a strict HFB method can always be easily recovered from those derived for a more general
bilinear functional simply by enforcing the antisymmetry of $\bar{v}^{\rho\rho}$ and by taking $\bar{v}^{\rho\rho}
= \bar{v}^{\kappa\kappa} = \bar{v}$.

\section{Particle number restoration}
\label{projectedmeanfield}

In order to restore good particle number and include the correlations associated with the corresponding
Nambu-Goldstone mode, it is necessary to extend the EDF
framework to a multi-reference formalism. This extension requires the explicit treatment of the fluctuations of
the gauge angle of the gap field. This is particularly crucial for situations where the symmetry breaking is weak
or even absent at the SR level, as it is the case for instance around closed shells or at high spin. The Variation
After Projection (VAP) method~\cite{sheikh00,anguiano01b,anguiano02a,sheikh02,stoitsov07,Rodriguez07a} is superior
in that respect to the Projection After Variation (PAV) one since the latter cannot compensate for the spurious
sharp phase transition occurring at the SR level in the weak symmetry-breaking
regime~\cite{ring80a,anguiano01b,anguiano02a,sheikh02,stoitsov07}. An intermediate treatment consists of performing a
projection after a SR+Lipkin-Nogami (HFBLN) calculation~\cite{heenen93a,anguiano02a,samyn04,niksic06b}. This
corrects for the principal defect of the PAV method as it guarantees the presence of pairing
correlations in the SR state in the weak-pairing regime. However, some doubts have been raised in the literature about the
quantitative reliability of this method~\cite{anguiano02a,Rodriguez05a}. The MR calculation could be extended
further to incorporate dynamical pairing correlations associated with fluctuations of the magnitude of an order
parameter that quantifies the amount of pairing correlations present in the SR state
\cite{Ripka69a,Siegal72a,Rodriguez05a,Bender06k}.

An operator that projects out an eigenstate of the particle number operator $\hat{N}$ with an eigenvalue $N$ from
any many-body wave function is provided by~\cite{Bay60a}
\begin{equation}
\label{eq:Pop}
\hat{P}^N
=  \frac{1}{2\pi} \int_{0}^{2\pi} \! d{\varphi} \; \,e^{i\varphi (\hat{N}-N)}
\, .
\end{equation}
For the purpose of the present paper, it is sufficient to consider the simple case of
particle number restoration after variation. For the sake of transparent notation
we discuss the formal framework assuming one type of particles only. The
extension to two types of particles is straightforward and will be mentioned only
whenever necessary. A normalized projected HFB state is given by
\begin{equation}
\label{PWF}
| \Psi^N \rangle
= \int_{0}^{2\pi} \! d{\varphi} \;
      \frac{e^{ - i \varphi N}}{2\pi \, c_{N}} \, | \Phi_{\varphi} \rangle
\, ,
\end{equation}
where the real and positive $c_{N}=\langle \Phi_0 | \Psi^N \rangle$ that reads
\begin{eqnarray}
\label{weight}
c^{2}_{N}
& = & \langle \Phi_0 | \hat{P}^N | \Phi_0 \rangle = \frac{1}{2\pi}  \int_{0}^{2\pi} \! d\varphi \, e^{-i \varphi N} \,
      \langle  \Phi_0 | \Phi_{\varphi} \rangle
\end{eqnarray}
provides the weight of the normalized projected state in the normalized SR state it is
projected from, whereas
\begin{equation}
\langle  \Phi_0 | \Phi_{\varphi} \rangle
= \prod_{\mu>0} \big( u_{\mu}^2 + v_{\mu}^2 \, e^{2 i \varphi} \big)
\end{equation}
denotes the overlap of a gauge-space rotated state with the unrotated one. The integration interval in Eq.~(\ref{PWF})
can be reduced to $[0,\pi]$ using symmetries of the integral whenever the SR state $| \Phi_{\varphi} \rangle$ has
a good number parity quantum number~\cite{ring70a,banerjee73a,ring80a}.

%
%
\subsection{Energy in the strict PNP-HFB approach}

In the strict PNP-HFB method, the energy is calculated as the expectation value of the Hamilton
operator in the normalized projected state $|\Psi^N \rangle$
\begin{equation}
\label{scalar6}
E^{N}
= \langle \Psi^{N} | \, \hat{H} \, | \Psi^{N} \rangle
=  \int_{0}^{2\pi} \!\!\! d\varphi \, \frac{e^{-i\varphi N}}{2\pi \, c^{2}_{N}} \,
   E[\varphi] \,  \langle  \Phi_0 | \Phi_{\varphi} \rangle
\, ,
\end{equation}
where we have used that $\hat{H}$ and $\hat{N}$ commute and that $\hat{P}^N$ is a projector $\hat{P}^N \hat{P}^N =
\hat{P}^N$. The energy kernel $E[\varphi]$ can be easily evaluated with the help of the Generalized Wick
Theorem (GWT)~\cite{onishi66,balian69a}, which in the canonical basis of $| \Phi_{0} \rangle$ gives
\begin{eqnarray}
\label{extwick} E[\varphi] & \equiv & \frac{\langle \Phi_{0} | \, \hat{H} \, | \Phi_{\varphi} \rangle}
                {\langle \Phi_{0} | \Phi_{\varphi} \rangle}
      \nonumber \\
& = & \sum_{\mu} t_{\mu\mu} \, \rho^{0\varphi}_{\mu\mu}
      + \tfrac{1}{2} \sum_{\mu\nu} \bar{v}_{\mu\nu\mu\nu}\, \rho^{0\varphi}_{\mu\mu} \, \rho^{0\varphi}_{\nu\nu}
      \nonumber \\
&   & \phantom{ \sum_{\mu} t_{\mu\mu} \, \rho^{0\varphi}_{\mu\mu} }
      + \tfrac{1}{4} \sum_{\mu\nu} \bar{v}_{\mu\bar{\mu}\nu\bar{\nu}} \,
        \kappa^{\varphi 0 \, \ast}_{\mu\bar{\mu}} \, \kappa^{0\varphi}_{\nu\bar{\nu}}
\, .
\end{eqnarray}
In this expression, the normal and anomalous \emph{transition} density matrices between the ket $| \Phi_{\varphi}
\rangle$ and the bra $\langle \Phi_{0} |$ are defined as
\begin{eqnarray}
\label{contractph}
\rho^{0\varphi}_{\mu \nu}
& \equiv & \frac{\langle \Phi_{0} | a^{\dagger}_{\nu} a_{\mu} | \Phi_{\varphi} \rangle}
           {\langle \Phi_{0} | \Phi_{\varphi} \rangle}
  =   \frac{v_{\mu}^2 \, e^{2 i \varphi} }
           {u_\mu^2 + v_{\mu}^2 \, e^{2 i\varphi} }  \, \delta_{\nu \mu} \, ,
      \\
\label{contracthh}
\kappa^{0\varphi}_{\mu \nu}
& \equiv & \frac{\langle \Phi_{0} | a_{\nu} a_{\mu} | \Phi_{\varphi} \rangle}
           {\langle \Phi_{0} | \Phi_{\varphi} \rangle}
  =   \frac{u_\mu v_{\mu} e^{2 i \varphi} }
           {u_\mu^2 + v_{\mu}^2 \, e^{2 i \varphi} } \, \delta_{\nu \bar{\mu}} \, ,
      \\
\label{contractpp}
\kappa^{\varphi 0 \, \ast}_{\mu \nu}
& \equiv & \frac{\langle \Phi_{0} | a^{\dagger}_{\mu} a^{\dagger}_{\nu} | \Phi_{\varphi} \rangle}
           {\langle \Phi_{0} | \Phi_{\varphi} \rangle}
  =   \frac{u_\mu v_{\mu}}{u_\mu^2 + v_{\mu}^2 \, e^{2 i \varphi} } \, \delta_{\nu \bar{\mu}} \, .
\end{eqnarray}
The functional kernel $E[\varphi]$ defined by Eq.~(\ref{extwick}) has the exact same form as the strict HFB energy
functional $E[\rho,\kappa,\kappa^{\ast}]$ given by Eq.~(\ref{EHFB0}) except that the SR density matrix and pairing
tensor~(\ref{intrdens1}-\ref{intrdens3})  have been replaced by the transition ones~(\ref{contractph}-\ref{contractpp}).
Also, the HFB functional is recovered from Eq.~(\ref{extwick}) for $\varphi = 0$, which amounts to connecting the
SR energy and MR energy kernels through $E[0] = E[\rho,\kappa,\kappa^{\ast}]$.

\subsection{Energy in the PNR energy functional approach}
\label{sect:efunc:pnr}

Difficulties arise when trying to construct the multi-reference energy kernel $\mathcal{E}[\varphi]$ within a true
functional framework and connect it to the single-reference one. At present, there is no ab-initio formalism to
derive MR energy functional kernels, of which the SR functional would be a special case, and one can only reverse
engineer the procedure and extend the SR energy density functional to the MR level by analogy with the strict
Hamiltonian case. Based on the strict HFB and
PNP-HFB methods described above, EDF practitioners have used a procedure where $\mathcal{E}[\varphi] \equiv  \mathcal{E}
[\rho^{0\varphi}, \kappa^{0\varphi}, \kappa^{\varphi 0 \, \ast}]$ is postulated to be the MR energy kernel that
corresponds to a given SR functional~\cite{heenen93a,anguiano01b,niksic06b,Robledo07a,doba05a}. In this case, the
MR energy corresponding to particle number restoration takes the form
\begin{equation}
\label{scalar2}
\mathcal{E}^{N}
\equiv \int_{0}^{2\pi} \! d\varphi \,
       \frac{e^{-i\varphi N}}
               {2\pi \, c^{2}_{N}} \,
       \mathcal{E}[\varphi] \,  \langle  \Phi_0 | \Phi_{\varphi} \rangle
\, ,
\end{equation}
where $\mathcal{E}[\varphi]$ denotes the set of MR energy functional kernels associated with each gauge angle
$\varphi$. A kernel $\mathcal{E}[\varphi]$ is a functional of the bra $\langle \Phi_0 |$ and of the ket
$| \Phi_{\varphi} \rangle$, in such a way that $\mathcal{E}^{N}$ depends only implicitly on the projected
state~\cite{duguet06b} and cannot be factorized into a form similar to the left-hand side of Eq.~(\ref{scalar6}).
We will call this procedure the "use of the GWT" below, although strictly speaking it is not the GWT that is
applied, but a formal analogy to the extension at play in the strict Hamiltonian case when using the GWT that is
exploited.

On the one hand, the standard strategy based on the GWT analogy to define the non-diagonal functional energy kernel
$\mathcal{E}[\varphi]$ from the single-reference functional replacing SR density matrices by the transition ones
guarantees that the MR energy functional passes all consistency requirements thought of so far~\cite{Robledo07a}.
On the other hand, this procedure is also at the origin of the divergences and finite steps discussed in
Ref.~\cite{anguiano01b,doba05a}. In Paper~I we proposed the general formalism appropriate for a remedy of these
problems. The remedy is valid for any type of multi-reference calculation but is limited to EDFs depending on
integer powers of the density matrices as is further elaborated on in Paper~III. The goal of the following sections
is to discuss the origin of the problem further and to illustrate the general regularization procedure in its
application to PNR.

We note in passing that in PNR and all other MR-EDF calculations the energy is the only observable that is
currently determined from a functional; all other observables that are routinely calculated
within such an approach are obtained as matrix elements of the corresponding operator between projected states,
such that they do not contain spurious contributions.

%
%

\section{Self-interaction and self-pairing}
\label{selfenergyselfpairing1}

%
%
\subsection{Single-Reference level}
\label{selfenergyselfpairingHFB}

\subsubsection{Self-interaction}
\label{sect:selfenergyselfpairingHFB:self-energy}

Microscopic methods for low-energy nuclear structure physics usually describe a self-bound nucleus in terms of
nucleons characterized by their experimental mass. In such an approach, a nucleon should not gain energy by
interacting with itself. Its so-called \emph{self-interaction} energy, which can be extracted from the one-orbital
limit of the interaction part of the energy functional $\mathcal{E}_{\mu} \equiv \mathcal{E}
\left[\rho^{\varphi\varphi}_{\mu\mu},0,0\right]$ in the canonical basis, should be strictly zero. This
requirement is, however, not fulfilled for most functionals used in electronic
DFT~\cite{perdew81a,lecturenotesFNM,koch01,Ull00aDFT,Leg02aDFT,Ruz07aDFT} or nuclear EDF
methods~\cite{stringari78a}. Energy functionals with higher-order density dependencies than those
discussed here might also exhibit multi-particle self-energies, not having the proper $n$-particle
limit of the energy functional~\cite{Ruz07aDFT}.

Let us consider the energy $\mathcal{E}_{\mu}$ of a single Fermion occupying
the canonical state $\phi_{\mu}$, divided by the probability
$\rho^{\varphi\varphi}_{\mu\mu} = v^{2}_{\mu}$ of this state to be
occupied in the auxiliary state $| \Phi_{0} \rangle$
\begin{equation}
\label{onefermion}
\frac{\mathcal{E}_{\mu}}{v^{2}_{\mu}}
=   t_{\mu \mu}
  + \tfrac{1}{2} \, \bar{v}^{\rho\rho}_{\mu \mu \mu \mu} \, v^2_\mu
\, .
\end{equation}
This expression shows that a self-interaction arises whenever the vertex $\bar{v}^{\rho\rho}$ is not
antisymmetric, $\bar{v}^{\rho\rho}_{\mu \mu \mu \mu} \neq 0$, which is impossible when calculating the exact
matrix element of a Hamilton operator, but happens for general energy density functionals. The total one-body
self-interaction energy is obtained summing all individual contributions $\mathcal{E}_{\mu}$.

\subsubsection{Self-pairing}

Beyond the well-known problem of spurious self-interactions, there exists a similar problem of spurious
\emph{self-pairing} processes which may arise whenever superfluidity is incorporated into an energy functional in
a DFT or EDF framework. The rationale behind it is that two Fermions occupying a pair of conjugated states should
not gain extra binding through the pairing interaction by scattering onto themselves. This requirement constrains
the two-particle limit of the theory and the contribution of a conjugated pair to the many-body energy. To the
best of our knowledge, the possibility of self-pairing has never been addressed before.

Self-pairing can be easily identified when isolating the energy of two Fermions occupying a pair of conjugated
states $\{\phi_{\mu}, \phi_{\bar{\mu}}\}$ in the canonical basis. We define the \emph{direct} interaction energy
of such a pair
by removing the one-body contributions defined through Eq.~(\ref{onefermion}) to $\mathcal{E}_{\mu\bar{\mu}}
\equiv \mathcal{E} \left[\left\{\rho^{\varphi\varphi}_{\mu\mu},\rho^{\varphi\varphi}_{\bar{\mu}\bar{\mu}}\right\},
\left\{\kappa^{\varphi\varphi}_{\mu\bar{\mu}},\kappa^{\varphi\varphi}_{\bar{\mu}\mu}\right\},
\left\{\kappa^{\varphi\varphi \, \ast}_{\mu\bar{\mu}},\kappa^{\varphi\varphi \,
\ast}_{\bar{\mu}\mu}\right\}\right]$ and by dividing the result by the probability $P^{\Phi}_{\mu\bar{\mu}}$ to
occupy the pair in the auxiliary state $| \Phi_{0} \rangle$
\begin{equation}
\label{enerpair}
\frac{\mathcal{E}_{\mu\bar{\mu}} - \mathcal{E}_{\mu} - \mathcal{E}_{\bar{\mu}}}
     {P^{\Phi}_{\mu\bar{\mu}}}
= \tfrac{1}{2}
  \left( \bar{v}^{\rho\rho}_{\mu \bar\mu \mu \bar\mu}
        +\bar{v}^{\rho\rho}_{\bar\mu \mu \bar\mu \mu}
  \right) \, v^2_\mu
  +\bar{v}^{\kappa\kappa}_{\mu \bar\mu \mu \bar\mu} \, u^2_{\mu}
\, .
\end{equation}
The probability $P^{\Phi}_{\mu\bar{\mu}}$ to occupy the pair
\begin{equation}
\label{twobodypropa}
P^{\Phi}_{\mu\bar{\mu}}
\equiv  \frac{\langle \Phi_{\varphi} | a^{\dagger}_{\mu} a^{\dagger}_{\bar{\mu}}
                                  a_{\bar{\mu}} a_{\mu} |\Phi_{\varphi} \rangle}
        {\langle \Phi_{\varphi} | \Phi_{\varphi} \rangle}
= v^{2}_{\mu}
\end{equation}
is equal to the probability of each state to be occupied, which is a particularity of fully paired
quasiparticle vacua, Eq.~(\ref{eq:transfo_phi}). In the strict HFB case where
$\bar{v}^{\rho\rho}_{\mu \bar\mu \mu \bar\mu}
 = \bar{v}^{\rho\rho}_{\bar\mu \mu \bar\mu \mu}
 = \bar{v}^{\kappa\kappa}_{\mu \bar\mu \mu \bar\mu}
 \equiv \bar{v}_{\mu \bar\mu \mu \bar\mu}$,
the two terms on the r.h.s.\ of Eq.~(\ref{enerpair}) combine into
\begin{equation}
\label{eq:no:self}
\frac{E_{\mu\bar{\mu}} - E_{\mu} - E_{\bar{\mu}}}
     {P^{\Phi}_{\mu\bar{\mu}}}
= \bar{v}_{\mu \bar\mu \mu \bar\mu}
\, ,
\end{equation}
using $u_\mu^2 + v^2_{\mu} = 1$. The same result is obtained in a strict HF method without explicit treatment of
pairing correlations. The equality of the two-body interaction energy~(\ref{eq:no:self}) in the HF and HFB case
means that a conjugated pair of states  $\{\mu,\bar{\mu}\}$ does not gain extra \emph{direct} binding by
scattering onto itself. Genuine pairing correlations originate from scattering to different pairs of conjugated
states and back.

For most of the standard SR energy density functionals used for nuclear structure calculations,
however, the three terms in Eq.~(\ref{enerpair}) can in general not be
recombined into a single one because the vertices entering $\mathcal{E}^{\rho\rho}$ and
$\mathcal{E}^{\kappa\kappa}$ are not related, either by construction or due to approximations.
Consequently,
the direct interaction energy of the conjugated pair is not equal to its zero-pairing limit
as it should be, which gives rise to a spurious self-pairing interaction where one has
a contribution to the energy functional from the scattering of a pair of conjugated
states onto itself.

\subsubsection{Further discussion}

In a composite system consisting of two particle species such as atomic nuclei, the like-particle self-interaction
for a given particle species is obtained as the one-particle limit of the interaction energy for this particle species,
while keeping the particle number of the other particle species unchanged. Otherwise self-interactions in the terms that
couple the two particle species will be missed.

The existence of spurious self-interactions was first recognized in Kohn-Sham DFT for electronic systems
\cite{perdew81a}. In this context, the construction of self-interaction-free functionals has been studied in some
detail, see Refs.~\cite{perdew81a,koch01,Ull00aDFT,Leg02aDFT,Ruz07aDFT} and references given therein. It turns out to be
not trivial at all knowing that the standard correction method is formulated within the frame of so-called
orbital-dependent energy density
functionals~\cite{engel03a,kummel08a} and significantly complexifies the calculations through the modification of both the
total energy and the single-particle equations of motion. The (unknown) exact Hohenberg-Kohn functional of DFT is of course
self-interaction free. The spurious terms arise when constructing approximate energy functionals that are
tractable for the use in actual calculations; i.e.\ self-interaction is one of the prices to pay for replacing the exact
many-body problem by a much simpler set of coupled one-body problems. It is of course desirable to work within a
theory that conserves the Pauli principle, but its restoration is mandatory only when its violation affects
observables of interest on a scale comparable with or larger than the precision desired and reachable within a
given method. The situation is thus similar to the necessity to restore other broken symmetries. As a matter
of fact, the merits of
self-interaction corrected energy functionals for electronic DFT are still debated from a phenomenological point
of view, as they improve some observables, but at the same time degrade others when compared to uncorrected
functionals; see Ref.~\cite{Ruz07aDFT} and references given therein.

The same remarks apply to self-pairing. Both self-interaction and self-pairing processes are actually rooted in
a violation of the Pauli principle at the level of the two-body (or even higher-order) density matrix in the
definition of the energy functional. It is important to stress that they are solely a shortcoming of common energy
functionals and not of the auxiliary states of reference used, as the latter are set up as antisymmetrized product
states. In particular, all observables other than the energy, which are customarily calculated as expectation
values of the corresponding operators, do not exhibit any explicit spurious contributions, although they
might be indirectly affected through the use of density matrices that are determined from the solution of a
variational equation that uses an energy functional containing spurious contributions as an input.

In the nuclear context, the possible contamination of nuclear energy density functionals by spurious self-energies
has been noticed before~\cite{stringari78a,Rutz99a,BenReiDisc99,bender03b}, but was never studied in quantitative
detail so far.

It has to be stressed that using self-interaction and self-pairing free energy functionals is not \emph{per se}
equivalent to the use of an effective Hamilton operator. Indeed, self-interaction, as usually characterized,
and self-pairing, as presently defined, probe only the exchange
symmetry of a particle in the canonical basis with itself and its conjugate partner, not the exchange symmetry
between all particles. Asking for a full restoration of the Pauli principle necessarily leads to using a
genuine Hamilton operator~\cite{stringari78a}.

\subsection{Multi-Reference level}
\label{selfenergyselfpairing2}

The appearance of self-interaction and self-pairing processes persists to MR calculations whereas new spurious
contributions particular to the MR level arise from the construction of non-diagonal energy kernels. The extension
of the self-interaction
and self-pairing concepts to the multi-reference framework, however, is not at all straightforward. For instance
the very notion of "occupied" orbitals is ill-defined for transition density matrices between arbitrary quasiparticle
vacua. In the case of
particle-number restoration, the situation is significantly simplified owing to the fact that all vacua entering
the PNR energy~(\ref{scalar2}) share the same canonical single-particle basis, which consequently also is
the canonical basis of the Bogoliubov transformation linking any pair of these vacua. As demonstrated in
Paper~I it is precisely the latter canonical basis of the transformation connecting a given pair of mixed
vacua that must be used to meaningfully identify self-interaction and
self-pairing contributions to the corresponding multi-reference energy kernel.

\subsubsection{"Naive" extension of self-interaction}

In the context of PNR multi-reference calculations, the energy of a single Fermion
occupying the canonical orbital $\phi_{\mu}$ divided by the probability
$\rho^{\Psi^{N}}_{\mu\mu}$ to occupy that orbital in the projected state
$| \Psi^{N} \rangle$ is given by
\begin{widetext}
\begin{equation}
\label{onefermionPNR}
\frac{\mathcal{E}^{N}_{\mu}}
     {\rho^{\Psi^{N}}_{\mu\mu}}
= t_{\mu \mu}
  + \tfrac{1}{2} \, \bar{v}^{\rho\rho}_{\mu \mu \mu \mu} \,
    \frac{1}{\rho^{\Psi^{N}}_{\mu\mu}}
    \int_{0}^{2\pi} d\varphi \, \frac{e^{-i\varphi N}}{2\pi \, c^{2}_{N}}
    \frac{v^{4}_{\mu} \, e^{4 i \varphi}}
         {u_\mu^2 + v_{\mu}^2 \, e^{2 i \varphi}}  \,
     \prod_{\nu > 0 \atop \nu \neq \mu}  (u_{\nu}^2 + v_{\nu}^2 \, e^{2 i \varphi})
\, .
\end{equation}

The one-body density matrix $\rho^{\Psi^{N}}$ of the projected state
\begin{eqnarray}
\label{probaonefermionPNR}
\rho^{\Psi^{N}}_{\mu\mu}
& \equiv & \frac{\langle \Psi^{N} | a^{\dagger}_{\mu} a_{\mu}| \Psi^{N} \rangle}
           {\langle \Psi^{N} | \Psi^{N} \rangle}
      = \int_{0}^{2\pi} \! \! d\varphi \,
      \frac{e^{-i\varphi N}}
           {2\pi \, c^{2}_{N}} \, \rho^{0\varphi}_{\mu\mu} \;
      \langle \Phi_0 | \Phi_\varphi \rangle
      = v^{2}_{\mu} \int_{0}^{2\pi} \! \! d\varphi \,
      \frac{e^{-i\varphi N}}
           {2\pi \, c^{2}_{N}} \,
      e^{2 i \varphi}
      \prod_{\nu > 0 \atop \nu \neq \mu}  ( u_{\nu}^2 + v_{\nu}^2 \, e^{2 i \varphi} )
\, ,
\end{eqnarray}
\end{widetext}
is diagonal in the canonical basis of the HFB state it is projected from, which means that the canonical basis of
the underlying HFB state is also the natural basis of the projected one.

As for the SR case, the energy~(\ref{onefermionPNR}) reduces to kinetic energy when antisymmetric
vertices $\bar{v}^{\rho\rho}$ are used. However, an important aspect specific to the MR case is that
the integrand appearing in Eq.~(\ref{onefermionPNR}) contains a potential (simple) pole for
$\varphi = \pi /2$ and $v_\mu^2 = u_{\mu}^2 = 1/2$, i.e.\ when the state $\mu$ is located at
the Fermi level and is not more than twofold degenerate in terms of occupation numbers $v_\mu^2$.
If the states present a higher degree of degeneracy, an additional factor in the norm overlap
will cancel out the dangerous denominator.

\subsubsection{"Naive" extension of self-pairing}

In multi-reference EDF calculations, the direct interaction energy of a
conjugated pair as defined above takes the form
\begin{widetext}
\begin{equation}
\label{enerpairproj1}
\frac{\mathcal{E}^{N}_{\mu\bar{\mu}} - \mathcal{E}^{N}_{\mu}
      - \mathcal{E}^{N}_{\bar{\mu}}}
     {P^{\Psi^{N}}_{\mu\bar{\mu}}}
= \int_{0}^{2\pi} d\varphi \,
  \frac{e^{-i\varphi N}}
       {2\pi \, c^{2}_{N} \, P^{\Psi^{N}}_{\mu\bar{\mu}}}
  \Big[ \tfrac{1}{2}
         \left( \bar{v}^{\rho\rho}_{\mu \bar{\mu} \mu \bar{\mu}}
               +\bar{v}^{\rho\rho}_{\bar{\mu} \mu \bar{\mu} \mu}
         \right) \, v^{2}_{\mu} e^{2 i \varphi}
        +\bar{v}^{\kappa\kappa}_{\mu\bar{\mu} \mu \bar{\mu}} \, u^{2}_{\mu}
  \Big] \,
  \frac{v^{2}_{\mu} \, e^{2 i \varphi}}
       {u_\mu^2 + v_{\mu}^2 \,
  e^{2 i \varphi}} \,
  \prod_{\nu > 0 \atop \nu \neq \mu} (u_{\nu}^2 + v_{\nu}^2 \, e^{2 i \varphi})
\, ,
\end{equation}
\end{widetext}
where
\begin{equation}
\label{densitymatrix}
P^{\Psi^{N}}_{\mu\bar{\mu}}
=   \frac{\langle \Psi^{N}| a^{\dagger}_{\mu} a^{\dagger}_{\bar{\mu}} a
                            _{\bar{\mu}} a_{\mu}
   | \Psi^{N} \rangle}{\langle \Psi^N | \Psi^N \rangle}
= \rho^{\Psi^{N}}_{\mu\mu}
\end{equation}
is the occupation probability of the pair ($\mu,\bar{\mu}$) in the projected HFB state. The probability
$P^{\Psi^{N}}_{\mu\bar{\mu}}$ is equal to the probability $\rho^{\Psi^{N}}_{\mu\mu}$ of each
state to be occupied as we assume the underlying SR state to be a fully-paired quasiparticle vacuum
with even number parity.

Using a genuine Hamilton operator, for which
$\bar{v}^{\rho\rho}_{\mu \bar\mu \mu \bar\mu}
 = \bar{v}^{\rho\rho}_{\bar\mu \mu \bar\mu \mu}
 = \bar{v}^{\kappa\kappa}_{\mu \bar\mu \mu \bar\mu}
 \equiv \bar{v}_{\mu \bar\mu \mu \bar\mu}$
the matrix elements entering Eq.~(\ref{enerpairproj1}) can be recombined in such a way that the potential pole
disappears~\cite{anguiano01b} and that the zero-pairing limit is again recovered
\begin{eqnarray}
\label{enerpairproj2}
\frac{E^{N}_{\mu\bar{\mu}} - E^{N}_{\mu} - E^{N}_{\bar{\mu}}}
     {P^{\Psi^{N}}_{\mu\bar{\mu}}} = \bar{v}_{\mu \bar{\mu} \mu \bar{\mu}} \, .
\end{eqnarray}
In the EDF formalism, however, the recombination of terms in  Eq.~(\ref{enerpairproj1}) that gives
Eq.~(\ref{enerpairproj2}) cannot be achieved anymore. In this case, the integrand in Eq.~(\ref{enerpairproj1})
contains the same kind of pole as the integrand in Eq.~(\ref{onefermionPNR}).

\subsection{Poles versus "true" self-interaction and self-pairing}
\label{truesisp}

In the previous section, we have shown how the self-interaction and self-pairing persist to the multi-reference
EDF framework in the case of particle-number restoration. What cannot be deduced from such an extension of the
single-reference case, Eqns.~(\ref{onefermion}) and~(\ref{enerpair}), to the multi-reference case,
Eqns.~(\ref{onefermionPNR}) and~(\ref{enerpairproj1}), is if self-interaction and self-pairing processes are
actually responsible for the poles. Indeed, recalling our general analysis of possible spurious terms in MR energy
density functionals from Paper~I, there are in fact two distinct levels of spuriosity contained in
Eqns.~(\ref{onefermionPNR}) and~(\ref{enerpairproj1}), which are of different origins.

The first level is a consequence of using effective vertices that are not antisymmetrized, and/or that are
different on the particle-hole and particle-particle channels. In the MR framework, such spurious
contributions appear in the diagonal energy kernels, which are equivalent to the self-interaction and
self-pairing contributions to the SR energy density functional discussed in
Sec.~\ref{selfenergyselfpairingHFB}, and also enter the off-diagonal kernels. Neither contain poles;
hence they cannot be at the origin of the divergences and steps which are the target of the present work.

In addition to that, a second level of spuriousity arises as a consequence of constructing non-diagonal
energy kernels in analogy with the generalized Wick theorem, although strictly speaking the GWT applies
only to matrix elements of operators. As a matter of fact, and as demonstrated in Paper~I, using a
SWT-motivated procedure rather than a GWT-motivated one does not lead to the second level of spuriosity.
Taking the example of a bilinear EDF, the use of the GWT instead of the SWT gives an additional contribution
of the form
\begin{widetext}
\begin{eqnarray}
\label{spurious:bi}
\mathcal{E}_{CG}^{N}
& \equiv & \int_{0}^{2\pi} \! \! d\varphi \,
      \frac{e^{-i\varphi N}}
           {2\pi \, c^{2}_{N}} \,
      \big(   \mathcal{E}^{\rho\rho}_{CG}[\varphi]
            + \mathcal{E}^{\kappa\kappa}_{CG}[\varphi]
      \big) \, \langle \Phi_{0} | \Phi_{\varphi} \rangle
      \\
& = & \sum_{\mu >0}
      \left[ \tfrac{1}{2}
           \left(  \bar{v}^{\rho\rho}_{\mu \mu \mu \mu}
                  +\bar{v}^{\rho\rho}_{\bar{\mu}\bar{\mu}\bar{\mu}\bar{\mu}}
                  +\bar{v}^{\rho\rho}_{\mu \bar{\mu} \mu \bar{\mu}}
                  +\bar{v}^{\rho\rho}_{\bar{\mu} \mu \bar{\mu} \mu}
           \right)
         - \bar{v}^{\kappa\kappa}_{\mu\bar{\mu}\mu\bar{\mu}}
      \right]
      \left(u_{\mu} v_{\mu}\right)^{4} \int_{0}^{2\pi} \! \! d\varphi \;
      \frac{e^{-i\varphi N}}{2\pi \, c^{2}_{N}} \,
      \frac{\big( e^{2 i \varphi} - 1 \big)^{2}}
           {u_\mu^2 + v_{\mu}^2 \, e^{2 i \varphi}} \,
      \prod_{\nu > 0 \atop \nu \neq \mu} (u_{\nu}^2 + v_{\nu}^2 \, e^{2 i \varphi})
      \nonumber
\end{eqnarray}
that is absent in a SWT-motivated procedure and which contains a pole clearly similar to those discussed
in connection with Eqns.~(\ref{onefermionPNR}-\ref{enerpairproj1}). Having identified the contribution
(\ref{spurious:bi}) caused by the use of the GWT, we defined in Paper~I the \emph{regularized} MR energy
and energy kernels, respectively, as
\begin{eqnarray}
\label{regularized:bi}
\mathcal{E}_{REG}^{N}
& \equiv & \mathcal{E}^{N}-\mathcal{E}_{CG}^{N}  \, ,\\
\mathcal{E}_{REG}[\varphi]
& \equiv & \mathcal{E}[\varphi]-\mathcal{E}_{CG}[\varphi] \, .
\end{eqnarray}
Removing $\mathcal{E}_{CG}^{N}$ from Eqns.~(\ref{onefermionPNR}) and~(\ref{enerpairproj1}),
one obtains the "true" MR self-interaction
\begin{eqnarray}
\label{eq:sitrue}
\mathcal{E}^N_{SI}
& \equiv &  \int_{0}^{2\pi} \! \! d\varphi \,
    \frac{e^{-i\varphi N}}{2\pi \, c^{2}_{N}} \;
   \mathcal{E}^{\rho \rho}_{SI}[\varphi] \, \langle \Phi_{0} | \Phi_{\varphi} \rangle
   \nonumber \\
& = & \sum_{\mu >0}
      \tfrac{1}{2} \left( \bar{v}^{\rho\rho}_{\mu \mu \mu \mu}
                          +\bar{v}^{\rho\rho}_{\bar{\mu}\bar{\mu}\bar{\mu}\bar{\mu}}
                   \right)  \,
      \int_{0}^{2\pi} \! \! d\varphi \,
      \frac{e^{-i\varphi N}}{2\pi \, c^{2}_{N}} \,
      \Big[ v^4_\mu  \left( {u_\mu}^2 + {v_{ \mu}}^2 e^{2i\varphi} \right)
            +  2 \, u^2_\mu v^4_\mu (e^{2i\varphi}-1)
      \Big]
      \prod_{\nu > 0 \atop \nu \neq \mu} (u_{\nu}^2 + v_{\nu}^2 \, e^{2 i \varphi})
\,  .
\end{eqnarray}
and the "true" self-pairing contribution
\begin{eqnarray}
\label{eq:sptrue}
\mathcal{E}^N_{SP}
& \equiv &
      \int_{0}^{2\pi} \! \! d\varphi \,
      \frac{e^{-i\varphi N}}{2\pi \, c^{2}_{N}} \;
       \mathcal{E}^{\kappa \kappa}_{SP}[\varphi] \, \langle \Phi_{0} | \Phi_{\varphi} \rangle
      \nonumber \\
& = & \sum_{\mu >0}
      \Big[  \bar{v}^{\kappa \kappa}_{\mu \bar \mu \mu \bar \mu}
             - \tfrac{1}{2} \big( \bar{v}^{\rho\rho}_{\mu \bar{\mu} \mu\bar{\mu}}
             + \bar{v}^{\rho\rho}_{\bar{\mu} \mu\bar{\mu} \mu} \big)
      \Big]
      \int_{0}^{2\pi} \! \! d\varphi \,
      \frac{e^{-i\varphi N}}
           {2\pi \, c^{2}_{N}} \,
      \Big[   u_\mu^2 \, v_\mu^2 \, \big( {u_\mu}^2 + {v_{ \mu}}^2 e^{2i\varphi} \big)
           + (u^4_\mu v^2_\mu -u^2_\mu v^4_\mu) \big( e^{2i\varphi}-1 \big)
      \Big]
      \nonumber \\
&   & \times
      \prod_{\nu > 0 \atop \nu \neq \mu } \big( u_{\nu}^2 + v_{\nu}^2 \, e^{2 i \varphi} \big)
 \, .
\end{eqnarray}
\end{widetext}
both of which belong to the first level of spuriosity and do not contain any dangerous poles.
The expressions (\ref{eq:sitrue}) and (\ref{eq:sptrue}) could also have been obtained directly from
Eqns.~(79) and~(80) of Paper~I.

\begin{figure}[t!]
\includegraphics{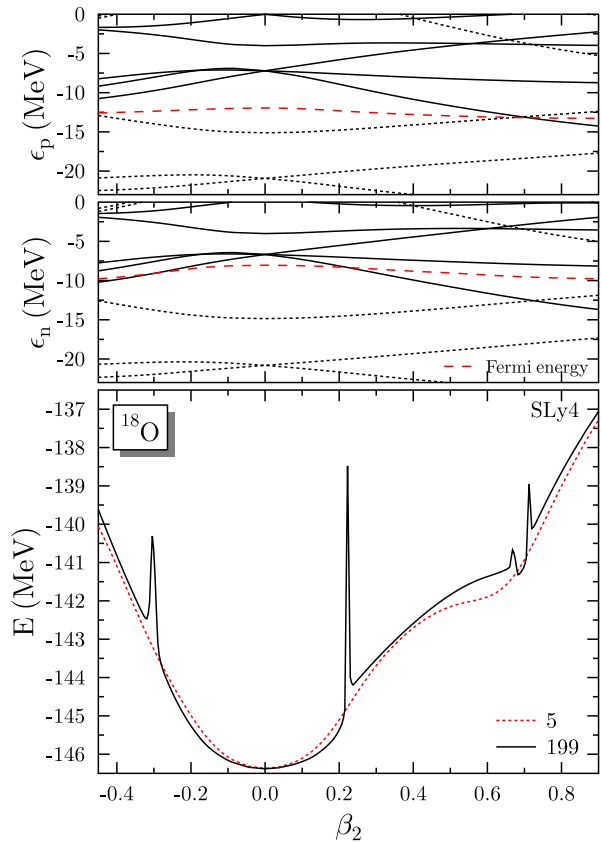}
\caption{\label{divergence}
(Color online)
Particle-number restored deformation energy surface of \nuc{18}{O} calculated with
SLy4 and a density-dependent pairing interaction and the corresponding
single-particle spectra of protons and neutrons as a function of the
axial quadrupole deformation for $L=5$ and 199 discretization
points of the integral over the gauge angle (lowest panel). There are clear
anomalies that appear when either a proton or neutron
single-particle level crosses the Fermi energy. The dimensionless quadrupole
deformation $\beta_2$ is defined in Eq.~(\ref{eq:beta2}).
}
\end{figure}

\subsection{Impact of the poles on PNR energies}

In the previous section, we demonstrated that the spurious contribution $\mathcal{E}^N_{CG}$
contains poles. Figure~\ref{divergence} illustrates, through a realistic calculation of the
particle-number restored deformation energy surface of \nuc{18}{O}, the impact of such poles for a functional
containing a fractional power of the density matrix. The SLy4 parameterization of the standard Skyrme EDF
is used in connection with a density-dependent pairing energy functional, which was used in many MR calculations
before~\cite{bender03a,duguet03c,bender04a,bender06a,bender03c,severyukhin07a,bender06b}. In practice, the
integral over the gauge angle appearing in Eq.~(\ref{scalar2}) is discretized into a sum using the Fomenko
expansion, as will be explained in Sec.~\ref{numerics} below. It is important to stress that all observables
calculated as operator matrix elements, e.g.\ particle number, quadrupole moment, radius, etc., are converged using
five integration points. The particle-number restored energy functional, however, does not converge. Instead, one
observes the development of several localized divergences as one increases the precision of the calculation, which
appear exactly where neutron or proton levels cross the Fermi energy; i.e.\ where their occupation probability is
$v^2 = 0.5$. In spite of the evidence for their appearance presented in
Refs.~\cite{donau98,anguiano01b,almehed01a,doba05a}, the divergences remained undetected so far in our PAV
calculations, because on the one hand the appearance of the divergence requires a
number of integration points far above the one used in practical calculations, and beyond what is tractable in
connection with other projections and mixing of different deformations, and because on the other hand the divergences are
sufficiently localized in deformation space that the area obviously affected by the pathology is smaller than the
typical distance of states commonly used when calculating energy surfaces and when mixing states with different
deformations.

At this point, three questions arise (1) do the divergences seen in Fig.~\ref{divergence} constitute the only
pathological manifestation of the poles? (2) Do divergences manifest for any type of functional, i.e.\
irrespective of the fact that it is bilinear, trilinear or contain non-integer powers of the density matrices?
(3) Is the spurious contribution isolated in Eq.~(\ref{spurious:bi}) responsible for all problems associated
with the poles; i.e.\ would removing it from PNR energy kernels properly regularize the MR EDF calculation?
Answering theses questions  will be the aim of Sec.~\ref{substraction}. Before discussing the results
obtained using the method proposed in Paper~I to regularize MR energy kernels, we discuss the
pathological manifestations of the poles in more detail through a complex plane analysis, following Ref.~\cite{doba05a}.

\section{Complex plane analysis}
\label{complexplane}

The integral over the real gauge angle can be reformulated as a contour integral in the complex plane, which
allows the analysis of the energy functional in terms of its poles within the integration contour~\cite{doba05a}.
In fact, particle-number projection was first introduced through such complex contour integrals~\cite{Bay60a,dietrich64a}.
It was only after Fomenko~\cite{fom70a} demonstrated that a simple trapezoidal rule gives a very efficient
discretization of integrals over the gauge angle that Eq.~(\ref{PWF}) became the standard way to formulate and
evaluate PNR observables.

\subsection{Analytic continuation}

To that aim, one introduces the complex variable $z=e^{i \varphi}$. As a result, quantities used in the PNR method involve an
integration over the unit circle $C_{1} \, (|z|=1)$~\footnote{We abusively replace the gauge angle $\varphi$ by the complex variable $z$ in all our expressions; i.e. SR states characterized by the gauge angle $\varphi$, $| \Phi_{\varphi} \rangle$, are extended into $| \Phi_{z} \rangle$ to denote SR states anywhere on the complex plane. In particular, the unrotated SR state, denoted as $| \Phi_0 \rangle$ when using $\varphi$ as a variable, is written as $| \Phi_1 \rangle$ when using $z$ as a more general variable.}
\begin{eqnarray}
\label{state3}
| \Psi^N \rangle
& = & \oint_{C_{1}} \frac{dz}{2i\pi c_{N}} \, \frac{1}{z^{N+1}} \, | \Phi_{z} \rangle
\, , \\
\label{projenergy3}
\mathcal{E}^{N}
& = & \oint_{C_{1}} \frac{dz}{2i\pi c^{2}_{N}} \, \frac{\mathcal{E} \left[z\right]}{z^{N+1}} \,
      \langle  \Phi_1 | \Phi_{z} \rangle
\, , \\
\label{denominator2}
c^{2}_{N}
& = & \oint_{C_{1}} \frac{dz}{2i\pi} \, \frac{1}{z^{N+1}} \, \langle  \Phi_1 | \Phi_{z} \rangle  \, ,
\end{eqnarray}
whereas the overlap now reads
\begin{eqnarray}
\label{overlap3} \langle  \Phi_1 | \Phi_{z} \rangle
& = & \prod_{\mu>0} \big( u_{\mu}^2 + v_{\mu}^2 \, z^{2} \big) \, .
\end{eqnarray}
Finally, the transition density matrix and pairing tensor extended to the complex plane become
\begin{eqnarray}
\label{contractphcomplex}
\rho^{1z}_{\mu \nu}
& = & \frac{v_{\mu}^2 \, z^{2}}{u_\mu^2 + v_{\mu}^2 \, z^{2}} \,
      \delta_{\nu \mu} \, ,
      \\
\label{contractppcomplex}
\kappa_{\mu \nu}^{1z}
& = & \frac{u_\mu v_{\mu}}{u_\mu^2 + v_{\mu}^2 \, z^{2} } \,
      \delta_{\nu \bar{\mu}}  \, ,
      \\
\label{contracthhcomplex}
\kappa_{\mu \nu}^{z1 \, \ast}
& = & \frac{u_\mu v_{\mu} \, z^{2} }{u_\mu^2 + v_{\mu}^2 \, z^{2} }
       \, \delta_{\nu \bar{\mu}}
\, .
\end{eqnarray}

\begin{figure}[t!]
\centerline{\includegraphics[width=6cm]{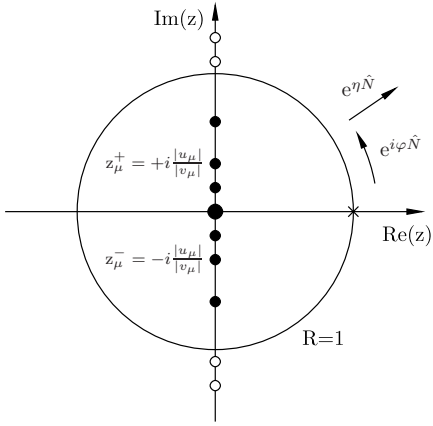}}
\caption{\label{poles}
Schematic view of the analytical structure of the transition densities defined in
Eqs.~(\ref{contractphcomplex}-\ref{contracthhcomplex}) and of the PNR functional energy
kernel $\mathcal{E}[\varphi]$ in the complex plane. Poles marked with filled circles are
within the standard circular integration contour of radius $R=1$, while those outside
are marked with open circles. The cross marks the location of the SR energy functional
at $z=1$. The operator $e^{i\varphi\hat{N}}$ produces a rotation in gauge space, while
$e^{\eta\hat{N}}$ is a shift transformation as defined in Eq.~(\ref{eq:shift}).}
\end{figure}

\subsection{Energy functional kernels}

Taking advantage of the  Cauchy residue theorem, going to the complex plane allows the
calculation of all quantities of interest in terms of poles of the integrand
located inside the integration contour. For the norm
\begin{equation}
c^{2}_{N} =  \mathcal{R}es (0)
  \left[  \frac{1}{z^{N+1}} \prod_{\mu >0}
          \big( u_{\mu}^2 + v_{\mu}^2 \, z^{2} \big)
  \right]
\end{equation}
or any other operator matrix elements between projected states, only the pole at $z=0$ contributes.

The situation is different for the PNR energy as additional poles at finite $z^{\pm}_{\mu} = \pm i
|u_{\mu}|/|v_{\mu}|$ enter the energy kernel $\mathcal{E} \left[z\right]$. Thus, Eq.~(\ref{projenergy3})
takes the form
\begin{eqnarray}
\label{polescontribution_a} \mathcal{E}^{N} & = & \sum_{z_{i}=0, |z^{\pm}_{\mu}| < 1} \! \! \frac{1}{c^{2}_{N}} \,
       \mathcal{R}es (z_{i}) \left[\frac{\mathcal{E}[z]}{ z^{N+1}}
      \prod_{\mu >0} \big( u_{\mu}^2 + v_{\mu}^2 z^{2} \big) \right]
      \nonumber \\
\end{eqnarray}
with contributions from the pole at the origin and from all pairs of "hole-like" poles at $z^{\pm}_{\mu}$. The
situation is schematically depicted on Fig.~\ref{poles}. The location of the pole associated to a given pair
$(\mu,\bar\mu)$ moves along the imaginary axis as the occupation $v^2_{\mu}$ changes
with deformation. When the corresponding pole crosses the unit circle, either entering or leaving the Fermi sea,
the integrand is non-analytical on the integration contour and the integral diverges.

The point has now come to realize that the divergences constitute the most obvious part of the problem,
but do not contain the entire problem. As can be seen from Eq.~(\ref{polescontribution_a}), the
poles at $|z^{\pm}_{\mu}| < 1$
contribute to the energy when using an energy functional that contains self-interactions and self-pairing.
On the other hand, only the pole at the origin contributes in the strict PNP-HFB/Hamiltonian framework as the
poles at $|z^{\pm}_{\mu}|$ do not exist in this case. Consequently, one has to ask the question
whether or not the contributions from the poles at $0 < |z^{\pm}_{\mu}| < 1$ to the projected energy are physical, in
particular when realizing that the contribution of a given pole can be many orders of magnitude larger than the total
energy gain from PNR~\cite{doba05a}. In addition, a pole
at finite $|z^{\pm}_{\mu}|$ entering or leaving the integration circle does not only provoke a divergence but also
provides the PNR energy with a finite step after the crossing is completed~\cite{doba05a}. Looking carefully at
the potential energy surface obtained using $L=199$ integration points, such a step can be seen in
Fig.~\ref{divergence}; i.e.\ compare the energy before and after the crossings at $\beta_{2} = +0.22$ and
$\beta_{2} = - 0.3$. As a matter of fact, the binding energy jumps from one potential energy surface to another.

\subsection{Spurious contributions}

In Section \ref{truesisp}, we have identified $\mathcal{E}_{CG}^{N}$ as the only
possible source of spurious poles. In order to obtain a deeper insight
to its content, we rewrite Eq.~(\ref{spurious:bi}) as
\begin{eqnarray}
\label{spuriouscomplex}
\lefteqn{
\mathcal{E}_{CG}^{N}
} \nonumber \\
& \equiv & \oint_{C_{1}} \! \frac{dz}{2i\pi c^{2}_{N}} \,
  \frac{\mathcal{E}_{CG} \left[z\right]}{z^{N+1}}
  \prod_{\mu >0} (u_{\mu}^2 + v_{\mu}^2 \, z^{2}) \nonumber \\
& = & \sum_{\mu > 0} \left[ \tfrac{1}{2}
      (   \bar{v}^{\rho\rho}_{\mu \mu \mu \mu}
        + \bar{v}^{\rho\rho}_{\bar{\mu}\bar{\mu}\bar{\mu}\bar{\mu}}
        + \bar{v}^{\rho\rho}_{\mu \bar{\mu} \mu \bar{\mu}}
        + \bar{v}^{\rho\rho}_{\bar{\mu} \mu \bar{\mu} \mu} )
      - \bar{v}^{\kappa\kappa}_{\mu\bar{\mu}\mu\bar{\mu}} \right] \,
      \nonumber \\
&   & \times
      \frac{\left(u_{\mu} v_{\mu}\right)^{4}}
           {2 i \pi c^{2}_{N}} \!
      \oint_{C_{1}} \! \frac{dz}{z^{N+1}} \,
      \frac{\left(z^{2}-1\right)^{2}}
           {u_\mu^2 + v_{\mu}^2 \, z^{2}}
      \prod_{\nu > 0 \atop \nu \neq \mu} (u_{\nu}^2 + v_{\nu}^2 \, z^{2})
\, ,
      \nonumber \\
\end{eqnarray}
and define in passing the spurious contribution $\mathcal{E}_{CG} \left[z\right]$ to the MR energy kernel
over the entire complex plane. From Eq.~(\ref{spuriouscomplex}), the spurious contribution of each pole to
the PNR energy can be
calculated. As for the total energy, the poles of the integrand are located at $z_0 = 0$ and $z^{\pm}_{\mu} = \pm
i |u_{\mu}|/|v_{\mu}|$. This has the important consequence that removing $\mathcal{E}_{CG}^{N}$ from
$\mathcal{E}^{N}$ does not only extract the contribution of the poles at $|z^{\pm}_{\mu}| < 1$ but also a spurious
contribution of each conjugated pair $(\mu,\bar\mu)$ to the physical pole at $z_0 = 0$. The latter could not have
simply been guessed from the analysis of the analytical structure of $\mathcal{E}[z]$ in the complex plane. As a matter of
fact, the spurious contribution from the pole at $z_0 = 0$ is absolutely essential for the internal consistency of
$\mathcal{E}_{CG}^{N}$. On the one hand, it was shown in Ref.~\cite{doba05a} that the energy associated with a
single pole at $|z^{\pm}_{\mu}| < 1$ can be gigantic (away from where it might be divergent). On the other hand, the
total spurious energy hidden in a PNR method cannot be larger than the energy gain from
particle number restoration itself, which is on the order of at most a few MeV. It is only the combined
contribution from the poles at $z_0 = 0$ and $z^{\pm}_{\mu}$, which nearly cancel each other, that will give
reasonable values to the total spurious energy $\mathcal{E}_{CG}^{N}$ as will be exemplified below.

The residue for the pair of poles at $|z^{\pm}_{\mu}|$ contained in Eq.~(\ref{spuriouscomplex}) can be evaluated
analytically
\begin{eqnarray}
\label{residuezmu}
\lefteqn{
\mathcal{R}e^{N}_{CG} (z^{\pm}_{\mu})
} \nonumber \\
&\equiv & \sum_{z_{i}=z^{\pm}_{\mu}} \, \mathcal{R}es (z_{i})
      \left[ \frac{\left(z^{2}-1\right)^{2} \prod_{\nu > 0 \atop \nu \neq \mu} (u_{\nu}^2 + v_{\nu}^2 \, z^{2})}
                  {v^{2}_{\mu} \, z^{N+1}
                   \left(z - i \frac{|u_{\mu}|}{|v_{\mu}|}\right)\left(z + i \frac{|u_{\mu}|}{|v_{\mu}|}\right)}\right]
 \nonumber \\
& = & - \frac{1}{v^{6}_{\mu}} \!
        \left(\frac{v_{\mu}}{u_{\mu}}\right)^{N+2}
        \frac{1 + (-1)^{N}}{2 \, i^{N}} \!
        \prod_{\nu > 0 \atop \nu \neq \mu} \frac{u^{2}_{\nu} v^{2}_{\mu} - v^{2}_{\nu} u^{2}_{\mu}}{v^{2}_{\mu}}
\, .
\end{eqnarray}
Note that $\mathcal{R}e^{N}_{CG} (z^{\pm}_{\mu})$ is zero if projecting on an odd particle number $N$ as
the underlying reference state~(\ref{eq:transfo_phi}) has been chosen to have an even number-parity quantum
number~\cite{ring70a,banerjee73a}. The generalization of the present discussion to the case one- (or $2n+1$)
quasiparticle states with an odd number-parity is straightforward, but not important for the purpose of this paper.

 The total contribution from the pair of poles
$0< |z^{\pm}_{\mu}| < 1$ to the PNR energy is then obtained by replacing the integral in
Eq.~(\ref{spuriouscomplex}) by $2 i \pi \, \mathcal{R}e^{N}_{CG} (z^{\pm}_{\mu})$, where $\mathcal{R}e^{N}_{CG}
(z^{\pm}_{\mu})$ is given by Eq.~(\ref{residuezmu}). We will discuss the individual contributions from the poles
in Sec.~\ref{applications} below. Note that calculating the residue of the pole at $z_{0}$ is much more
involved because it is a pole of order $N+1$. Its residue can
in fact be calculated analytically through a recursive formula, which, however, involves a sum over such a large
number of terms that it is of no practical use and is not reported here. In any case, one can access the spurious
contribution from the pole $z_{0}$ by subtracting the analytic expression of Eq.~(\ref{residuezmu}) from a numerical
evaluation of the full expression given by Eq.~(\ref{spurious:bi}).

\subsection{Properties under shift transformation}
\label{shift}

The interpretation of the poles at $z^{\pm}_{\mu} \neq 0$ becomes clearer when looking at the properties of the PNR energy
functional under a so-called shift transformation~\cite{doba05a}. In the present paper, we choose a slightly
 different definition of the shift transformation from the one used in Ref.~\cite{doba05a}
\begin{equation}
\label{eq:shift}
| \Phi_{\varphi -i \eta} \rangle
\equiv e^{\eta \hat{N}} \, | \Phi_{\varphi} \rangle
\, ,
\end{equation}
such that the shift transformation operator $e^{(\eta + i\varphi) \hat{N}}$ used in~\cite{doba05a}
is the product of ours~(\ref{eq:shift}) and a rotation in gauge space\footnote{Starting
from a circular contour, the additional rotation in the definition of Ref.~\cite{doba05a} does not make any
difference. The situation would have been different if we had started from a non-circular contour.}.
In contrast to a gauge-space rotation that is unitary, the shift transformation~(\ref{eq:shift})
is non-unitary and changes the norm of the product state.

In the complex plane, the shift transformation~(\ref{eq:shift}) corresponds to a radial shift
of $z$ from $z = e^{i \varphi}$ to $z' = e^{\eta} \, e^{i \varphi}$, see Fig.~\ref{poles}. Thus, projecting
a shifted HFB state on particle number amounts to changing the radius of the integration circle from $R=1$ to
$R = e^{\eta}$~\cite{doba05a}
\begin{eqnarray}
\hat{P}^{N} \, | \Phi_{\varphi -i \eta} \rangle
& = & \oint_{C_1} \frac{dz'}{2i\pi} \frac{1}{(z')^{N+1}} \, | \Phi_{Rz'} \rangle
      \nonumber \\
& = & \oint_{C_R}  \frac{dz}{2i\pi} \frac{R^N}{z^{N+1}} \, | \Phi_z \rangle
\, ,
\end{eqnarray}
where we have made the substitution $z' = e^{i\varphi}$ in the first line and the
substitution $z = Rz'$ in the second one. Both expressions will turn out to be
useful below. The overlap between the non-normalized projected SR state and its counterpart
shifted along the real axis is given by
\begin{equation}
\label{weight:R}
c^{2}_{N}(R) \equiv \langle \Phi_1 | \hat{P}^{N} | \Phi_{R} \rangle
= c^{2}_{N} \, R^N
\end{equation}
with $c^{2}_{N}$ as defined through Eq.~(\ref{weight}); i.e.\ $c^{2}_{N} \equiv c^{2}_{N}(1)$.

All normalized projected matrix elements are shift invariant if the operator $\hat{O}$ in question commutes with
$\hat{N}$. Just as the exact ground-state energy, its approximation obtained
through the particle-number restored expectation value of the Hamilton operator is shift invariant. On the other
hand, this is not the case for standard particle-number restored energy density functionals~\cite{doba05a}. The
violation of shift invariance is obviously a consequence of the presence of the poles at finite $z^{\pm}_{\mu}$
contained in the PNR energy kernel constructed on the basis of the GWT. For a given spectrum of poles
$z^{\pm}_{\mu}$ the energy $\mathcal{E}^N$ changes by a finite quantity whenever the integration circle crosses a pair of poles
$|z^{\pm}_{\mu}|$ in the course of a shift transformation. As a result, the PNR-EDF is shift invariant only over a
finite range of values of the shift parameter $\eta$~\cite{doba05a}. This result clearly points to the unphysical
nature of these poles.

\subsection{Sum rules}
\label{sumrule}

One might wonder where the energy that is added/removed when crossing a pole with the integration contour comes
from/goes to. In the present section, two different sum rules involving PNR energies $\mathcal{E}^{N}$ extracted
from a given SR functional are carefully derived and discussed to answer such a question.

\subsubsection{Radius-weighted sum rule} \label{delires1}

As it is introduced in Ref.~\cite{doba05a}, we first discuss the characteristics of the radius-weighted
sum rule $\sum c_{N}^{2}(R) \, \mathcal{E}^{N}(R)$, although we already insist here that the physical sum rule
of interest is the non-radius-weighted one discussed in Sec.~\ref{delires2} below. The number
$R$ appearing in the sum rule is taken to be real even though it is possible to formulate the sum rule
using an arbitrary complex number of norm $R$~\cite{doba05a}. Our conclusions will be insensitive to
this detail.

First, let us recall how such sum rules arise in the operator- and wave-function-based context.
Inserting the complete set of normalized particle-number projected states\footnote{The fact that one does not need to sum over $N < 0$ can be seen as a consequence of the fact that
$| \Psi^N \rangle = 0$ for $N < 0$ as a result of the disappearance of the physical pole at $z=0$ in the
contour integral of Eq.~(\ref{state3}). Note that the normalized projected-state on $N=0$ is $| \Psi^0 \rangle = | 0 \rangle$.}
\begin{equation}
\label{eq:complete}
\sum_{N \geq 0} | \Psi^N \rangle \langle \Psi^N |
= \sum_{N \geq 0} \hat{P}^N
= 1
\end{equation}
into an unprojected shifted matrix element of an operator $\hat{O}$
that commutes with $\hat{N}$ gives
\begin{eqnarray}
\label{eq:operator:sumrule}
\langle \Phi_1 | \hat{O}  | \Phi_R \rangle
& = & \langle \Phi_1 | \hat{O} e^{\eta \hat{N}} | \Phi_1 \rangle \nonumber \\
& = & \sum_{N \geq 0}
      \langle \Phi_1 | \hat{O} e^{\eta \hat{N}} | \Psi^N \rangle \,
      \langle \Psi^N | \Phi_1 \rangle
      \nonumber \\
& = & \sum_{N \geq 0}
      c_N^2(R) \,
      O^N
      \, ,
\end{eqnarray}
where we have used that $e^{\eta \hat{H}} | \Psi^N \rangle = R^N | \Psi^N \rangle$ and define
$O^N = \langle \Phi_1 | \hat{O} |\Psi^N \rangle / \langle \Phi_1 |\Psi^N \rangle$.
Equation~(\ref{eq:operator:sumrule}) expands the shifted SR matrix element
$O[R]\equiv \langle \Phi_1 | \hat{O} | \Phi_R \rangle$ in terms of average values
$O^N$ of the operator in all normalized projected states. Applied to the Hamilton
operator, Eq.~(\ref{eq:operator:sumrule}) reads
\begin{equation}
\label{eq:energy:sumrule}
E[R]
= \sum_{N > 0} c_N^2(R) \, E^N
\, ,
\end{equation}
and provides for $\eta= 0$ $(R=1)$ that the strict HFB energy decomposes into strict
PNP-HFB energies (with $N>0$) weighted by the probability to find the normalized
projected states into the SR state. In Eq.~(\ref{eq:energy:sumrule}), the sum
could be further reduced to $N > 0$ as the contribution from the term $N=0$ is strictly
zero, i.e. $c_0^2 \, E^0 = E[z\!=\!0] \, \prod_{\nu > 0} u_{\nu} = 0$. Such a result relies on the fact that only the physical pole at $z=0$ contributes to the integral providing $E^N$.

Let us now come to the EDF context and lay out some specificities that are crucial
to provide a meaningful discussion of sum rules. (i) In Eq.~(\ref{eq:energy:sumrule}),
it was not necessary to specify the integration contour used to calculate $E^N$ as
the latter is shift invariant. In the EDF context where the shift invariance might
be broken, it is mandatory to specify the contour employed. Consequently, the notation
$\mathcal{E}^{N}(R)$ is used whenever necessary to characterize that a circular
contour $C_{R}$ of radius $R$ is employed to calculate PNR energies. (ii) There is no
equivalent to "inserting a complete set of states" in the EDF context as one directly
postulates the PNR energy under the form of a functional built from one-body transition
density matrices and integrated over the gauge angle, and not from the expectation value of
a Hamilton operator in projected many-body wave functions. As a consequence, the existence
of a sum rule similar to the one discussed for operators is neither obvious nor trivial.
By contrast to the above derivation, one has to start from the weighted sum over PNR
energies and see if and how it recombines in the same manner as for an operator matrix
element. To obey a sum rule analogous to the one provided by Eq.~(\ref{eq:energy:sumrule})
can thus be demanded as a consistency requirement for MR energy density functionals.
To recover the SR energy from such a sum rule, it is a necessary condition
(but not sufficient) that the MR energy kernel $\mathcal{E} [z]$ is set up such that
it gives back the SR energy functional $\mathcal{E} [\rho, \kappa, \kappa^{\ast}]$
for $z=1$, as assumed throughout this paper. (iii) The sum rule considered in the
present section actually differs from the one discussed in Ref.~\cite{doba05a}.
Indeed, it is mandatory in the EDF context to make the sum running over both positive
\emph{and} negative "particle numbers". As will be shown below, the latter is crucial
to establish the expected sum rule when individual particle-number restored energies
$\mathcal{E}^{N}$ are not shift invariant, i.e. when MR energy kernels $\mathcal{E}[z]$
possess spurious poles at finite $z^{\pm}_{\mu}$. Indeed, the product $c_{N}^{2}(R) \,
\mathcal{E}^{N}(R)$ is different from zero in this case for $N\leq0$ because, although
the physical pole at $z=0$ disappears from the integrand as it should, the poles at
finite $z^\pm_\mu$ contribute. This is certainly the most direct proof of the non-physical
nature of such poles and non-regularized energy functionals. In the context of the
real-space derivation of Ref.~\cite{doba05a}, obtaining the appropriate sum rule calls
for using the correct Fourier decomposition of the periodic delta function over \emph{all}
irreducible representations of $U(1)$ including those characterized by negative integers
$N$; i.e., $\sum_{N=-\infty}^{+\infty} e^{-i\varphi N} = 2 \pi \, \delta_{2\pi} (\varphi)$.
In the following we proceed in the complex plane to establish the needed sum rules.

First, the change of variable $z=R \, z'$ is performed in order to recover an
integration over the unit circle
\begin{equation}
\label{sumrulecomplexplane1b}
\sum_{N=-\infty}^{+\infty} c_{N}^{2}(R) \, \mathcal{E}^{N}(R)
= \sum_{N=-\infty}^{+\infty}
  \oint_{C_{1}} \frac{dz}{2i\pi} \,
  \frac{\mathcal{E} [Rz]}{z^{N+1}} \,
  \langle \Phi_{1} | \Phi_{Rz} \rangle
\,.
\end{equation}
We recall that $\mathcal{E}^{N}(R)$ is proportional to $1/c_{N}^{2}(R)$, Eq.~(\ref{scalar2}).
As a consequence, $c_{N}^{2}(R) = 0$ alone is not a sufficient condition that the contribution of
a given $N$ to the l.h.s.\ of Eq.~(\ref{sumrulecomplexplane1b}) vanishes, as
$c_{N}^{2}(R) \mathcal{E}^{N}(R)$ might remain finite. We will come back to this below.

To invert the summation and the integral in Eq.~(\ref{sumrulecomplexplane1b}) and perform the
summation explicitly, the power series must be (uniformly) converging on the integration
contour. To ensure this property, one has to separate the sums over positive and negative $N$
and use the (local) shift invariance of $\mathcal{E}^{N}$ to scale the integration radius
appropriately in each of the two terms thus generated. Using two infinitesimal shift
transformations characterized by $\eta_{+}>0$ $(\eta_{-}<0)$ for $N>0$ ($N\leq 0)$,
the right-hand-side of Eq.~(\ref{sumrulecomplexplane1b}) splits into two geometric
series converging separately and uniformly on the corresponding integration contours
$C_{1^{+}}$ $(C_{1^{-}})$. Performing the summation of both geometric series, one
obtains
\begin{eqnarray}
\label{sumrulecomplexplane2}
\lefteqn{
\sum_{N=-\infty}^{+\infty} c_{N}^{2}(R) \, \mathcal{E}^{N}(R)
} \nonumber \\
& = & \left[ \oint_{C_{1^{+}}} \! - \oint_{C_{1^{-}}}\right]
      \frac{dz}{2i\pi} \frac{ \mathcal{E} [Rz]}{z(z-1)} \, \langle \Phi_{1} |
      \Phi_{Rz} \rangle
\,  .
\end{eqnarray}

The physical pole at $z=0$, which is of order $N+1$ in $\mathcal{E}^{N}$, has transformed
into two simple poles at $z=0$ and $z=1$ in both integrals in Eq.~(\ref{sumrulecomplexplane2}). Note in passing that the pole at $z=0$
would have not appeared if we had grouped the component $N=0$ to the sum over positive numbers. The
pole at $z=1$ is on the unit circle and is thus located inside of $C_{1^{+}}$, but outside of $C_{1^{-}}$.
Thus, it contributes to the first integral only in Eq.~(\ref{sumrulecomplexplane2}) and provides the
sum rule with the contribution $\mathcal{E}[R] \, \langle \Phi_{1} | \Phi_{R} \rangle$ which represents
the transition kernel involving the original HFB state $| \Phi_{1} \rangle$ and the
state $| \Phi_{R} \rangle$ shifted along the real axis to $z=R$.

In the strict PNP-HFB method, this is the only
contribution to Eq.~(\ref{sumrulecomplexplane2}) as the residue of the simple pole at $z=0$, which corresponds to the contribution from the $N=0$ component, is zero for the reason explained earlier. In any case, such a pole contributes to both integrals in Eq.~(\ref{sumrulecomplexplane2}) such that any finite residue would have canceled out anyway. Thus, the sum rule~(\ref{eq:energy:sumrule}) is recovered.

The question is
whether this still holds in the EDF context  As a matter of fact, the contribution from the
poles of $\mathcal{E}[z]$  at $z^{\pm}_{\mu}$ depends on the original contour $C_{R}$ and on the infinitesimal
shift transformations leading to Eq.~(\ref{sumrulecomplexplane2}). If the shift transformations are such
that no pole appears in between the two contours $C_{1^{-}}$ and $C_{1^{+}}$, all poles with $|z^{\pm}_{\mu}|<R$
contribute to both integrals and cancel out in Eq.~(\ref{sumrulecomplexplane2}) whereas all poles
with $|z^{\pm}_{\mu}|>R$ do not contribute to either of them. This proves that, except for the
ill-defined case of a pair of poles sitting on the original integration circle $C_{R}$, one can always
perform two infinitesimal shift transformations to prove that
\begin{equation}
\label{sumrulecomplexplane3}
\sum_{N=-\infty}^{+\infty} c_{N}^{2}(R) \, \mathcal{E}^{N}(R)
=  \mathcal{E}[R] \, \langle \Phi_{1} | \Phi_{R} \rangle
\,  .
\end{equation}
Equation~(\ref{sumrulecomplexplane3}) thus expresses that the expected sum rule is found to be
valid, even for contaminated and yet uncorrected EDFs, i.e.\ using energy kernels constructed
on the basis of the GWT, at the price of including the contributions from unphysical
components ($N\leq0$).

Applying the same derivation as above to the spurious contribution isolated in
Eq.~(\ref{spuriouscomplex}), one obtains
\begin{widetext}
\begin{eqnarray}
\label{sumrule11}
\sum_{N=-\infty}^{+\infty} c_{N}^{2}(R) \, \mathcal{E}^{N}_{CG}(R)
& = & \mathcal{E}_{CG}[R] \, \langle \Phi_{1} | \Phi_{R} \rangle
      \\
& = & \left(R^{2}-1\right)^2\sum_{\mu > 0} \left[ \tfrac{1}{2}
      \left(   \bar{v}^{\rho\rho}_{\mu \mu \mu \mu}
             + \bar{v}^{\rho\rho}_{\bar{\mu}\bar{\mu}\bar{\mu}\bar{\mu}}
             + \bar{v}^{\rho\rho}_{\mu \bar{\mu} \mu \bar{\mu}}
             + \bar{v}^{\rho\rho}_{\bar{\mu} \mu \bar{\mu} \mu}
      \right)
      - \bar{v}^{\kappa\kappa}_{\mu\bar{\mu}\mu\bar{\mu}} \right] \,
      \frac{\left(u_{\mu} \, v_{\mu}\right)^4}{u^{2}_{\mu}+R^{2}\,v^{2}_{\mu}}  \,
      \prod_{\nu > 0 \atop \nu \neq \mu} \left(u^{2}_{\nu} + R^2 \, v^{2}_{\nu}\right)
\, , \nonumber
\end{eqnarray}
\end{widetext}
which is zero for $R=1$ as $z=1$ is the only point in the complex plane where the
GWT-related spurious contributions to the MR energy kernel is zero.

It is crucial to analyze further the cancellation of the contribution of spurious poles in
Eqs.~(\ref{sumrulecomplexplane2}-\ref{sumrulecomplexplane3}). Indeed, such a cancellation relies on the original
summation over \emph{both} positive and negative "particle numbers" in the definition of the sum rule. If one sums
over positive particle numbers only, all pairs of poles situated inside $C_{R}$ contribute to the sum rule. This
is puzzling as it is clearly unphysical to consider negative "particle numbers". Indeed, one necessarily has $c^2_{N}(R) \, E^{N}(R)=0$ for $N \leq 0$ when employing a
genuine Hamiltonian. However,
the product $c_{N}^{2}(R) \, \mathcal{E}^{N}(R)$ is different from zero for $N\leq0$ \emph{if} $\mathcal{E}[z]$
possesses poles at finite $|z^{\pm}_{\mu}|<R$. This is to our opinion the most direct way of stating the
non-physical nature of those poles. In any case, and as proven above, one can at least recover a sum rule
for uncorrected functionals at the price of summing over both positive and negative particle numbers.
If summing over positive values only, one obtains, using our example of a bilinear functional,

\begin{eqnarray}
\label{sumrulecomplexplane3ter}
\lefteqn{
\sum_{N>0} c_{N}^{2}(R) \, \mathcal{E}^{N}(R) - \mathcal{E}[R] \, \langle \Phi_{1} | \Phi_{R} \rangle
} \nonumber \\
& = & \sum_{|z^{\pm}_{\mu}| < R} \! \!
      \mathcal{R}es (z^{\pm}_{\mu}/R) \left[\frac{\mathcal{E}[Rz]}{z(z-1)}
      \prod_{\mu >0} \big( u_{\mu}^2 + v_{\mu}^2 R^{2} z^{2} \big) \right]
    \nonumber \\
& = & \sum_{\mu > 0  \atop |z_\mu^\pm| < R} \!
  \left[ \tfrac{1}{2}
  \left(   \bar{v}^{\rho\rho}_{\mu \mu \mu \mu}
         + \bar{v}^{\rho\rho}_{\bar{\mu}\bar{\mu}\bar{\mu}\bar{\mu}}
         + \bar{v}^{\rho\rho}_{\mu \bar{\mu} \mu \bar{\mu}}
         + \bar{v}^{\rho\rho}_{\bar{\mu} \mu \bar{\mu} \mu}
  \right) - \bar{v}^{\kappa\kappa}_{\mu\bar{\mu}\mu\bar{\mu}}
  \right]
   \nonumber \\
&   & \quad \times
   \frac{u^{2}_{\mu} \, R^{2} \, v^{2}_{\mu}}
        {u^{2}_{\mu}+R^{2}\,v^{2}_{\mu}}  \,
   \prod_{\nu > 0 \atop \nu \neq \mu}
  \frac{u^{2}_{\nu} v^{2}_{\mu} - v^{2}_{\nu} u^{2}_{\mu}}{v^{2}_{\mu}}
   \,  ,
\end{eqnarray}
which shows that the physical sum rule ($N>0$) is broken by a finite amount that relates directly to the presence of
spurious poles at finite $z^{\pm}_{\mu}$ inside the original integration circle $C_{R}$. Note again that
the simple pole at $z=0$ does not contribute as its residue is zero. Equation~(\ref{sumrulecomplexplane3ter})
proves that the sum rule derived in Ref.~\cite{doba05a} is incorrect for the cases of interest. In particular,
computing Eq.~(\ref{sumrulecomplexplane3ter}) for $R=1$ provides the non-zero amount by which the decomposition
of the SR EDF into its \emph{physical} PNR components ($N>0$) is broken, already for the standard integration circle.
However, as we will show in Sec.~\ref{sect:application:sumrules} below, the contribution from $N \leq 0$ is
several orders of magnitude smaller than the contribution from $N > 0$ in realistic cases, such that it might
pass as numerical noise to the unsuspecting eye.

Subtracting Eq.~(\ref{sumrule11}) from~(\ref{sumrulecomplexplane3}) provides the quantity
$\sum_{-\infty}^{+\infty} c_{N}^{2}(R) \, [\mathcal{E}^{N}(R)-\mathcal{E}^{N}_{CG}(R)]$
by which the sum rule is modified when regularizing the MR energy kernels. One observes that the
non-physical components are zero, i.e.\ $c_{N}^{2}(R) \, \mathcal{E}_{REG}^{N}(R)=0$ for $N\leq0$,
and that the sum rule matches the \emph{regularized} kernel at $z=R$ $\mathcal{E}_{REG}[R] \,
\langle \Phi_{1} | \Phi_{R} \rangle$.

\subsubsection{Non-radius-weighted sum rule} \label{delires2}

The sum rule~(\ref{sumrulecomplexplane3}) is of particular interest when the unit circle $C_{1}$ is used
as an integration contour to define PNR energies. Indeed, Eq.~(\ref{sumrulecomplexplane3}) reduces in this
case to
\begin{equation}
\label{sumrulecomplexplane4}
\sum_{N=-\infty}^{+\infty} c_{N}^{2} \, \mathcal{E}^{N}(R\!=\!1)
= \mathcal{E}[z\!=\!1]
= \mathcal{E}[\rho,\kappa,\kappa^{\ast}] \, ,
\end{equation}
which expresses that the SR EDF decomposes into PNR energies obtained for all possible
"particle numbers" $N \gtrless 0$. This decomposition actually relies on the (required) connection between
the SR EDF and the MR energy functional kernel; i.e.\ $\mathcal{E}[z=1] = \mathcal{E}[\rho,\kappa,\kappa^{\ast}]$.
Equation~(\ref{sumrulecomplexplane4}) is valid prior to any regularization of the PNR energy kernel,
as long as the sum runs over both positive and negative particle numbers. The null sum rule~(\ref{sumrule11})
at $R=1$ shows that regularizing the PNR EDF method through the removal of $\mathcal{E}^{N}_{CG}$ from
$\mathcal{E}^{N}$ consists, for this radius, of reshuffling contributions among different particle-number restored
energies, in such a way that the decomposition of the SR EDF into its \emph{physical} PNR components ($N>0$) is
fulfilled. Note that the regularized sum rule matches the SR EDF precisely because the regularization does
not modify the energy kernel $\mathcal{E}[z]$ for $z=1$.

Still, the radius-weighted sum rule considered in Sec.~\ref{delires2} and in Ref.~\cite{doba05a} does not
allow us to study the shift invariance of Eq.~(\ref{sumrulecomplexplane4}), which is the real question of interest.
Indeed, what matters is whether or not the standard decomposition of the SR EDF into $c^{2}_N$-weighted  PNR
energies is valid independently on the radius of integration chosen initially to compute $\mathcal{E}^{N}$.
In a Hamiltonian and wave function based framework, such an invariance reflects the trivial identity
\begin{eqnarray}
\langle \Phi_1 | \hat{H} | \Phi_1 \rangle
& = & \sum_{N>0} \frac{\langle \Phi_1 | \hat{H} | \Psi^N \rangle}
                      {\langle \Phi_1 | \Psi^N \rangle} \,
                 |\langle \Phi_1 | \Psi^N \rangle|^2
      \nonumber \\
& = & \sum_{N>0} \frac{\langle \Phi_1 | \hat{H} e^{i\eta \hat{N}} | \Psi^N \rangle}
                      {\langle \Phi_1 | e^{i\eta \hat{N}} | \Psi^N \rangle} \,
                 |\langle \Phi_1 | \Psi^N \rangle|^2
\, .
\end{eqnarray}
Translated to the functional framework, this amounts to considering
the non-radius-weighted sum rule
\begin{equation}
\label{sumrulecomplexplane5}
\sum_{N=-\infty}^{+\infty} c_{N}^{2}(1) \, \mathcal{E}^{N}(R)
= \sum_{N=-\infty}^{+\infty}
  \oint_{C_{R}} \frac{dz}{2i\pi} \,
  \frac{\mathcal{E} [z]}{z^{N+1}} \,
  \langle \Phi_{1} | \Phi_{z} \rangle
\, ,
\end{equation}
where $c_{N}^{2}(1)=c_{N}^{2}$ and where the circle of integration $C_{R}$ is the one chosen to
calculate PNR energies. Again, the power series must be split into two parts to perform the summation
over particle numbers explicitly. The initial circle of integration $C_{R}$ being
above/below the unit circle, one needs to perform a finite shift transformation to bring the circle associated
with negative/positive particle numbers on the other side of the unit circle, in
order to make the corresponding series convergent. If particle-number restored energies are shift invariant, one
can proceed without any difficulty and obtain the trivial result that the sum rule $\sum_{N=-\infty}^{+\infty}
c_{N}^{2} \, \mathcal{E}^{N}(R)=\mathcal{E}[\rho,\kappa,\kappa^{\ast}]$ is valid independently on the original
radius $R$. This is of course the case for a Hamiltonian- and wave-function-based
PNR method which, once again, would only require the summation over positive particle numbers in the first place.

Of course, problems arise if particle-number restored energies are not invariant as the shifted circle
crosses a spurious pole at $z^{\pm}_{\mu}$, i.e.\ if there are poles $z^{\pm}_{\mu}$ located in between $C_{R}$ and
$C_{1}$. Indeed, proceeding to the required shift transformation brings an extra contribution to the
sum rule in this case. Exemplifying the problem for a bilinear functional and an initial radius $R>1$,
one obtains
\begin{widetext}
\begin{eqnarray}
\label{spuriouscomplexquartic}
\sum_{N=-\infty}^{+\infty} c_{N}^{2} \, \mathcal{E}^{N}(R)
& = & \left[
      \oint_{C_{R}}-\oint_{C_{1^{-}}}\right]
      \frac{dz}{2i\pi} \frac{ \mathcal{E} [z]}{z(z-1)} \, \langle \Phi_{1} | \Phi_{z} \rangle
      + 2i\pi\sum_{N=-\infty}^{0} \sum_{1<|z^{\pm}_{\mu}| < R} \! \!
        \mathcal{R}es (z^{\pm}_{\mu}) \left[\frac{\mathcal{E}[z]}{z^{N+1}}
        \prod_{\mu >0} \big( u_{\mu}^2 + v_{\mu}^2 z^{2} \big) \right]
      \nonumber \\
& = & \mathcal{E}[\rho,\kappa,\kappa^{\ast}] + \sum_{N=-\infty}^{+\infty} c_{N}^{2} \, \mathcal{E}^{N}_{CG}(R)
\, ,
\end{eqnarray}
with
\begin{eqnarray}
\label{sumrule11a}
\sum_{N=-\infty}^{+\infty} c_{N}^{2} \, \mathcal{E}^{N}_{CG}(R)
& = & \! \! \sum_{\mu > 0  \atop 1<|z_\mu^\pm| < R} \!
      \left[ \tfrac{1}{2}
      \left(   \bar{v}^{\rho\rho}_{\mu \mu \mu \mu}
            + \bar{v}^{\rho\rho}_{\bar{\mu}\bar{\mu}\bar{\mu}\bar{\mu}}
            + \bar{v}^{\rho\rho}_{\mu \bar{\mu} \mu \bar{\mu}}
            + \bar{v}^{\rho\rho}_{\bar{\mu} \mu \bar{\mu} \mu}
      \right)
      - \bar{v}^{\kappa\kappa}_{\mu\bar{\mu}\mu\bar{\mu}} \right]
      u^{2}_{\mu}\,v^{2}_{\mu}  \,
      \prod_{\nu > 0 \atop \nu \neq \mu} \frac{u^{2}_{\nu} v^{2}_{\mu} - v^{2}_{\nu} u^{2}_{\mu}}{v^{2}_{\mu}}
      \nonumber   \\
&   & + \! \! \sum_{\mu > 0  \atop 1<|z_\mu^\pm| < R} \!
       \left[ \tfrac{1}{2}
       \left(   \bar{v}^{\rho\rho}_{\mu \mu \mu \mu}
         + \bar{v}^{\rho\rho}_{\bar{\mu}\bar{\mu}\bar{\mu}\bar{\mu}}
         + \bar{v}^{\rho\rho}_{\mu \bar{\mu} \mu \bar{\mu}}
         + \bar{v}^{\rho\rho}_{\bar{\mu} \mu \bar{\mu} \mu}
       \right)
      - \bar{v}^{\kappa\kappa}_{\mu\bar{\mu}\mu\bar{\mu}} \right]
      \left(u_{\mu} v_{\mu}\right)^{4}  \sum_{N=-\infty}^{0} \mathcal{R}e^{N}_{CG} (z^{\pm}_{\mu})
\, ,
\end{eqnarray}
\end{widetext}
where $\mathcal{R}e^{N}_{CG} (z^{\pm}_{\mu})$ is given by Eq.~(\ref{residuezmu}) and where the sums run over all
pairs of poles located in between the unit circle $C_{1}$ and the integration circle $C_{R}$. Note that, in agreement
with Eq.~(\ref{sumrule11}), the sum rule~(\ref{sumrule11a}) is zero for $R=1$ as no pole resides between $C_{R}$
and $C_{1}$ in this case. However, it is easy to see from Eq.~(\ref{residuezmu}) that
$\sum_{N\leq 0} \mathcal{R}e^{N}_{CG} (z^{\pm}_{\mu})$ is a diverging geometric series of common ratios
$|z^{\pm}_{\mu}|>1$ for $R>1$; i.e.\ the sum rule is broken by a diverging amount as soon as poles are located
in between the integration circle $C_{R}$ and the unit circle $C_{1}$. One can check that the situation is similar
if $R<1$ and the conclusion identical. Regularizing the PNR EDF through the removal of $\mathcal{E}^{N}_{CG}(R)$
amounts to transferring the second term in the right-hand side of Eq.~(\ref{spuriouscomplexquartic}) to the
left-hand side. Doing so restores the physical value ($\mathcal{E}[\rho,\kappa,\kappa^{\ast}]$) and the shift
invariance of the sum rule as the shift invariance of each individual PNR energy $\mathcal{E}^{N}(R)$ is actually
restored. As $c_{N}^{2} \, \mathcal{E}_{REG}^{N}(R)=0$ for $N\leq0$, the sum rule is in fact restored and made
shift invariant by summing over positive particle numbers only
\begin{equation}
\label{sumrulefinal}
\sum_{N>0} c_{N}^{2} \, \mathcal{E}^{N}_{REG}(R)
=  \mathcal{E}[\rho, \kappa, \kappa^{\ast}]
\, .
\end{equation}
Last but not least, it is of interest to look at the non-regularized sum rule obtained by summing
over physical components only ($N>0$). In this case, the physical sum rule calculated for $R>1$ is
broken by a finite amount
\begin{widetext}
\begin{eqnarray}
\label{spuriouscomplexbisbis}
\sum_{N=1}^{+\infty} c_{N}^{2} \, \mathcal{E}^{N}(R)
& = & \oint_{C_{R}}
      \frac{dz}{2i\pi} \, \frac{ \mathcal{E} [z]}{z(z-1)} \, \langle \Phi_{1} | \Phi_{z} \rangle
  =   \mathcal{E}[\rho,\kappa,\kappa^{\ast}] +  \sum_{N=1}^{+\infty} c_{N}^{2} \, \mathcal{E}^{N}_{CG}(R)
\, ,
\end{eqnarray}
with
\begin{eqnarray}
\label{spuriouscomplexbisbisbis}
\sum_{N=1}^{+\infty} c_{N}^{2} \, \mathcal{E}^{N}_{CG}(R)
& = & \sum_{\mu > 0  \atop |z_\mu^\pm| < R}
      \left[ \tfrac{1}{2}
      \left(   \bar{v}^{\rho\rho}_{\mu \mu \mu \mu}
             + \bar{v}^{\rho\rho}_{\bar{\mu}\bar{\mu}\bar{\mu}\bar{\mu}}
             + \bar{v}^{\rho\rho}_{\mu \bar{\mu} \mu \bar{\mu}}
             + \bar{v}^{\rho\rho}_{\bar{\mu} \mu \bar{\mu} \mu}
      \right)
      - \bar{v}^{\kappa\kappa}_{\mu\bar{\mu}\mu\bar{\mu}} \right]
      u^{2}_{\mu}\,v^{2}_{\mu} \,
      \prod_{\nu > 0 \atop \nu \neq \mu} \frac{u^{2}_{\nu} v^{2}_{\mu} - v^{2}_{\nu} u^{2}_{\mu}}{v^{2}_{\mu}}
\, ,
\end{eqnarray}
\end{widetext}
where the sum runs over all pairs of poles located inside the integrations circle $C_{R}$. This time, however, and as
already made clear above, the sum rule~(\ref{sumrulecomplexplane4}) is \emph{not} even recovered for $R=1$ as the last
term of Eq.~(\ref{spuriouscomplexbisbis}) does not go to zero. Regularizing the PNR EDF through the removal of
$\mathcal{E}^{N}_{CG}$ amounts to transferring the second term in the right-hand side of Eq.~(\ref{spuriouscomplexbisbis})
to the left-hand side. Once again, doing so restores the physical value, i.e.\ $\mathcal{E}[\rho,\kappa,\kappa^{\ast}]$,
and the shift invariance of the sum rule.

\subsubsection{Main conclusions}
\label{delires3}

The first conclusion is that the decomposition of the SR energy $\mathcal{E}[\rho,\kappa,\kappa^{\ast}]$ into its
\emph{physical} ($N>0$) particle-number restored components is (i) always fulfilled for
a Hamiltonian- and wave-function-based method, whatever the chosen integration circle is, while it is
(ii) broken by an amount  that depends on the chosen integration contour for an EDF-based PNR method
if MR energy kernels $\mathcal{E}[z]$ contain poles at finite $z^{\pm}_{\mu}$, but (iii) recovered for
any value of $R$ after regularizing $\mathcal{E}^{N}$ through the removal of $\mathcal{E}^{N}_{CG}$.

The second conclusion is that the decomposition of $\mathcal{E}[\rho,\kappa,\kappa^{\ast}]$ involving
\emph{unphysical} components ($N\leq0$) is (i) always fulfilled in a Hamiltonian- and wave-function-based
PNR method as unphysical components do not contribute anyway (ii) fulfilled in the EDF context if
integrating over the unit circle $C_{1}$, even for MR energy kernels $\mathcal{E}[z]$ plagued by poles at
finite $z^{\pm}_{\mu}$ (iii) fulfilled for any integration circle $C_{R}$ by the regularized EDF-based
PNR method, noticing in addition that unphysical components do not contribute anymore.

\section{Applications}
\label{applications}
\label{substraction}

\subsection{General remarks}

As seen in Sec.~\ref{selfenergyselfpairing1} there are two distinct classes of spurious contributions
to a multi-reference energy density functional. The first one represents the "true" self-interaction and
self-pairing processes which already appear at the single-reference level. It does not provide MR energy
kernels with poles; hence, it does not cause divergences or steps in the PNR energy and does not break its
shift invariance. The second one is due to the use of the GWT out of its context to define MR energy functional
kernels from an underlying SR EDF that contains self-interaction and self-pairing contributions.

As outlined in Sec.~\ref{sect:selfenergyselfpairingHFB:self-energy}, correcting consistently for the standard
(true) self-interaction $\mathcal{E}^N_{SI}$, Eq.~(\ref{eq:sitrue}), is not an easy task; the correction
enters the variational equations already on the single-reference level and leads to a state-dependent
single-particle field \cite{perdew81a,Ull00aDFT,Leg02aDFT,Ruz07aDFT}. The same would hold regarding the
correction for true spurious self-pairing $\mathcal{E}^N_{SP}$, Eq.~(\ref{eq:sptrue}). For that reason,
and because such spurious contributions are not responsible for divergences and steps in the PNR energy,
we concentrate here on $\mathcal{E}^N_{CG}$, Eq.~(\ref{spurious:bi}) which is at the origin of the
specific and dramatic pathologies  encountered in PNR EDF calculations. Note that subtracting
$\mathcal{E}^{N}_{CG}$ from the PNR energy will also modify the variational equations of a VAP calculation.
Here, we confine ourselves to an analysis of the poles and of their impact on the particle-number restored
energy after the variation. In this case, $\mathcal{E}^N_{CG}$ is easily subtracted \emph{a posteriori}.

There is one important limitation to the applicability of the regularization method proposed in Paper~I and
applied in the present work. Although it is straightforward to extend Eq.~(\ref{spurious:bi}) to an EDF
depending on any integer powers of the density matrices, this is not the case for EDFs depending on non-integer
powers of the densities. This is a significant limitation, considering that most successful modern functionals use
density dependencies of non-integer power\footnote{An exception is the relativistic functional~\cite{bue02a}
used in the MR calculations of Nik{\v{s}}i{\'c} \etal\ \cite{niksic06b}.
}.
Indeed, this allows them to provide
a good description of the most important nuclear matter properties with a very small number of terms and coupling
constants to be adjusted phenomenologically~\cite{bender03b}. Also the widely used Slater approximation to the Coulomb
exchange term falls into the category of a density dependent term of non-integer power. We analyze the
spurious contributions to such category of functionals in Paper~III, complementing the study of Dobaczewski
\etal\ \cite{doba05a}. In the present work, however, we use instead the particular early parameterization
SIII~\cite{beiner75a} of the Skyrme EDF that contains only bilinear and trilinear terms in the normal density
matrix. We complement the SIII energy functional with a density-independent local pairing
functional that is bilinear in either the neutron or proton anomalous density matrix. For the Coulomb
energy functional, we only consider the direct term and neglect the approximate exchange term that was considered
in the fit of SIII. As a consequence, all calculated nuclei will be underbound by a few MeV, but this is of no
importance for the purpose of the present paper. Having said that, it is clear that the construction of high-precision
\emph{correctable} EDFs, i.e. only containing integer powers of the density matrices, represents an important task
for the future\footnote{In practice, one will have to restrict the form to rather low orders in the density matrices. For example, the EDF recently proposed by Baldo \emph{et al}.\ \cite{baldo08a} includes terms up to fifth power in the total density $\rho(\vec{r})$, which clearly lead to self-interaction terms~\cite{stringari78a} that will require a regularization containing quadruple sums over single-particle states, which will be too costly in realistic calculations.}
.

The calculation of the various contributions to the correction $\mathcal{E}_{CG}^{N}$ is outlined in
Appendix~\ref{app:functional}. The trilinear terms in the SIII functional are motivated by a local zero-range
three-body force which excludes terms of third order in the same nucleon density; it only contains terms of
the kind $\rho^2_n (\vec{r}) \rho_p(\vec{r})$ and $\rho^2_p (\vec{r}) \rho_n(\vec{r})$.
From a practical point of view, the absence of a genuine term of third power in the same density
matrix has the advantage that we do not have to invoke the corresponding correction term outlined in Paper~I.
Instead, the correction of the trilinear terms has the structure of the one of bilinear terms times the projected
density of the other species as outlined in Appendix~\ref{tri-linearcorrection}.

\subsection{Numerical Implementation}
\label{numerics}

In practice, the integrals over gauge angles are discretized
with a simple $n$-point trapezoidal formula
\begin{equation}
\label{eq:fomenko}
\frac{1}{\pi} \int_{0}^{\pi} d\varphi \; f(e^{i\varphi})
\Rightarrow \frac{1}{L} \sum_{l=1}^{L} f \left( e^{i\frac{\pi l}{L}} \right)
\end{equation}
where we assume the projection of a state with even number parity
on even particle number to reduce the integration interval to $[0,\pi]$.
As was shown by Fomenko~\cite{fom70a}, this simple scheme eliminates exactly
all components from the SR state which differ from the desired particle number
$N$ by up to $\pm 2(L\!-\!1)$ particles. Although the spread in particle number
is large compared to the total particle
number, already small values for $L$, ranging from 5 in light nuclei to 13 in
heavy ones, are sufficient to obtain a converged projected state.

It is customary to use an odd number of discretization points $L$ in the
interval $[0,\pi]$ to avoid numerical problems that may appear at $\phi = \pi/2$.
This practice does not relate to the real divergences of the energy functional
contained in $\mathcal{E}_{CG}^N$ that we discuss here, but avoids the implicit
division of $u_\mu^2 + v_\mu^2 \, e^{i \pi/2}$ contained in an operator kernel by
the same factor in the normalization factor $c_N^2$
when evaluating projected operator matrix elements (as, for example, particle number,
deformation or radii), which numerically will not give the analytical result 1 when
$u_\mu^2$ comes very close to $v_\mu^2$. Of course, the numerical representation
of the pole contained in the energy functional would not be very precise
in this case either.

With a small modification, the discretization~(\ref{eq:fomenko}) can also be used
to represent complex contour integrals with an arbitrary radius $R$
\begin{equation}
\label{eq:fomenkoc} \oint_{C_{R}} \! \frac{dz}{2i\pi} f(z) = \int_{0}^{\pi} \! \frac{d\varphi}{\pi} \, f( R \;
e^{i\varphi}) \Rightarrow \frac{1}{L} \sum_{l=1}^{L} f \left( R \;  e^{i\frac{\pi l}{L}} \right) .
\end{equation}
which we will use to examine the properties of the energy functional under shift
transformations.

For all results shown below, the SR calculations used as a starting point
were performed with an approximate particle-number
projection before variation within the Lipkin-Nogami approach to ensure that pairing
correlations are present in all SR states. Otherwise, pairing correlations
would collapse in the SR state whenever there is a large gap in the single-particle spectrum
around the Fermi surface.

The dependence of various quantities on axial quadrupole deformation is shown
in function of the dimensionless deformation of the mass density distribution
$\beta_2$ defined as
\begin{equation}
\label{eq:beta2}
\beta_2
= \sqrt{\frac{5}{16\pi}} \, \frac{4\pi}{3 R^2 A} \,
  \langle 2z^2 - y^2 - x^2 \rangle
\, ,
\end{equation}
where $R = 1.2 \, A^{1/3}$ fm.

\subsection{\nuc{18}{O}}

As a first example we discuss \nuc{18}{O}. It has the advantage that the density
of single-particle levels around the Fermi energy is sufficiently low that the impact
of the spurious contribution brought by each single-particle level to the projected
energy can be studied separately without having them interfere too much. The
integration radius $R_q=1$ is used until we come to discussing shift invariance.

\subsubsection{Convergence of Operator Matrix Elements}

\begin{figure}[t!]
\includegraphics{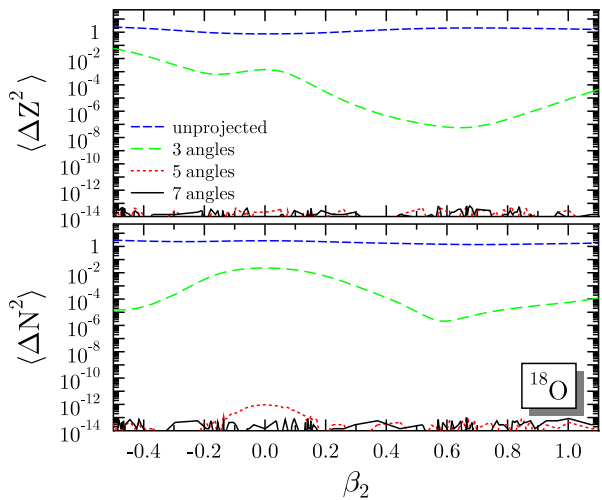}
\caption{\label{fig:o18:e:disp}
(Color online)
Dispersion of the proton and neutron number of the unprojected SR state
and the particle-number projected SR using 3, 5 or 7 discretization
points of the gauge-space integrals as a function of their deformation.
For 5 points the projected state is sufficiently converged, for 7 and more
points (not shown) the dispersion cannot be distinguished from numerical noise.
}
\end{figure}

Before we enter the discussion of the energy functional, we demonstrate the convergence
of the particle-number projection method for observables that are calculated as
expectation values of the corresponding operators in the projected states.
In the context of particle-number projection, the most sensitive observable is the
dispersion of particle number
$\langle \Delta N^2 \rangle = \langle \hat{N}^2 \rangle - \langle \hat{N} \rangle^2$,
a two-body operator that provides a measure for the quality of the
particle-number projected state as it has to be zero for an eigenstate
of the particle-number operator. For an (unprojected) SR state,
$\langle \Delta N^2 \rangle$ is proportional to its spread in particle-number space
\cite{Flocard97a}. One can see in Fig.~\ref{fig:o18:e:disp} that the Fomenko discretization converges quickly, already
$L=5$ gives excellent results for \nuc{18}{O}, and for $L \geq 7$ the dispersion of
particle number cannot be distinguished from numerical noise.

\begin{figure}[t!]
\includegraphics{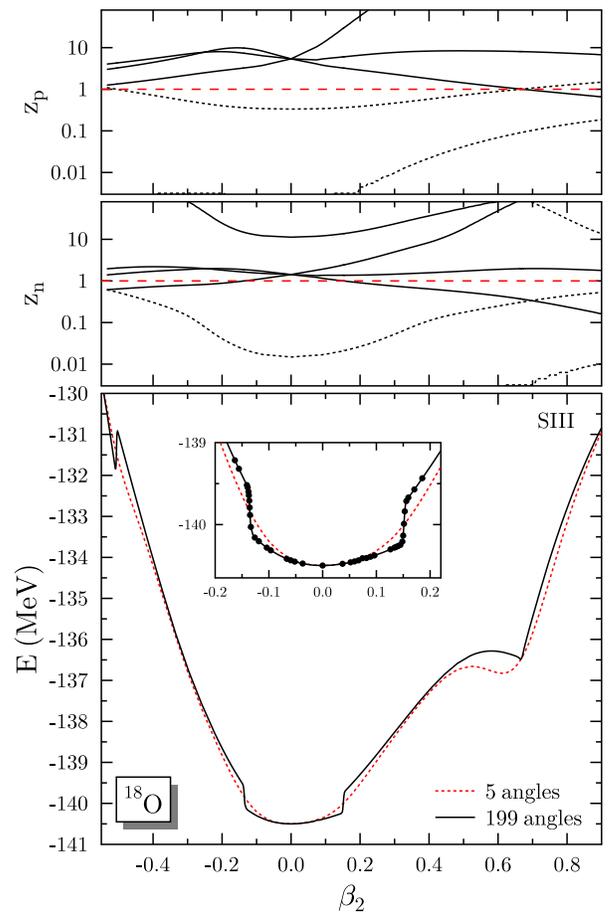}
\caption{\label{fig:o18:e:pole}
(Color online)
Spectrum of poles $z_\mu = |u_\mu/v_\mu|$ for protons (top panel) and neutrons
(middle panel), which for levels in the vicinity of the Fermi energy resembles
a stretched and slightly distorted Nilsson diagram. The dashed red line at $z=1$
denotes the radius of the standard integration contour $R=1$.
The bottom panel shows the particle-number projected quadrupole deformation energy
for $L=5$ and 199 discretization points for the integral in gauge space.
The insert shows a close-up of the steps at small deformation.
}
\end{figure}

\subsubsection{Regularized PNR Energy}

Unlike any operator expectation value, particle-number restored energies
do not converge when increasing the number of discretization points in the gauge-space
integrals, as already demonstrated in Fig.~\ref{divergence} for the parameterization SLy4.
Figure~\ref{fig:o18:e:pole} shows the projected deformation energy curve of \nuc{18}{O},
now calculated with SIII. What appears to be a smooth
deformation energy curve when calculating it with $L=5$,
develops steps and discontinuities when increasing the number of discretization
points to 199, i.e.\ when one starts to resolve the poles at finite $z_\mu^\pm$
close to the integration contour~\cite{doba05a}. For example, at small prolate
and oblate deformation $\beta_2 \approx \pm 0.15$, the energy jumps from a
lower deformation curve around the spherical point to a higher-lying one at
larger deformation. Using a small number of discretization points provides a curve
that smoothly interpolates between the two energy curves distinguished with $L=199$.
Figure~\ref{fig:o18:e:pole} also displays, as a function of the deformation, the poles
at $|z_\mu^\pm| = |u_\mu/v_\mu|$ that enter uncorrected energy kernels for protons
and neutrons. We follow Dobaczewski \etal\ \cite{doba05a} and plot $z^\pm$ instead of
a Nilsson diagram of single-particle energies, as divergences and steps appear where
poles cross the integration contour. Note again that the radius of the latter can be
chosen to be different from the standard value $R_q=1$ that is equivalent to the Fermi energy.

In Fig.~\ref{fig:o18:e:pole}, however, we do not yet make use of the freedom
to modify the integration contour and use the standard values $R_p = R_n = 1$.
It can be seen that the two steps developing at
$\beta_2 \approx \pm 0.15$ coincide with a pair of neutron levels originating from the
spherical $\nu$ $d_{5/2^+}$ shell that enters the integration contour either at
the prolate or the oblate deformation. It is noteworthy that the steps are not completely sharp
even when using $L=199$ points for the calculation, as can be seen from the markers in the
insert in the lowest panel. There also is a step at $\beta_2 = -0.5$ that
coincides with a pair of proton levels from the $\pi$ $p_{1/2^-}$ shell leaving the
integration contour. A particular case is the discontinuity at $\beta_2 = 0.7$
that coincides with the crossing of two different pairs of proton levels right on the integration contour.

It is worth noting that no divergence is seen in the PNR energy surface displayed in Fig.~\ref{fig:o18:e:pole}.
This is at variance to Fig.~\ref{divergence}. Indeed, SIII corresponds to a specific functional form such
that poles at $z= z^{\pm}_{\mu}$ are simple poles. This is due to the fact that the trilinear terms
entering SIII do not contain products of three density matrices referring to the same isospin. As
explained in Paper~III, this leads to a finite Cauchy principal value as the poles cross the integration
circle. Divergences appear only for poles of higher order.

\begin{figure}[t!]
\includegraphics{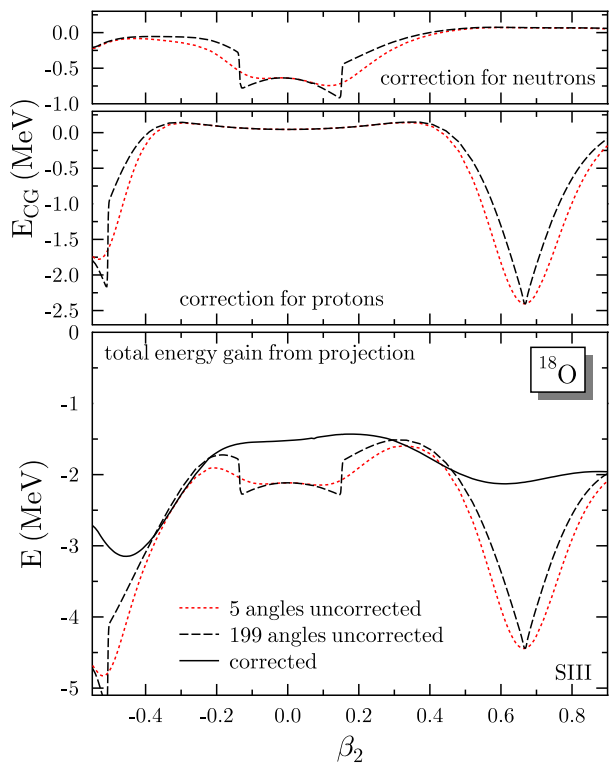}
\caption{\label{fig:o18:gain:c}
(Color online)
Correction for neutrons (top panel) and protons (middle panel) and
energy gain from projection without and with correction for \nuc{18}{O}
as a function of quadrupole deformation for 5 and 199 discretization
points for the integrals in gauge space. The corrected energy gain
in independent on the discretization of the integrals when 5 or more
angles are used. All panels share the same energy scale.
}
\end{figure}

The effects of particle-number restoration on the energy is partly masked in
Fig.~\ref{fig:o18:e:pole} by the genuine evolution of the energy with deformation. To obtain a clearer
picture, we show in the lower panel of Fig.~\ref{fig:o18:gain:c} the energy gain from
particle number restoration, obtained as the difference between the MR and SR energy functionals
for a given deformation of the SR state. For a cleaner comparison, the LN correction is removed from
the SR energy. The steps and discontinuities already seen in Fig.~\ref{fig:o18:e:pole} appear when
increasing $L$ from 5 to 199. The two upper panels show the correction $\mathcal{E}_{CG}^{N}$,
Eq.~(\ref{spurious:bi}), separately for protons and neutrons. The lower panel also shows the energy gain for the
regularized PNR energy surface $\mathcal{E}_{REG}^{N}$ obtained by subtracting the neutron and proton corrections
$\mathcal{E}_{CG}^{N}$ from the uncorrected PNR energy $\mathcal{E}^{N}$ for a given value of $L$.
The correction has many interesting and appealing features
\begin{itemize}
\item
The regularized PNR energy $\mathcal{E}_{REG}^{N}$ is independent on the discretization of the integral;
it is identical, within the numerical accuracy, for $L=5$ and 199. As a result, only one curve is shown
in Fig.~\ref{fig:o18:gain:c}.
\item
The previous result confirms that the entire dependence of the (uncorrected) PNR
energy on the discretization of the gauge space integral is contained in
$\mathcal{E}_{CG}^{N}$.
\item
Looking separately at protons and neutrons, the corresponding correction $\mathcal{E}_{CG}^{N}$ is
largest when a pole of a given nucleon species is close to the integration contour ($R=1$ here).
However, the correction is different from zero for the deformations in between; i.e.\ the spurious
nature of the poles is also felt when being away from divergences and steps.
\item
All terms in the energy functional (central, spin-orbit, pairing, Coulomb, etc)
contribute to $\mathcal{E}_{CG}^{N}$, with slightly
different magnitudes and different signs, so one has to strictly correct for all
of them. This is not unexpected as the source of the spuriosity we focus on here
is the weight the matrix elements $\bar{v}^{\rho\rho}$ and $\bar{v}^{\kappa\kappa}$ are
multiplied with in Eq.~(\ref{spurious:bi}), not the matrix elements themselves.
\item
The correction depends strongly on the deformation and will have a non-negligeable
impact on the topology of the deformation energy curve. The regularized
energy gain from projection is a much smoother function of deformation
than the uncorrected one, meaning that regularized particle-number restoration
will provide potential energy surfaces with less pronounced structures
than uncorrected PNR.
\item
The correction $\mathcal{E}_{CG}^{N}$ is of the order of 1 MeV. Of course it
has to be smaller than the energy gain from particle number restoration, which is a few MeV.
For \nuc{18}{O} however (and when calculated with SIII), the spurious contribution to the
uncorrected energy can be as large as $50 \, \%$ of the total energy gain at some deformations.
Also, one MeV error on the mass is larger than the targeted accuracy from EDF methods. In addition,
and as exemplified below, the correction to the mass varies from nucleus to nucleus.
\item
The regularized PNR energy gain can be both larger and smaller than the uncorrected one. In all
cases we have looked at so far, however, an increase obtained from the correction rests always
very small, while a reduction from correction might be quite substantial, but this might
not always be the case.
\end{itemize}

\begin{figure}[t!]
\includegraphics{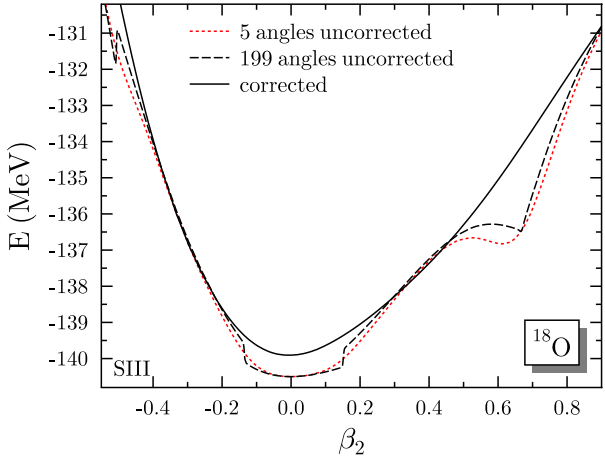}
\caption{\label{fig:o18:e:pes:c}
(Color online)
Corrected (solid line) and uncorrected (dotted and dashed lines)
particle-number projected quadrupole deformation energy for \nuc{18}{O}, calculated
with $L=5$ and 199 discretization points of the integral in gauge space.
The corrected curves are identical.
}
\end{figure}

The corrected deformation energy surface of \nuc{18}{O} is shown in Fig.~\ref{fig:o18:e:pes:c}
together with the uncorrected ones obtained with $L=5$ and $199$ as was
already displayed in Fig~\ref{fig:o18:e:pole}.
It is striking to see that the corrected PNR energy surface has less structure than
the uncorrected ones; its curvature changes now monotonically
and the shoulder at $\beta_2 = 0.6$, that always appears as a secondary minimum
in SR calculations without pairing for oxygen isotopes, disappears completely.
The latter does not mean \emph{a priori} that a regularized PNR plus configuration
mixing calculation will not give a collective state located at this deformation
anymore as it was obtained for \nuc{16}{O}~\cite{bender03c} and \nuc{20}{O}~\cite{severyukhin07a}
using SLy4. This question needs to be addressed in the near future by performing regularized MR
calculations including quadrupole shape configuration mixing.

\subsection{Detailed analysis of spurious contributions}

\subsubsection{Contributions of individual poles}

After discussing the behavior of the contaminated and regularized PNR energies of a nucleus as a
function of its quadrupole deformation, it is instructive to investigate the contribution
$\varepsilon_{\mu}$ of each canonical pair $(\mu, \bar{\mu})$ to the unphysical energy
$\mathcal{E}_{CG}^{N}$ that contaminate uncorrected MR energies $\mathcal{E}^{N}$. Formally,
each pair of single-particle levels provides a spurious contribution $\varepsilon^0_\mu$ through
the pole at $z=0$, in addition to the contribution $\varepsilon^\pm_\mu$ associated with the
unphysical poles at finite $z_\mu^\pm=\pm i|u_{\mu}/v_{\mu}|$, if the latter are located inside
of the integration contour of radius $R_q$. In the end, one can rewrite
Eq.~(\ref{spuriouscomplex}) as
\begin{equation}
\label{eq:spurious:decomp}
\mathcal{E}_{CG}^{N}
\equiv   \sum_{\mu > 0} \varepsilon_\mu
\equiv   \sum_{\mu > 0} \varepsilon^0_\mu
  + \sum_{\mu > 0 \atop |z_\mu^\pm| < R} \varepsilon^\pm_\mu
\, .
\end{equation}
The total contribution $\varepsilon_\mu$ is calculated numerically through
Eq.~(\ref{spurious:bi}) and might depend on the number of discretization
points $L$ used for the gauge-space integral. The partial contribution
$\varepsilon^\pm_\mu$ can be evaluated using the analytical expression
for the residue of the poles, Eq.~(\ref{residuezmu}), which does not depend
on the discretization of the gauge-space integrals. Finally,
$\varepsilon^0_\mu$ is equal to $\varepsilon_\mu$ when $|z_\mu^\pm| > R$,
while for $|z_\mu^\pm| < R$ it can be estimated through
$\varepsilon^0_\mu = \varepsilon_\mu - \varepsilon^\pm_\mu$.
As $\varepsilon^\pm_\mu$ is calculated analytically
while $\varepsilon_\mu$ is obtained numerically, the values obtained for
$\varepsilon^0_\mu$ might not be very precise when $|z_\mu^\pm| \approx R$.

It turns out that only a few pairs of levels located close to the
Fermi level give a non-zero contribution to $\mathcal{E}_{CG}^{N}$.
The relative size and behavior of these contributions as the spectrum of poles
changes can be understood by analyzing Eqns.~(\ref{spuriouscomplex})
and~(\ref{residuezmu}) for a few idealized cases.
For this discussion, the combination of matrix elements entering the expression of
$\mathcal{E}_{CG}^{N}$ can be ignored. The values of the matrix elements
depend of course on the actual pair of conjugated states they refer to and thereby scale
the contribution of a given level to $\mathcal{E}_{CG}^{N}$. However, the matrix elements
do not show a particular dependence on $\mu$ that determines the limit of
$\varepsilon_\mu$ for completely occupied or unoccupied levels. Therefore it is
sufficient to concentrate on the occupation-number dependent weight-factors in
Eqns.~(\ref{spuriouscomplex}) and~(\ref{residuezmu}).

\begin{figure}[t!]
\includegraphics{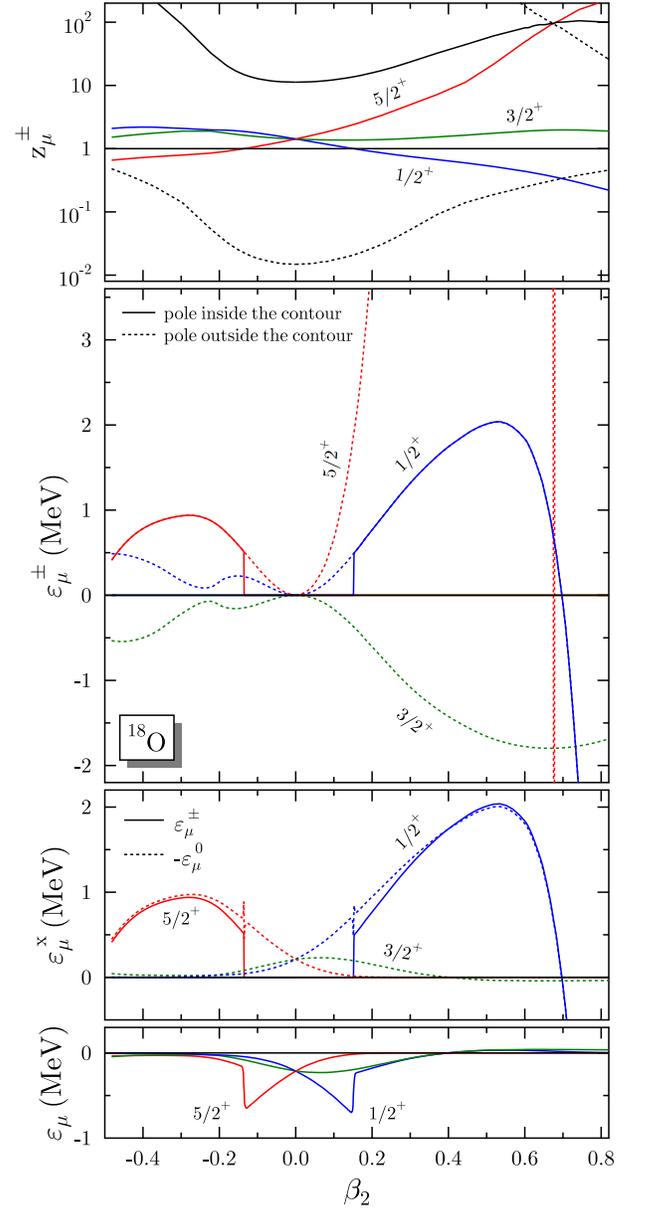}
\caption{\label{fig:o18:res:n}
(Color online)
Spurious energy from the single-particle orbits that correspond to the
spherical neutron $d_{5/2^+}$ level in \nuc{18}{O} as a function of
quadrupole deformation (see text).
}
\end{figure}

Figure~\ref{fig:o18:res:n} separates the various contributions to
$\mathcal{E}_{CG}^{N}$ for the three pairs of canonical orbits that originate
from the spherical neutron $d_{5/2^+}$ level in \nuc{18}{O}. The top panel of
Fig.~\ref{fig:o18:res:n} displays the location of the three poles of interest
on the imaginary axis. Those three pairs of poles are explicitly labeled by the
$j_z$ quantum number denoting the projection of the angular momentum on the
symmetry axis. Other poles are left unmarked. The three other
panels show $\varepsilon_\mu$, $\varepsilon^0_\mu$
and $\varepsilon^\pm_\mu$ for the three pairs of $d_{5/2^+}$ levels only, as
these entirely determine the neutron contribution to
$\mathcal{E}_{CG}^{N}$ for the deformations shown\footnote{At large oblate
and prolate deformation, the $\varepsilon^\pm_\mu$ of the
other levels approaching $z=1$ are of the same order as those shown, but
make the plot difficult to read and do not add crucial information.}.

The second panel from the top shows $\varepsilon^\pm_\mu$. Solid lines
denote $\varepsilon^\pm_\mu$ when the corresponding pole is inside
the integration contour ($R_n =1$ here), while dotted lines denote
$\varepsilon^\pm_\mu$ when the pole is outside. Only the former of the
two contributes to $\mathcal{E}_{CG}^{N}$.
As $\varepsilon^\pm_\mu$ is usually finite when the corresponding
pole crosses the integration contour, its size determines
the step left in the PNR deformation energy curve.
To understand how $\varepsilon^\pm_\mu$ changes as a function of the location
of the corresponding pole $z^\pm_\mu$ within the spectrum of the other poles, Eq.~(\ref{residuezmu})
has to be analyzed further.  The product over $\nu \neq \mu$ in this expression can
be estimated by first considering that there are $k_r$ pairs of levels with
$|z^{\pm}_{\xi}| \ll |z^{\pm}_{\mu}|$, such that their contribution to the
product can be approximated by
\begin{eqnarray}
\prod_{\xi = 1}^{k_r}
\frac{u^{2}_{\xi} v^{2}_{\mu} - v^{2}_{\xi} u^{2}_{\mu}}{v^{2}_{\mu}}
& \approx & (-)^{k_r} |z^{\pm}_{\mu}|^{2k_r} \prod_{\xi = 1}^{k_r} v_\xi^2
\, .
\end{eqnarray}
For a small number $k_f$ of pairs of levels, $|z^{\pm}_{\nu}|$ is of the same order as
$|z^{\pm}_{\mu}|$, such that the full factor in the product has to be kept. Finally, all remaining
levels are such that $|z^{\pm}_{\mu}| \ll |z^{\pm}_{\lambda}|$ and the product can again be simplified
\begin{eqnarray}
\prod_{\lambda = k_r+k_f+1}^{\infty}
\frac{u^{2}_{\lambda} v^{2}_{\mu} - v^{2}_{\lambda} u^{2}_{\mu}}{v^{2}_{\mu}}
& \approx & \prod_{\lambda = k_r+k_f+1}^{\infty} u_\lambda^2
\, .
\end{eqnarray}
In practical calculations one works with a limited number of pairs of levels $k_t$ in the basis. This cutoff,
however, has no consequence for the contribution $\varepsilon^\pm_\mu$ from a pair of levels
($\mu, \bar\mu$) below the cutoff, as for all reasonable cutoffs the discarded pairs of levels
contribute a factor 1 to the product in Eq.~(\ref{residuezmu}). Altogether one obtains
\begin{eqnarray}
\label{residuezmu2}\varepsilon^\pm_\mu &\propto& \lefteqn{ u_\mu^4 v_\mu^4 \; \mathcal{R}e^{N}_{CG} (z^{\pm}_{\mu})
} \nonumber \\
& \approx & (-)^{k_r+N/2+1} \;
              u_\mu^2 \,
              |z^{\pm}_{\mu}|^{2k_r - N} \! \! \!
              \prod_{\xi = 1 \atop |z^{\pm}_{\xi}| \ll |z^{\pm}_{\mu}|}^{k_r} \! \! \! v_\xi^2
              \nonumber \\
&         & \times \! \! \!
              \prod_{\nu = k_r+1 \atop \nu \neq \mu}^{k_r+k_f+1} \! \! \!
              u_\nu^2 \left( 1 - \frac{|z^{\pm}_{\mu}|^2}{|z^{\pm}_{\nu}|^2} \right) \!
              \prod_{\lambda = k_r+k_f+1 \atop |z^{\pm}_{\mu}| \ll |z^{\pm}_{\lambda}|}^{k_t} \! \! \! u_\lambda^2
\end{eqnarray}
where we assume even particle number $N$. Equation~(\ref{residuezmu2}) allows for the complete explanation of the
global behavior of $\varepsilon^\pm_\mu$ seen in Fig.~\ref{fig:o18:res:n}.

First, for a bilinear functional as discussed here,
$\varepsilon^\pm_\mu$ is zero whenever the pair of levels ($\mu, \bar\mu$)
is degenerate with another pair $(\nu, \bar{\nu})$, i.e. $|z^{\pm}_{\mu}| = |z^{\pm}_{\nu}|$, as in this case
the middle product in Eq.~(\ref{residuezmu2}) contains a factor zero. In
fact, this is a direct consequence of the disappearance of the pole at
$z^{\pm}_{\mu}$ in the PNR energy kernel, as the dangerous remaining denominator is now
canceled by an additional factor in the norm kernel\footnote{This results holds for any bilinear functional in the density matrix of
a given isospin, even if it is multiplied with the densities of the other one.
When allowing for higher-order functionals, however, a term of order
$n$ in the density matrix can generate a pole at $z^{\pm}_{\mu}$ of order (at most) $(n-1)$. In order for $\varepsilon^{\pm}_{\mu}$ to be 0,
one needs the pole at $z^{\pm}_{\mu}$ to disappear altogether, which requires
$(n-1)$ additional factors from the norm kernel to cancel the denominator
$(u^{2}_{\mu}+ v^{2}_{\mu} \, z^{2})^{-(n-1)}$. Thus, the pair of interest
($\mu$, $\bar{\mu}$) needs to be degenerated (at least) with $(n-1)$ other pairs
for $\varepsilon^{\pm}_{\mu}$ to be 0. As a consequence, $\varepsilon^{\pm}_{\mu}$
will \emph{not} be 0 at a simple level crossing when working with a trilinear
(or higher-order) energy functional in the same isospin.}. This alone
already indicates that the contribution $\varepsilon^\pm_\mu$ of a given pair of
levels might fluctuate rapidly when the spectrum of poles $|z^{\pm}_{\mu}|$ changes as a
function of a collective coordinate.
The $(-)^{k_r}$ factor in Eq.~(\ref{residuezmu2}), whose sign depends on the number of pairs of levels
$k_r$ located below the pair ($\mu$, $\bar\mu$), makes $\varepsilon^\pm_\mu$ to change sign through
a crossing with another pair.  Figure~\ref{fig:o18:res:n} contains
several such examples. The downsloping $j_z = 1/2^+$ substate from the $d_{5/2}$ spherical shell
crosses with an upsloping level at large prolate deformation. There, $\varepsilon^\pm_\mu$ changes its sign as $k_r$
changes by 1 through the crossing. At spherical
deformation, where the three pairs of $d_{5/2}$ levels are degenerate, each of them crosses with the
two others and $k_r$ changes either by 2 (for the $j_z = 1/2^+$
and $j_z = 5/2^+$) or 0 (for the $j_z = 3/2^+$), such that the corresponding
$\varepsilon^\pm_\mu$
do not change their sign. A very particular case is the  subsequent crossing of the
upsloping $j_z = 5/2^+$ level with two other levels within a very small interval around
$\beta_2 \approx 0.63$. As the three levels do not cross at exactly the same deformation,
 $\varepsilon^\pm_\mu$ changes its sign twice in a tiny
deformation interval, oscillating between values far outside the vertical energy interval shown,
that cannot be resolved by what appears as a single vertical (red) dotted line in the
plot at $\beta_2 = 0.67$.

Second, let us consider the case of a pair ($\mu, \bar\mu$) that is well separated from all others. Thus, there remains only
two categories of "other" states in Eq.~(\ref{residuezmu2}), $k_r$ pairs of levels $(\xi,\bar{\xi})$ with
$|z^{\pm}_{\xi}| \ll |z^{\pm}_{\mu}|$ and $k_t-k_r-1$  pairs of levels $(\lambda,\bar{\lambda})$  with
$|z^{\pm}_{\lambda}| \gg |z^{\pm}_{\mu}|$. One has still to
distinguish between the two cases where $|z^{\pm}_{\mu}|$ is larger or smaller than 1.

We start with the case $|z^{\pm}_{\mu}| = |u_\mu / v_\mu| > 1$ for which the $u_\mu^2$ factor in
Eq.~(\ref{residuezmu2}) rapidly converges towards 1 as $|z^{\pm}_{\mu}|$ increases. In this case,
the number of pairs below the pair $(\mu, \bar{\mu})$ is larger than half the particle number;
i.e.\ $k_r > N/2$. For $k_r=N/2+1$, $\left| \varepsilon^\pm_\mu \right|$ grows linearly with
$|z^{\pm}_{\mu}|$ for $|z^{\pm}_{\mu}| > 1$, for $k_r = N/2+2$ it grows quadratically etc, but
always only until it approaches another level, where $\left| \varepsilon^\pm_\mu \right|$ goes
back to 0 as a consequence of the degeneracy as described above. After the
crossing, however, $\left| \varepsilon^\pm_\mu \right|$ grows
again, although one of the $u_\lambda^2 \approx 1$ factors in Eq.~(\ref{residuezmu2}) has changed into a $v^2_\xi
\ll 1$ factor at the crossing. At the same time, the number of pairs $k_r$ below the pair ($\mu, \bar\mu$) has
grown by one, such that after the crossing there is an additional $|z_\mu^\pm|^2 = u^2_\mu/v^2_\mu$ factor, that
overcompensates the effect of the occupation factor $v^2_\xi$ from the level just crossed, as
$v^2_\xi > v^2_\mu$ and $v^2_\mu < 1/2$ give $v^2_\xi \, u^2_\mu / v^2_\mu > 1$. For the simultaneous crossing
with more than one level, the net effect is the product of the change brought by each crossed level. For poles far
from the Fermi level, the values of $\varepsilon^\pm_\mu$ can be very large. For example, the
$\varepsilon^\pm_\mu$ of the $j_z = 5/2^+$ level reaches about 550 MeV around $\beta_2 = 0.42$ where the
corresponding pole $|z^{\pm}_{\mu}|$ is well isolated in the spectrum, drops below zero and rises immediately back
when it crosses a pair from a higher-lying spherical $j$ shell, and quickly rises to values larger than $10^5$ MeV,
dropping back to zero right away as the pole crosses the next pair, and quickly gaining a value again several
orders of magnitude larger. The sheer size of these values that quickly grow beyond any physical scale that
appears in the EDF description of nuclei clearly shows that $\varepsilon^\pm_\mu$ alone cannot be a meaningful
quantity in a well-defined theory. The only reason why the $\varepsilon^\pm_\mu$ of these high-lying levels with
$|z^\pm_{\mu}| \gg 1$ do not make $\mathcal{E}_{CG}^{N}$ incommensurably large is that the corresponding poles are outside of the
standard integration circle and thus do not contribute. We will come back to this when discussing PNR with shifted contour integrals below.

For a sufficiently isolated level below the Fermi level,
$|z^{\pm}_{\mu}| = |u_\mu / v_\mu| < 1$, $\left| \varepsilon^\pm_\mu \right|$ also tentatively grows when $|z^{\pm}_{\mu}|$
goes towards 0. This is now a consequence of the fact that $k_r \leq N/2$, such that
$\varepsilon^\pm_\mu$ scales with powers of the inverse of
$|z^{\pm}_{\mu}|$. At each crossing with a lower lying pair of levels, the additional $u^2_\lambda \ll 1$
factor is overcompensated by the additional $|z^{\pm}_{\mu}|^{-2}$ factor
from the decreasing number of pairs $k_r$ below . Again, $\varepsilon^\pm_\mu$
goes to 0 at level crossings and changes its sign depending on the number
of pairs crossed.

An important consequence of Eq.~(\ref{residuezmu2}) and the discussion
above is that the $\varepsilon^\pm_\mu$ of an isolated pair is smallest when
there are exactly $k_r = N/2$ pairs of other levels below it, which is
usually the case for a level with its pole $z^{\pm}_{\mu}$ close to
the Fermi level. A side effect is that the spurious step due to a pair crossing
the integration contour remains rather small when the latter is chosen as the
unit circle. This is to put in perspective with the rather small spurious steps
observed in Fig.~\ref{fig:o18:gain:c} and contaminating the unregularized PNR energy
computed using a unit integration circle. We will see in the following that
the situation would have been more dramatic if we had used different contours.

As discussed in Sec.~\ref{complexplane}, poles at finite $z_\mu^\pm$
entering or leaving the integration contour are the origin of the spurious steps in PNR energy surfaces,
as the corresponding (usually finite) $\varepsilon^\pm_\mu$ is suddenly
added to or removed from $\mathcal{E}_{CG}^{N}$, respectively. In the
second panel of Fig.~\ref{fig:o18:res:n}, contributions from poles inside
or outside the standard integration contour of radius $R = 1$ are plotted
as solid or dotted lines, respectively, to make this distinction.
The third panel from the top also shows $\varepsilon^\pm_\mu$ with solid
lines, but now only when it actually contributes to $\mathcal{E}_{CG}^{N}$.
The dotted lines represent $-\varepsilon^0_\mu$,
such that the distance between the curves for $\varepsilon^\pm_\mu$ and
$-\varepsilon^0_\mu$ provides the
total contribution $\varepsilon_\mu$ from the pair $(\mu, \bar{\mu})$ to $\mathcal{E}_{CG}^{N}$~\footnote{The spikes
of $\varepsilon^0_\mu$ at the deformations where the contribution from
$\varepsilon^\pm_\mu$ to $\mathcal{E}_{CG}^{N}$ jumps to 0 are of numerical
origin.}.

The results for the neutron levels depicted in Fig.~\ref{fig:o18:res:n}
suggest that $\varepsilon^\pm_\mu$ converges towards $-\varepsilon^0_\mu$
when $z^\pm_\mu$ goes to zero, i.e.\ for deeply-bound levels far below
the Fermi energy, such that the total contribution $\varepsilon_\mu$ is zero
for deeply-bound levels. When $z^\pm_\mu$ approaches the Fermi energy from
below, $\varepsilon^\pm_\mu$ and $-\varepsilon^0_\mu$ slowly grow apart.
Still, for all examples we have looked at, $\varepsilon^0_\mu$ and
$\varepsilon^\pm_\mu$ remain of similar size, but opposite sign, and
have a similar dependence on deformation around
the Fermi energy, $z^\pm_\mu \approx 1$. They do not cancel exactly
when the pole at $z_\mu^\pm$ approaches the Fermi level
but the difference between $\varepsilon^\pm_\mu$ and $-\varepsilon^0_\mu$
remains much smaller than the individual contributions and provides the finite
and smoothly varying spurious energy $\mathcal{E}^{N}_{CG}$ between the steps.
For levels far above the Fermi level, $\varepsilon^0_\mu$ goes to zero.
Also, the pole $z^\pm_\mu$ is beyond the integration contour and $\varepsilon^\pm_\mu$ does not contribute to $\varepsilon_\mu$
either. Consequently, levels far above the Fermi energy do not
contribute to $\mathcal{E}_{CG}^{N}$ for standard integration contours at $R_q=1$.

The behaviors described above can be understood as limiting cases of the factor $u^4_\mu v^4_\mu$
times the contour integral in Eq.~(\ref{spuriouscomplex}). Omitting
unimportant prefactors, one obtains for $|z^\pm_\mu| \to 0$, that is for
$u^{2}_{\mu} \approx 0$ and $v^{2}_{\mu} \approx 1$, that
\begin{eqnarray}
\label{limitsintegral1}
\varepsilon_{\mu}
& =       & \varepsilon^0_\mu + \varepsilon^\pm_\mu
            \nonumber \\
& \propto & u^{4}_{\mu} \, v^{4}_{\mu}
            \oint_{C_1} \! \frac{dz}{2 i \pi} \frac{1}{z^{N+1}}
            \frac{(z^{2} - 1)^{2}}
                 {v^{4}_{\mu} \, z^{4}}
            \prod_{\nu > 0} (u_{\nu}^2 + v_{\nu}^2 \, z^{2})
            \nonumber \\
& \propto & |z^{\pm}_{\mu}|^{4} \,
            \left( c^{2}_{N} - 2 \, c^{2}_{N+2} + c^{2}_{N+4}\right)
            \nonumber \\
& \to     & 0
\, ,
\end{eqnarray}
while for $|z^\pm_\mu| \to \infty$, that is for $u^{2}_{\mu} \approx 1$ and $v^{2}_{\mu} \approx 0$, one has
\begin{eqnarray}
\label{limitsintegral2}
\varepsilon_{\mu} &=& \varepsilon^0_\mu \nonumber \\
& \propto & u^{4}_{\mu} \, v^{4}_{\mu} \!
            \oint_{C_1} \! \frac{dz}{2 i \pi } \frac{1}{z^{N+1}}
            \frac{(z^{2} - 1)^{2}}{u^{4}_{\mu}}
            \prod_{\nu > 0} (u_{\nu}^2 + v_{\nu}^2 \, z^{2})
            \nonumber \\
& \propto & |z^{\pm}_{\mu}|^{-4} \,
            \left( c^{2}_{N-4} - 2 \, c^{2}_{N-2} + c^{2}_{N}\right)
            \nonumber \\
& \to     & 0
\, .
\end{eqnarray}
where the $c_{N}$ denote in both cases the amplitudes of the normalized projected
states with particle number $N$ in the SR state, Eq.~(\ref{polescontribution_a}),
all of which are usually non-zero and independent of $\mu$.
The key element to obtain both limits is that the integral over the gauge angle
becomes simply proportional to $v^{-4}_{\mu} \approx 1 $ or $ u^{-4}_{\mu} \approx 1$, respectively.
As a result, the prefactor $u_{\mu}^4 \, v_{\mu}^{4}$ dominates and drives
$\varepsilon_{\mu}$ towards zero in both cases. As a consequence, one indeed finds
as a general rule that
\begin{alignat}{3}
\label{residuezmu4}
\varepsilon_{\mu} = \varepsilon^{0}_{\mu}   & \to 0   \qquad                   & \text{for} & \qquad z^{\pm}_{\mu} & \to \infty  \, ,
                       \\
\label{residuezmu5}
\varepsilon_{\mu}  =\varepsilon^{\pm}_{\mu}+  \varepsilon^{0}_{\mu}   & \to  0                           & \text{for} & \qquad z^{\pm}_{\mu} & \to 0 \, ,
\end{alignat}
as suggested by the numerical results in Fig.~\ref{fig:o18:res:n}.

Unlike $\varepsilon^\pm_\mu$, the contribution $\varepsilon^0_\mu$ to
the physical pole at $z=0$ is not a priori suppressed for degenerate levels and might have
a non-zero value. For deep-hole states, this seems contradictory with the previous proof that
$\varepsilon_\mu = \varepsilon^\pm_\mu + \varepsilon^0_\mu$ goes to zero. In fact, when the pair $(\mu,\bar{\mu})$ crosses
another one $(\zeta,\bar{\zeta})$, not only the pole at $z^{\pm}_{\mu}$
is removed but the residue of the pole at $z=0$ is strongly affected by
the disappearance of the corresponding denominator. As a result,
$\varepsilon^{0}_{\mu}$ also goes towards zero as $\varepsilon^{\pm}_{\mu}$
goes to zero. Indeed, $\varepsilon_{\mu}$ right at the crossing behaves as
\begin{eqnarray}
\label{limitsintegral3}
\varepsilon_{\mu} &=&  \varepsilon^{0}_{\mu} \nonumber \\
& \propto & u_{\mu}^4 \, v_{\mu}^{4} \!
            \oint_{C_{1}} \! \frac{dz}{2 i \pi} \frac{1}{z^{N+1}} \,
            (z^2 - 1)^2 \!
            \prod_{\nu > 0 \atop \nu \neq \mu, \zeta} (u_{\nu}^2 + v_{\nu}^2 \, z^{2})
            \nonumber \\
& =       & u_{\mu}^4 \, v_{\mu}^{4}
            \Big( c^{2}_{N-4}[\mu,\zeta] - 2 \, c^{2}_{N-2}[\mu,\zeta] + c^{2}_{N}[\mu,\zeta] \Big)
\, ,
\end{eqnarray}
where $c^{2}_{N}[\mu,\zeta]$ denotes a modified norm obtained by removing the contributions
of both pairs ($\mu$, $\bar{\mu}$) and ($\zeta$, $\bar{\zeta}$) from the usual norm kernel
\begin{eqnarray}
 \label{denominator3}
c^{2}_{N}[\mu,\zeta]
& \equiv & \oint_{C_{1}} \! \frac{dz}{2i\pi} \, \frac{1}{z^{N+1}} \,
           \prod_{\nu > 0 \atop \nu \neq \mu, \zeta} \big( u_{\nu}^2 + v_{\nu}^2 \, z^{2} \big)
\, .
\end{eqnarray}
Considering either rather deep-hole or highly-lying single-particle states, the prefactor $(u_{\mu} \, v_{\mu})^{4}$
appearing in Eq.~(\ref{limitsintegral3}) makes $\varepsilon_{\mu} = \varepsilon^{0}_{\mu}$ to be small.

The bottom panel of Fig.~\ref{fig:o18:res:n} shows the total contribution $\varepsilon_\mu$ of each selected
pairs to $\mathcal{E}_{CG}^{N}$. One can now clearly see that there is more to the spurious energy than just
the steps and the divergences (the latter of which do not appear for the particular functional used here).
The poles $z^\pm_{\mu}$ associated to the $j_z=3/2^+$ pair remain outside the integration
contour for all deformations. Thus, it does not produce a step as the corresponding
$\varepsilon^\pm_{\mu}$ never contributes to $\mathcal{E}_{CG}^{N}$. Still,
this level gives a small contribution $\varepsilon^0_{\mu}$ to the spurious
energy through the pole at $z=0$, which happens to be slightly larger for
prolate deformations than for oblate ones.

Starting on the oblate side, only the pole at $z=0$ contributes at first to the
spurious energy from the $j_z=1/2^+$ pair of levels. The corresponding $\varepsilon^0_\mu$
increases slowly from zero with increasing $\beta_2$. The moment the corresponding
poles $z^\pm_{\mu}$ enter the integration contour at $\beta_2 = 0.15$,
$\varepsilon^\pm_{\mu}$ suddenly contributes to the spurious energy. We already
saw that the finite value of $\varepsilon^\pm_{\mu}$ at this point determines
the step. As $\varepsilon^\pm_{\mu}$ approaches $-\varepsilon^0_{\mu}$ when the $1/2^+$
levels become more and more occupied with increasing prolate deformation, the total contribution
$\varepsilon_{\mu}$ of the $1/2^+$ pair to  $\mathcal{E}_{CG}^{N}$ now decreases after the
step. With the total contribution $\varepsilon_{\mu}$ increasing on one side of the step and
suddenly decreasing on the other, the curvature of the spurious energy is different on both sides
of a step. As a consequence, removing $\mathcal{E}_{CG}^{N}$ modifies the improper curvature
of the uncorrected deformation energy surface that is visible in Fig.~\ref{fig:o18:e:pes:c},
even when the steps themselves are not numerically resolved. The regularized deformation energy
surfaces show much less structure; in \nuc{18}{O} to the extreme that the curvature of the corrected
energy surface is now positive for all  deformations as shown in Fig.~\ref{fig:o18:e:pes:c}. The
contribution from the $5/2^+$ levels to the spurious energy behaves very much as the one from
the $1/2^+$ levels with oblate and prolate sides exchanged. The sum of the three individual
contributions gives the neutron correction shown in the top panel of Fig.~\ref{fig:o18:gain:c}
for $L=199$.

\begin{figure}[t!]
\includegraphics{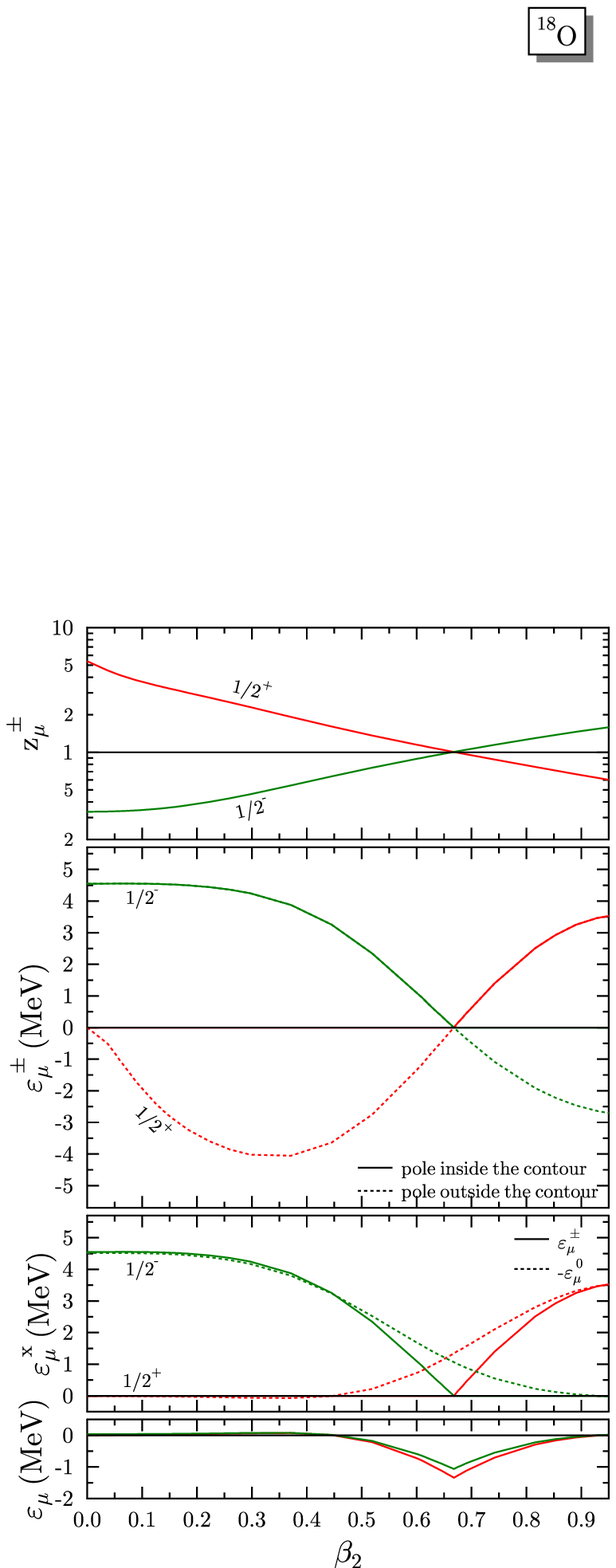}
\caption{\label{fig:o18:res:p}
(Color online)
Same as Fig.~\ref{fig:o18:res:n}, but for the proton $1/2^+$ and $1/2^-$
levels that give the dominant contributions to the proton correction at prolate
deformation and cross at the Fermi energy at $\beta_2 = 0.67$.
}
\end{figure}

We have seen that for a bilinear functional, the steps are always the consequence
of a pair of single poles $z_\mu^\pm$ crossing the integration contour and have the size
of the corresponding $\varepsilon^\pm_\mu$ at that crossing. The steps cannot add
up for a bilinear functional as, for degenerate poles with $\mu \neq \mu'$, $z^\pm_\mu = z^\pm_{\mu'}$
directly lead to $\varepsilon^\pm_\mu = \varepsilon^\pm_{\mu'}=0$.
This does not mean, however, that there is no spurious contribution to the PNR
energy when two poles cross the integration contour simultaneously, as the corresponding
$\varepsilon^0_\mu$ and $\varepsilon^0_{\mu'}$ are in general non-zero. In fact,
Fig.~\ref{fig:o18:gain:c} demonstrates clearly that, in our calculation of \nuc{18}{O} with
SIII, the spurious energy  $\mathcal{E}_{CG}^{N}$ is largest exactly where two proton
levels cross at the Fermi energy at $\beta_2 = 0.67$. The contribution of these two pairs of
levels to $\mathcal{E}_{CG}^{N}$, which also happen to be the only proton levels that
have a finite contribution at prolate deformation, is analyzed in Fig.~\ref{fig:o18:res:p}.
There is a number of interesting differences with Fig.~\ref{fig:o18:res:n}:
(i) The contribution $\varepsilon^\pm_\mu$ does not vanish at spherical shape for the $1/2^-$
levels for a bilinear functional. Indeed, the spherical $p_{1/2}$ shell is only doubly degenerate,
which does not suppress the corresponding $\varepsilon^\pm_\mu$. In fact, only $s_{1/2}$ and
$p_{1/2}$ levels with poles $z^\pm_\mu$ below the integration contour provide non-zero
$\varepsilon^\pm_\mu$ at spherical shape.
(ii) Both pairs cross right at the Fermi energy at $\beta_2 = 0.67$. For the standard
choice $R_p = 1$, their poles $z^\pm_\mu$ thus cross on
the integration contour. As a result, $\varepsilon^\pm_\mu$ from both pairs are zero, and change sign at the crossing.
(iii) The derivative of
$\varepsilon^\pm_\mu$ is not zero for both pairs when they simultaneously cross
the Fermi energy. By contrast, $\varepsilon_\mu^0$ slowly approaches zero such that the total contribution
$\varepsilon_\mu$ is quite large for the two proton pairs.
When the poles $z^\pm_\mu$ approach the integration contour from below, the distance
between $\varepsilon^\pm_\mu$ and $\varepsilon_\mu^0$ grows for both pairs. After the poles have crossed the contour, only
the $\varepsilon_\mu^0$ contribute. Finally, the total contribution $\varepsilon_\mu$ from each pair that crosses with
another at the integration contour is largest at the crossing, and decreases towards
zero on both sides. The sum of the two individual contributions gives the proton correction
shown in the middle panel of Fig.~\ref{fig:o18:gain:c} for $L=199$;
all other proton levels are too far away from the Fermi level to provide any visible contribution.

One can take advantage of the fact that only a very limited number of levels
actually contributes to $\mathcal{E}_{CG}^{N}$ in order to reduce the
numerical effort. Evaluating the necessary matrix elements $\bar{v}^{\rho\rho}$
and $\bar{v}^{\kappa \kappa}$ only for those levels for which the weight is
significantly different from zero is particularly welcome for the expensive
contribution from the Coulomb interaction.

\subsubsection{Shift Invariance}
\label{sect:appl:shift}

\begin{figure}[t!]
\includegraphics{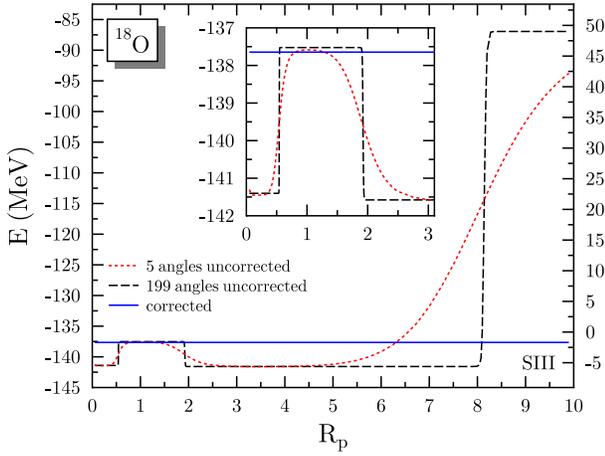}
\caption{\label{fig:o18:e:cont}
(Color online)
Projected energy for \nuc{18}{O} at the deformation $\beta_2 = 0.371$ as a function
of the radius $R_p$ of the integration contour calculated without and with correction using
5 and 199 angles. The energy scale on the left gives the absolute energy, the scale on the
right the energy gain from projection. The insert magnifies
the curves around $R_p = 1$. The regularized PNR energy in independent on the
discretization of the integrals when 5 or more angles are used.
The integration contour for projection on neutron number is $R_n = 1$ in all cases.
}
\end{figure}

In their recent paper, Dobaczewski \etal\ \cite{doba05a} pointed out that the
(uncorrected) PNR energy density functional is not shift invariant,
i.e.\ PNR energies depend on the radius chosen for the contour integral
in the complex plane. As outlined in Secs.~\ref{shift} and~\ref{sumrule},
the source of violation of the shift invariance is the contribution
$\varepsilon^\pm_\mu$ from the poles at $z^{\pm}_{\mu}$ inside the integration contour $C_{R}$ to the spurious
energy $\mathcal{E}_{CG}^{N}$ in Eq.~(\ref{eq:spurious:decomp}).
Each time a pole $z^{\pm}_{\mu}$ enters or leaves the integration contour
when changing its radius, $\mathcal{E}_{CG}^{N}$ changes by the amount
$\varepsilon^\pm_\mu$. This is illustrated in Fig.~\ref{fig:o18:e:cont}
for \nuc{18}{O} at $\beta_2 = 0.371$. The radius
of the contour used for neutrons is held fixed at $R_n = 1$, while the radius
of the contour used for the protons is varied. The three steps visible in
Fig.~\ref{fig:o18:e:cont} correspond to the three proton poles located at $0.1 < z^\pm_\mu <10$
visible in Fig.~\ref{fig:o18:e:pole} for the deformation of interest.

An interesting feature of the steps is that their size grows as the integration contour is shifted
away from $R_p=1$~\cite{doba05a}, i.e.\ away from the Fermi level. The reason is easy to understand
from the discussion of Eq.~(\ref{residuezmu2}) given in the previous section: $\varepsilon^\pm_\mu$
increases as $|z^{\pm}_{\mu}|$ moves away from 1 (as long as it is separated from other poles) and
as the difference between the number $k_r$ of pairs of states below $(\mu,\bar{\mu})$ and half the
number of particles $N/2$ one is restoring grows.

Using the small number of $L=5$ discretization points for the gauge-space integral does not
resolve the steps in the uncorrected PNR energy; only with much larger $L$ one obtains
sharp steps. By contrast, and as seen in Fig.~\ref{fig:o18:e:cont}, the regularized PNR energy
is constant within a numerical precision of the order 1 keV as $R_p$ is modified and $L$
increased beyond 5.

\begin{figure}[t!]
\includegraphics{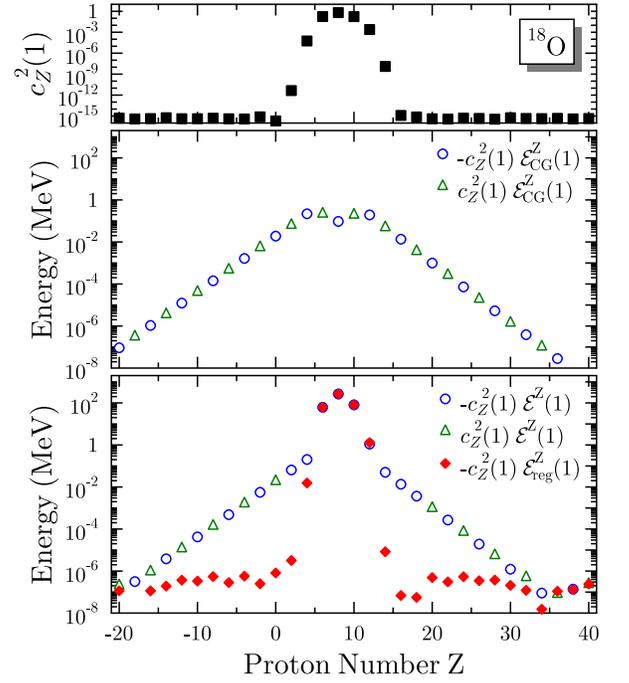}
\caption{\label{fig:decomposition:r1.0}
(Color online)
Weight $c_Z^2(R_p\!=\!1) = |\langle \Psi^Z | \Phi_1 \rangle|^2$ of the normalized proton-number
projected states in the SR HFB state (upper panel), the weighted spurious energy
$c_{Z}^{2}(R_p\!=\!1) \, \mathcal{E}^{Z}_{CG}(R_p\!=\!1)$ (middle panel), the non-regularized
weighted PNR energies $c_{Z}^{2}(R_p\!=\!1) \, \mathcal{E}^{Z}(R_p\!=\!1)$ and regularized
$c_{Z}^{2}(R_p\!=\!1) \, \mathcal{E}^{Z}_{REG}(R_p\!=\!1)$ (lower panel). All results are
obtained using the same SR state calculated for \nuc{18}{O} at a deformation of
$\beta_2 = 0.371$ as auxiliary state. The neutron number is not restored.
}
\end{figure}

\subsubsection{Distribution of weighted PNR energies}

As a next step, we analyze how the spurious energy $\mathcal{E}^{N}_{CG}(R)$ affects
the distribution of non-normalized PNR energies $c_{N}^{2}(R) \, \mathcal{E}^{N}(R)$
and $c_{N}^{2}(1) \, \mathcal{E}^{N}(R)$ as a function of the particle number one restores.
Of course, restoring other particle numbers than the one that the underlying SR state was
constrained in average to is not very useful for practical applications. The purpose of
the exercise, however, is to shed further light on the nature of the spurious energy
$\mathcal{E}^{N}_{CG}(R)$, especially through testing sum rules associated with such
a decomposition over $N$. For the latter test to be meaningful, and as motivated in
Sec.~\ref{sumrule}, it is essential to include zero and negative particle
numbers in the analysis.

Starting with a SR calculation for \nuc{18}{O}, the average proton and neutron number are small enough
that non-zero values of the quantities of interest for negative particle numbers can be unambiguously
detected in the tail of the distribution when performing a numerical calculation. Of course, a SR state
with even number-parity quantum number, as assumed here, can only be projected on even particle number,
such that the weight $c_N^2(R)$ and any operator matrix elements are obviously zero for odd $N$. In
addition, the contributions to $\mathcal{E}^{N}(R)$ from the spurious poles, see Eq.~(\ref{residuezmu}),
and from the physical pole\footnote{The Laurent series centered at $z=0$ of the integrand in
Eq.~(\ref{projenergy3}) does only contain even powers for odd $N$. As a result, such a pole does not
contribute to $\mathcal{E}^{N}(R)$.} are zero for odd $N$ when restoring particle number from a SR
state with an even-number parity quantum number. As a consequence, we can limit ourselves here to
looking at even particle numbers.

For the sake of transparency, and to avoid double sums over $N$ and $Z$ as well as the interference
of the corresponding terms when analyzing the sum rules, we limit ourselves to the restoration
of proton number in this section and in the following one. We start with the same SR state
calculated for \nuc{18}{O} with $\beta_2 = 0.371$ as in Fig.~\ref{fig:o18:e:cont} but \emph{without}
restoring neutron number, which is constrained to an average value of $N=10$. The restoration of
proton number is performed using $L=199$ integration points. In what follows, we discuss the
\emph{interaction} part of the EDF only, i.e.\ the EDF without kinetic energy and without the
one-body center of mass correction used in connection with SIII. Both are expectation values
of one-body operators and therefore free of spurious contributions. As before, the Coulomb
exchange term is omitted from the energy functional.

First, we discuss the standard case with an integration contour at $R_p = 1$. The upper panel
of Fig.~\ref{fig:decomposition:r1.0} displays the distribution of the weights
$c_Z^2(R_p\!=\!1) = |\langle \Psi^Z | \Phi_{z_p=1} \rangle|^2$, Eq.~(\ref{weight}),
of the normalized proton-number projected states in the SR state.
As expected, $c_Z^2(1)$ is peaked at $Z=8$ and falls off quickly to numerical noise.
Components with $Z > 14$ and $Z < 2$ cannot be numerically distinguished from
zero. In the former case and for $Z=0$ it is a consequence of these proton numbers being
too far from the average proton number such that $c_Z^2(1)$ becomes too small to
be distinguished from zero within the numerical precision of our code, while
for $Z<0$ the proton-number projected states $| \Psi^Z \rangle$ are
strictly zero for analytical reasons.

The lower panel of Fig.~\ref{fig:decomposition:r1.0} shows the interaction part of weighted
PNR energies before and after applying the regularization method. The distribution of
absolute values of $c_{Z}^{2}(1) \, \mathcal{E}^{Z}(1)$ does not follow the distribution
of the weights $c_Z^2(1)$ displayed in the upper panel. Instead, it has a long tail that
spreads visibly to $Z = -20$ and $Z = 34$, before it cannot be distinguished from
numerical noise anymore. In these tails, the values of $c_{Z}^{2}(1) \, \mathcal{E}^{Z}(1)$ have
alternating signs, which is clearly unphysical.
Only the regularized quantity $c_{Z}^{2}(1) \, \mathcal{E}^{Z}_{REG}(1)$
falls off in the same manner as $c_Z^2(1)$ does and is numerically zero for $Z \leq 0$.
This underlines again the spurious nature of $\mathcal{E}^{Z}_{CG}$, that is shown
separately in the middle panel of Fig.~\ref{fig:decomposition:r1.0}. In the present example,
$c_{Z}^{2}(1) \, \mathcal{E}^{Z}_{CG}(1)$ has alternating signs throughout the entire interval
of $Z$. This must not always be the case; in some other examples we have looked at, this happens
only for particle numbers that are a at least a few units away from the average particle number
of the underlying SR state.

For $Z \leq 0$, non-zero $c_{Z}^{2}(1) \, \mathcal{E}^{Z}(1)$ are entirely spurious with
$\mathcal{E}^{Z}(1)=\mathcal{E}^{Z}_{CG}(1)$; i.e.\ they originate entirely from spurious poles at
finite $z^\pm_\mu$. The same situation applies to the tail of the distribution
of $c_{Z}^{2}(1) \, \mathcal{E}^{Z}(1)$ for large positive $Z$.

\begin{figure}[t!]
\includegraphics{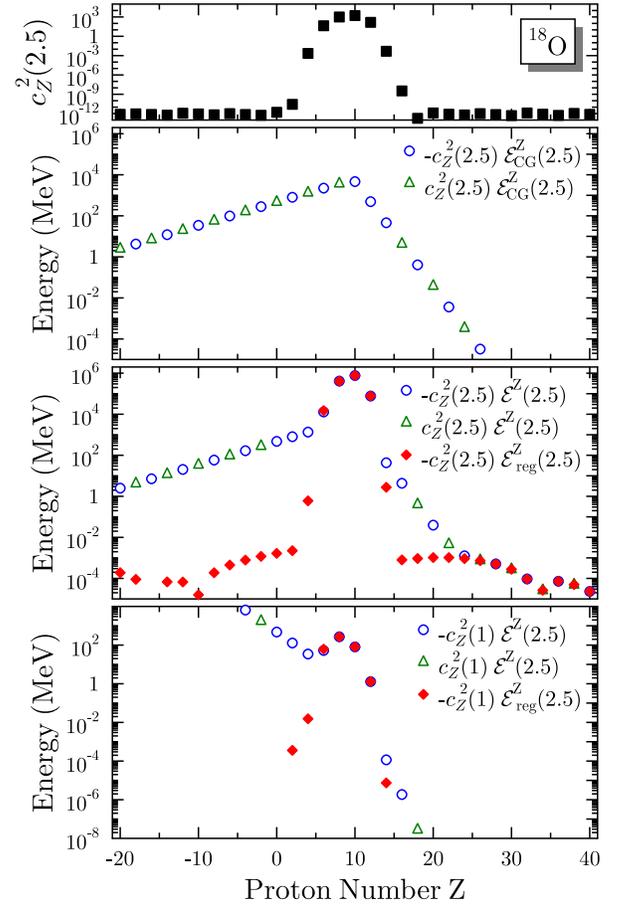}
\caption{\label{fig:decomposition:r2.5}
(Color online)
Weight $c_Z^2(R_p\!=\!2.5) = |\langle \Psi^Z | \Phi_{R_p} \rangle|^2$ of the normalized proton-number
projected states into the radially shifted SR HFB state at $R_p=2.5$ (upper panel), the weighted spurious
energy $c_{Z}^{2}(R_p\!=\!2.5) \, \mathcal{E}^{Z}_{CG}(R_p\!=\!2.5)$ (upper middle panel), the
non-regularized $\mathcal{E}^{Z}(R_p\!=\!2.5)$ and regularized $\mathcal{E}^{Z}_{REG}(R_p\!=\!2.5)$
PNR energies weighted by $c_{Z}^{2}(R_p\!=\!2.5)$ (lower middle panel) and by $c_{Z}^{2}(R_p\!=\!1)$
(lower panel). All results are obtained using the same SR state calculated for \nuc{18}{O} at a
deformation of $\beta_2 = 0.371$ as auxiliary state. The neutron number is not restored.
}
\end{figure}

As a second example, we show in the three upper panels of Fig.~\ref{fig:decomposition:r2.5} the same
quantities as in Fig.~\ref{fig:decomposition:r1.0}, but obtained employing an integration contour
of radius $R_p = 2.5$. By contrast to before ($R_p = 1$), the poles $z_\mu^\pm$ from the $1/2^+$
substate of the $\pi$ $d_{5/2^+}$ shell are located inside the integration contour, see
Fig.~\ref{fig:o18:res:p}. As a result, the spurious contribution $\varepsilon^\pm_\mu$ from those
poles increases $\mathcal{E}^{Z}$ by about 4 MeV when projecting on $Z=8$ using a non-regularized
functional, see Fig.~\ref{fig:o18:e:cont}. We analyze now if and how the energy restored on other
proton numbers are affected compared to using $R_p=1$.

Compared to Fig.~\ref{fig:decomposition:r1.0}, the distribution of weights $c_Z^2(2.5)$ is distorted
by the additional $R^Z_p=(2.5)^Z$ factor such that absolute values change by several orders of magnitude,
and the maximum of the distribution is shifted to $Z=10$. The main difference to the case using the
standard  integration contour $R_p=1$ is that the distribution of the spurious energy
$c_Z^2(2.5) \, \mathcal{E}_{CG}^Z(2.5)$ is distorted in a different manner than the distribution of
the norm, such that it falls off quicker for $Z > 8$, but much slower for $Z < 8$, including negative
$Z$. Again, only the distribution of the regularized MR energy functional $\mathcal{E}_{REG}^Z(2.5)$
follows that of the weights $c_Z^2(2.5)$.

The lowest panel of Fig.~\ref{fig:decomposition:r2.5} shows the contributions to the non-radius-weighted
energy sum rule discussed in Sec.~\ref{delires2}. The distribution $c_{Z}^{2}(1) \, \mathcal{E}^{Z}(R_p)$
is even more distorted than for the contributions to the radius-weighted sum rule shown in the panel above.
For $R_p > 1$, $c_{Z}^{2}(1) \, \mathcal{E}^{Z}(R_p)$ falls off much quicker than
$c_{Z}^{2}(R_p) \, \mathcal{E}^{Z}(R_p)$ for $Z > 8$, but much slower for $Z < 8$. For negative values of
$Z$ the missing factor $(2.5)^Z$ makes $c_{Z}^{2}(1) \, \mathcal{E}^{Z}(R_p)$ to grow so fast that it will
be impossible to safely evaluate numerically the sum rules including negative particle numbers.

For $R_p < 1$, a case not discuss here, the situation is reversed such that $c_{Z}^{2}(1) \, \mathcal{E}^{Z}(R_p)$
falls off faster than  $c_{Z}^{2}(R_p) \, \mathcal{E}^{Z}(R_p)$ for $Z < 8$, but slower for $Z > 8$, now with
the impossibility to safely evaluate the sum rule when including positive $Z$.

To summarize, the contamination of the PNR EDF by spurious contributions originating from the use of
the GWT impacts the decomposition of the (shifted) SR functional energy (kernel) into weighted PNR
energies with different particle numbers such that energy is shifted out of the physical subspace
corresponding to positive particle numbers. The impact of this finding on the fulfillment of basic
sum rules is examined in the next section.

\subsubsection{Sum rules}
\label{sect:application:sumrules}

Now we turn to the sum rules, which are obtained by summing the weighted PNR energies shown in
Figs.~\ref{fig:decomposition:r1.0} and~\ref{fig:decomposition:r2.5}. Numerical summation is
performed on a subset of even proton numbers in the interval $-20 \leq Z \leq 40$.

Again we begin with the case $R_p = 1$, for which the radius-weighted and non-radius-weighted sum
rules are identical. The SR energy\footnote{We recall that quoted energies are without the kinetic
and center-of-mass correction energies.} that sets the reference is given by
\begin{equation}
\label{eq:num:E:SR:1.0}
\mathcal{E} [\rho,\kappa,\kappa^*]
=  -410.3403 \; \text{MeV}
\, .
\end{equation}
In agreement with Eq.~(\ref{sumrulecomplexplane2}), the sum of $c_{Z}^{2}(1) \, \mathcal{E}^{Z}(1)$
over positive \emph{and} negative $Z$ reproduces this value better than 0.1 keV
\begin{equation}
\label{eq:num:sumall:E:1.0}
\sum_{Z =-\infty}^{+\infty} c_{Z}^{2}(1) \, \mathcal{E}^{Z}(1)
= -410.3403 \; \text{MeV}
\, .
\end{equation}
When limiting the sum to "physical" proton numbers $Z > 0$, however, we
obtain instead
\begin{equation}
\label{eq:num:sumpos:E:1.0}
\sum_{Z > 0 } c_{Z}^{2}(1) \, \mathcal{E}^{Z}(1)
= -410.3550 \; \text{MeV}
\, .
\end{equation}
With $14.7$ keV, the numerical difference between Eq.~(\ref{eq:num:sumall:E:1.0})
and~(\ref{eq:num:sumpos:E:1.0}), which constitutes the breaking of the physical
sumrule, is quite small. Using the standard integration contour of $R_p=1$, we
find similar values for other deformations in \nuc{18}{O}, whereas for
heavier nuclei this quantity becomes rapidly suppressed, such that it cannot be
unambiguously detected in a numerical calculation anymore.

The largest individual sum-rule breaking contribution is that for $Z=0$, for which
we obtain
\begin{equation}
\label{eq:E:Z=0:1.0}
c_{Z}^{2}(1) \, \mathcal{E}^{Z=0}(1)
= c_{Z}^{2}(1) \, \mathcal{E}^{Z=0}_{CG}(1)
= 0.0189 \; \text{MeV}
\, ,
\end{equation}
which is slightly larger than the entire sum over $Z \leq 0$. This is not
unexpected in view of the alternating signs of the contributions pointed out in the previous section.

For $Z \leq 0$, non-zero $c_{Z}^{2}(1) \, \mathcal{E}^{Z}(1)$ are of course entirely spurious,
such that they equally contribute to the sum rule of $c_{Z}^{2}(1) \, \mathcal{E}^{Z}_{CG}(1)$.
For $R_p = 1$, the right-hand-side of Eq.~(\ref{sumrule11}) is zero, such that the sum of
$c_{Z}^{2}(1) \, \mathcal{E}^{Z}_{CG}(1)$ over all $Z$ is zero as well, which we do find
numerically
\begin{equation}
\label{eq:sum:ECG:all:1.0}
\sum_{Z =-\infty}^{+\infty} c_{Z}^{2}(1) \, \mathcal{E}^{Z}_{CG}(1)
= 0.0000 \; \text{MeV}
\, .
\end{equation}
Although the alternating sign of $c_{Z}^{2}(1) \, \mathcal{E}^{Z}_{CG}(1)$ with $Z$ indicates
that a cancelation effect is at play, the result of Eq.~(\ref{eq:sum:ECG:all:1.0}) is not so obvious
when looking at the middle panel of Fig.~\ref{fig:decomposition:r1.0}. Summing up
$c_{Z}^{2}(1) \, \mathcal{E}^{Z}_{CG}(1)$ for positive values of $Z$ gives $-0.0146$ MeV, which
precisely is the difference between Eqns.~(\ref{eq:num:E:SR:1.0}) and~(\ref{eq:num:sumpos:E:1.0}).

The regularized energy $c_{Z}^{2}(1) \, \mathcal{E}^{Z}_{REG}(1)$ is numerically zero for $Z \leq 0$
as any meaningful particle-number restored observable should be. The same holds
for those large positive values of $Z$ where $c_Z^2 > 0$.
As a consequence, the sum over $c_{Z}^{2}(1) \, \mathcal{E}^{Z}_{REG}(1)$ can be limited to
"physical" particle numbers. The numerical value for this sum
\begin{eqnarray}
\sum_{Z =-\infty}^{+\infty} c_Z^2(1) \mathcal{E}^Z_{REG}(1)
& = & \sum_{Z > 0} c_Z^2(1) \mathcal{E}^Z_{REG}(1)
      \nonumber \\
& = & -410.3403 \; \text{MeV}
\end{eqnarray}
gives back the SR energy, Eq.~(\ref{eq:num:E:SR:1.0}), within 0.1 keV
as expected from Eq.~(\ref{sumrulefinal}).

When shifting one of the states to $R_p = 2.5$, the norm kernel is
$\langle \Phi_1 | \Phi_{2.5} \rangle = 2816.9760$, and the corresponding
transition energy kernel is $\mathcal{E}[2.5] = -830.2386$ MeV.
The reference for the radius-weighted sum rule is thus provided by
\begin{equation}
\mathcal{E}[2.5] \, \langle \Phi_1 | \Phi_{2.5} \rangle
= -1266844 \, \text{MeV}
\, ,
\end{equation}
where we limit ourselves again to seven digits. Summing $c_{Z}^{2}(2.5) \, \mathcal{E}^{Z}(2.5)$
over all $Z$ reproduces this value with the same precision, while summing over positive $Z$ only
gives $-1266546$~MeV, which differs from the above value by $-298$~MeV, which is of similar order
as in case of the the unshifted integration contour.

In the case of shifted contours, the non-radius-weighted sum rule is more interesting to look at.
As became clear from the bottom panel of Fig.~\ref{fig:decomposition:r2.5}, the sum over all $Z$
cannot be evaluated numerically. Let us anyway focus on the sum rule over positive $Z$ only; i.e.\
Eq.~(\ref{spuriouscomplexbisbis}). In this case, summing $c_{Z}^{2}(1) \, \mathcal{E}^{Z}(2.5)$ gives
$-309.4217$~MeV which indeed decomposes into $\mathcal{E}[\rho,\kappa,\kappa^{\ast}] = -410.3403$~MeV
plus the sum-rule breaking term obtained (either numerically or analytically through
Eq.~(\ref{spuriouscomplexbisbisbis})) by summing $c_{Z}^{2}(1) \, \mathcal{E}^{Z}_{CG}(2.5)$ over
$Z>0$ and which equates $+100.9189$~MeV. Thus, one realizes that the most essential non-radius-weighted
sum rule performed over physical components ($Z>0$) is broken and not shift invariant. In particular,
the breaking term can be very large as soon as the integration radius differs from 1. Of course, the
small (non-zero) value of that sum-rule breaking term obtained from using the unit circle as an integration
contour has masked the contamination of energy functionals with spurious terms for many years. Indeed,
practitioners naturally interpreted that very tiny breaking as due to numerical noise.

\subsubsection{Energies of physical systems}

\begin{figure}[t!]
\includegraphics{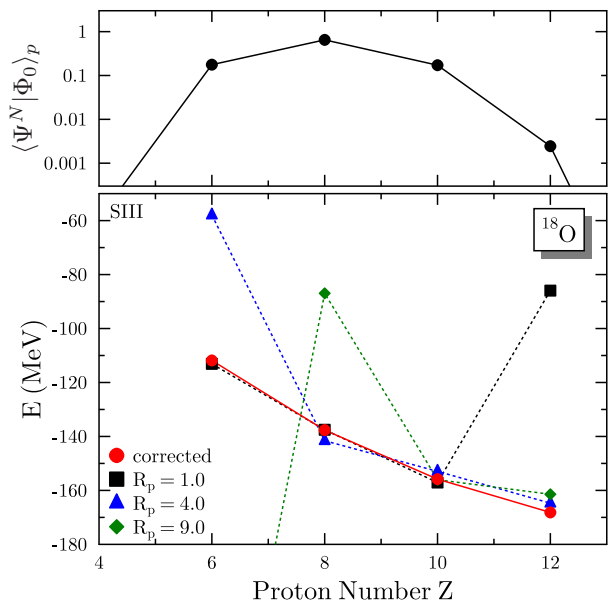}
\caption{\label{fig:o18:e:sumrule}
(Color online)
Weight of the normalized state projected on various values of $Z$ in the SR vacuum
(top panel) and decomposition of the energy into $Z$ components for three different radii of the
integration contour for protons (bottom panel) for \nuc{18}{O} at a deformation of $\beta_2 = 0.371$.
All states are projected on the same neutron number $N=10$ with an integration contour
of radius $R_n = 1$, using $L=199$ integration points for both protons and neutrons.
Corrected PNR energies are the same for all values of $R_p$ within numerical accuracy.
}
\end{figure}

After looking into the contributions to the sum rules, we now turn our
attention to normalized PNR energies pertaining to the physical subspace,
i.e.\ addressing only those particle numbers that give a non-zero norm.
Figure~\ref{fig:o18:e:sumrule} shows PNR energies (now again completed by
kinetic energy and c.m.\ correction) for three values of the integration contour
radius $R_p$. With each step in the uncorrected projected energy of the $Z=8$
component seen for $R_p = 1.9$ and $R_p = 8.2$ in Fig.~\ref{fig:o18:e:cont}, the energy
of all other $Z$ components changes as well. For each radius of the integration
contour there is at least one $Z$ component that has an obviously unphysical
uncorrected PNR energy.

The breaking of the physical sum rule for the non-regularized PNR EDF
discussed above is much smaller then the energy scale of the changes we
observe in Fig.~\ref{fig:o18:e:cont} when shifting $R_p$. Still, we can argue
with the help of the sum rules for the regularized and non-regularized
PNR EDF that any small spurious energy in a $Z$ component with large
weight $c_Z^2$ might have to be compensated by a very large spurious
energy in a $Z$ component with small weight, as it happens in
Fig.~\ref{fig:o18:e:sumrule} for the $Z=12$ component at $R_p = 1.0$ and
the $Z=6$ component at $R_p = 4.0$. As a consequence, the moderate energy scale found
for the spurious energy along a deformation energy surface when projecting on the
same nucleon number that SR vacua were constrained to does not apply
to the spurious energies entering other $Z$ components.
While this usually has no particular consequences for particle restoration calculations
where one is in most cases interested in projecting out the one particle number
that the SR HFB state was constrained to and which can be expected to have a comparatively
small contamination of spurious energy, the spurious redistribution of energy might
seriously compromise angular-momentum restoration, where one is often interested in producing the
entire spectrum of low-lying states.

\subsection{\nuc{76}{Kr}}

\begin{figure}[t!]
\includegraphics{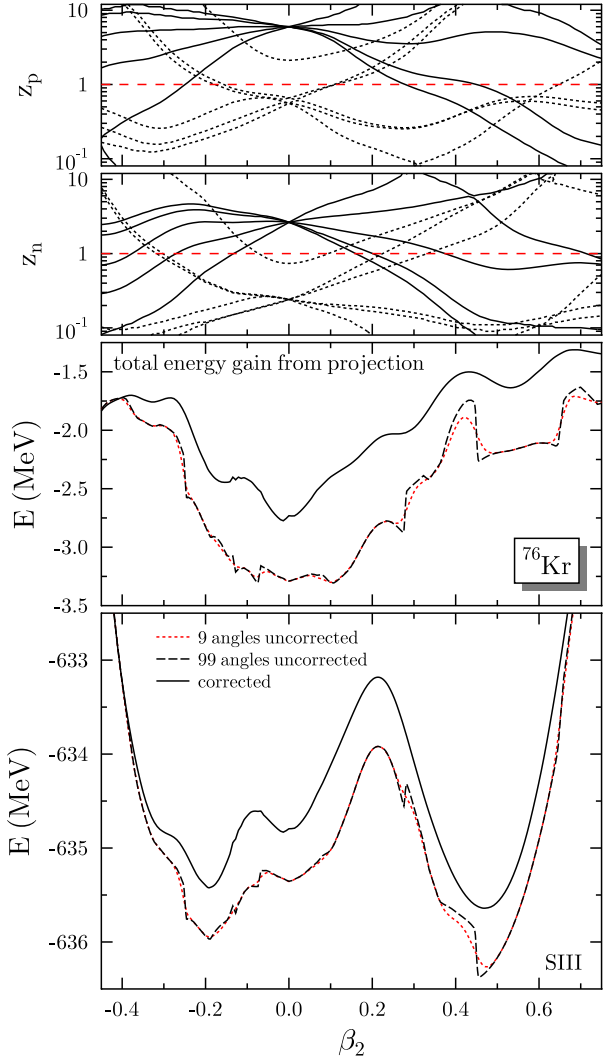}
\caption{\label{fig:kr76:e:pole}
(Color online)
Spectrum of poles $z_\mu = |u_\mu/v_\mu|$ for protons and neutrons, the
uncorrected and corrected energy gain from projection and the particle-number
projected quadrupole deformation energy for $L=9$ and 99 discretization
points of the integral in gauge space for \nuc{76}{Kr}.
}
\end{figure}

With the next example \nuc{76}{Kr}, a medium heavy nucleus located in a region
of shape coexistence, we examine how the spurious contributions to the
particle-number restored energy evolve when increasing the density of
single-particle levels.
This nucleus is one out of the series of neutron-deficient Kr isotopes that were
recently studied~\cite{bender06b} with GCM mixing
of quadrupole deformed axial particle-number and angular-momentum restored states using
SLy6.\footnote{The deformation energy surface obtained with SIII also displays shape
coexistence, although its topography is quite different from the one obtained with
SLy6. With SIII, the deformed minima are much more pronounced and lower in energy
compared to the spherical one. However, this is irrelevant for the present discussion.
}

Figure~\ref{fig:kr76:e:pole} shows the location of the poles at $z^\pm_\mu$ for protons
and neutrons, the energy gain from PNR and the absolute PNR energy as a function of
quadrupole deformation, both with and without correction and both calculated with $L=9$
and 99 discretization points of the gauge-space integrals. We have checked that all
observables calculated as operator matrix elements are converged for $L=9$. The main
difference to \nuc{18}{O} is the much larger overall density of poles. This has two
consequences. (i) It increases the number of poles crossing the integration contour
when deforming the nucleus and thus the number of steps. (ii) Poles crossing the Fermi
level are hardly isolated from other poles which limits the size of the steps through
the factors entering the middle product in Eq.~(\ref{residuezmu2}). As a consequence,
most of the steps visible in Fig.~\ref{fig:kr76:e:pole} are much smaller than those
found for \nuc{18}{O} in Fig.~\ref{fig:o18:e:pole}. Notable exceptions are the ones
on both sides of the prolate minimum at $\beta_2 \approx 0.43$, which indeed
correspond to the crossings of proton levels that are well separated from other poles.
The correction is not of the same magnitude in the various minima; in fact, the
variation of the correction between the various minima is of the same order as the
difference in total energy of the latter. Correcting for spurious energies might have
a visible impact on the excitation spectrum of this nucleus obtained from a GCM mixing
over quadrupole shapes of particle number and angular momentum restored states.

\subsection{\nuc{186}{Pb}}

\begin{figure}[t!]
\includegraphics{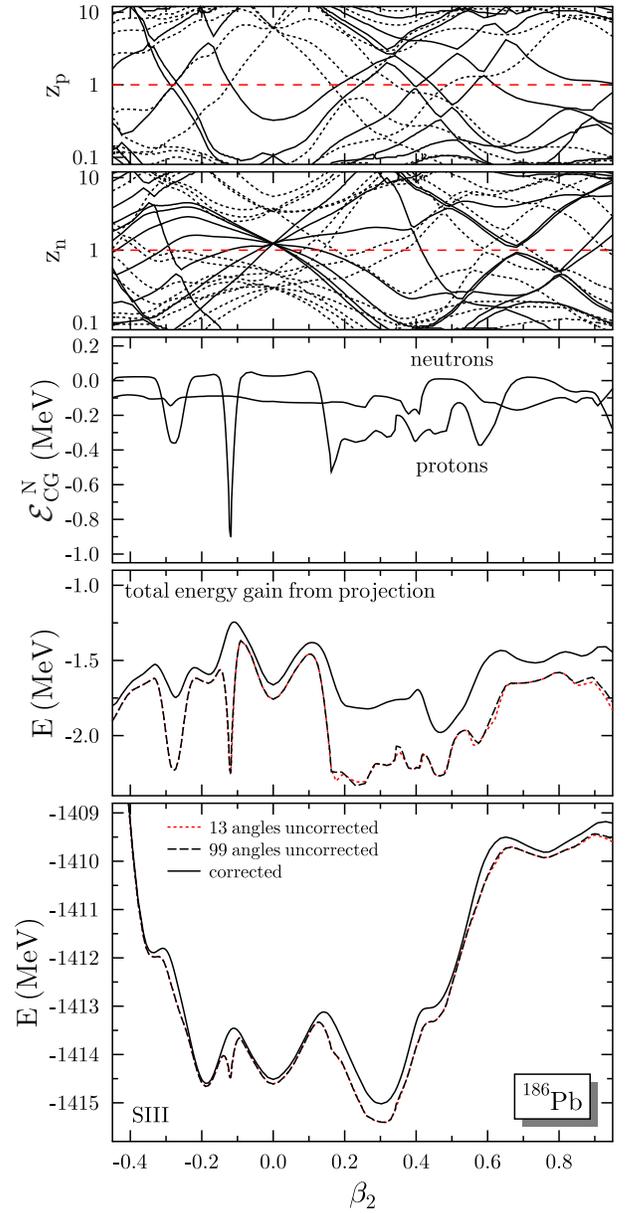}
\caption{\label{fig:pb186:e:pole}
(Color online)
Spectrum of poles $z_\mu = |u_\mu/v_\mu|$ for protons and neutrons, the correction
to the particle number restored EDF separately for protons and neutrons, the
uncorrected and corrected energy gain from projection, and the particle-number
projected quadrupole deformation energy without and with correction
for $L=13$ and 99 discretization points of the integral in gauge space for
\nuc{186}{Pb}.
}
\end{figure}

As the last example, we present in Fig.~\ref{fig:pb186:e:pole} results obtained
for \nuc{186}{Pb}, a nucleus exhibiting triple shape coexistence of spherical, oblate
and prolate states studied earlier in Refs.~\cite{duguet03c,bender04a} in a method
that includes particle-number restoration using the Skyrme EDF SLy6\footnote{ The
deformation energy surface obtained with SIII is at variance with the experimental
finding that the ground state is spherical with low-lying prolate and oblate bands
seen as excitations~\cite{duguet03c,bender04a}. However, this is irrelevant for the present
discussion.}. In this heavy nucleus, the number of neutron poles $z^\pm_\mu$ in the vicinity of the
Fermi level is even larger than for \nuc{76}{Kr}. When crossing the standard integration
contour $R_n=1$, those poles generate many steps which are, however, almost always of tiny
size due to the closeness of other poles; the sole exception being the step at
$\beta_2 = 0.4$. This is different for protons. As a consequence of the magic proton
number $Z=82$, the density of proton poles around the Fermi level is quite low for most
deformations, such that the few proton poles that cross in these regions have a much
larger impact. This is illustrated by the second panel in
Fig.~\ref{fig:pb186:e:pole} that shows the correction $\mathcal{E}_{CG}^{N}$ separately
for protons and neutrons. The narrow peak at small oblate deformation
$\beta_2 = 0.11$ is not a divergence, but stems from the crossing of two proton levels
at the Fermi energy in analogy to the structure found in \nuc{18}{O} around
$\beta_2 = 0.67$. In both cases the double-crossing is a direct consequence of the shell
closure: With all other levels being too far above or below to have occupation numbers
significantly different from 0 or 1, the constraint on the average particle-number dictates that
two pairs of levels in the gap have an occupation of $v^2_{\mu} = 1/2$ simultaneously.
Interestingly, the uncorrected deformation energy curve does
not change much when increasing the number of integration points from $L = 13$ to 99.
As for \nuc{18}{O} and \nuc{76}{Kr}, the correction varies strongly with deformation, has a
different value in the various minima, and, most importantly, is on the same energy scale
as the energy difference between those minima.

\section{Summary, conclusions and outlook}
\label{conclusions}

In the present paper, we introduce the notion of spurious self-pairing. It appears as a generalization
of spurious self-interaction processes, a well-known problem in electronic density-functional
theory \cite{perdew81a,koch01,Ull00aDFT,Leg02aDFT,Ruz07aDFT}, to systems with pairing
correlations that are modeled within EDF approaches using independent quasiparticle BCS states as
auxiliary states of reference. Self-interaction and self-pairing processes appear for
any energy functional that uses different vertices in the particle-hole and particle-particle channels,
and/or not fully antisymmetric vertices; e.g., as due to density-dependencies. Neither self-interaction nor self-pairing
appear when the many-body energy is strictly calculated as the expectation value of a Hamilton operator. Both are
a price to pay when replacing the exact nuclear many-body problem by a system of coupled
one-body problems in a EDF calculation, modeling higher-order in-medium correlations
through a simple energy functional depending on one-body densities and currents.
On the single-reference level, self-pairing gives a spurious contribution to the pairing field
(and therefore influences all quantities it determines) and to the total binding energy.

Energy density functionals extended to perform multi-reference calculations, i.e.\ symmetry
restoration or GCM-type configuration mixing, also contain unphysical contributions: First,
the previously discussed self-interaction and self-pairing processes that continuously extend
from SR energy functional to off-diagonal energy kernels, as well as a second and much more
dangerous  category of spuriosity that appears when the off-diagonal kernels are evaluated
on the basis of the generalized Wick theorem. The use of a Wick theorem to evaluate a functional
energy kernel that does not originate from a genuine Hamilton operator is not justified.
Relying on the generalized Wick theorem to construct off-diagonal functional energy kernels
has the unexpected particularity to provide previously discussed self-interaction
and self-pairing contributions with unphysical weights that contain poles leading
to divergences~\cite{anguiano01b} and steps in the energy~\cite{doba05a}. The latter have been
noticed recently in the context of particle-number restoration whenever a single-particle level
crosses the Fermi energy. As demonstrated in Paper~I \cite{lacroix06a}, the weights of
self-interaction and self-pairing terms obtained on the basis of the standard Wick theorem are
different and do not present any problematic contributions. This feature can be exploited to
unambiguously isolate the dangerous spuriosities and set-up a correction scheme that regularizes
unphysical divergences and steps in MR energy kernels~\cite{lacroix06a}. In the present paper, we have
applied this correction scheme to the simplest and formally most transparent MR case of particle-number
restoration after variation.

The complex-plane analysis performed in the present work reveals that each conjugated pair of single-particle
levels $(\mu,\bar{\mu})$ provides an unphysical contribution to the physical pole at $z=0$, in addition to
generating unphysical poles at $z^\pm_\mu = \pm i |u_\mu|/|v_\mu|$. The latter cause the steps
as they cross the integration contour~\cite{doba05a}. The unphysical poles are also at the origin of the
breaking of the shift invariance of PNR energies~\cite{doba05a}. However, removing only the contribution
from the poles at $z^\pm_\mu$ to the energy functional kernel leads to unphysical results. Instead, the
spurious contribution from a given pair of single-particle levels to the pole at $z=0$ has to be removed
simultaneously, as both are very large, of opposite sign, and nearly cancel.

The correction scheme proposed in Paper~I does indeed remove both contributions; thereby it eliminates
the divergences and steps and restores the shift invariance of PNR energies $\mathcal{E}^{N}$ as well
as standard sum rules that they can be expected to fulfill. The correction to  $\mathcal{E}^{N}$ is
of the order of 1 MeV, and in most cases reduces the energy gain from PNR. On the one hand, the correction
is sufficiently small that PNR EDF results published earlier are not meaningless. On the other hand, in extreme cases the correction might be as large as $50 \, \%$ of the energy gain from PNR, which casts some doubt on the reliability of published calculations performed within the EDF framework. The correction is also of the same order as the rms error of the mass residuals reached with the best available particle-number
restored EDF mass fits~\cite{samyn04}. The correction varies rapidly with deformation and affect
significantly the structure of complex nuclei presenting soft deformation energy surfaces and
coexisting minima.

In the present paper, we do not attempt to correct for the "true" self-interaction and self-pairing
processes that contaminate the single-reference energy density functionals. This amounts to modify
the underlying functional which we postpone to later works. In addition, a self-consistent
correction is very cumbersome, as documented in the literature for self-interaction in the context
of electronic DFT \cite{perdew81a,koch01,Ull00aDFT,Leg02aDFT,Ruz07aDFT}.

Particle-number restoration is not the only type of MR-EDF calculation where using the GWT as a basis
to construct non-diagonal functional energy kernels causes problems. In fact, any symmetry restoration
or GCM-type configuration mixing calculation is expected to be contaminated with similar spurious
contributions; e.g., anomalies were encountered in Ref.~\cite{doba06a} in angular-momentum restoration
calculations of cranked states without pairing and using a Skyrme EDF. The correction scheme proposed
in Paper~I can be applied to any type of MR-EDF calculation. However, all others but particle-number
restoration require the numerical construction of the canonical basis of the Bogoliubov transformation
connecting the two different quasi-particle bases associated with the two vacua entering the construction
of the functional energy kernel~\cite{lacroix06a}. Work towards the numerical implementation of such
a scheme is underway.

In the present study, we have limited ourselves to particle number restoration after variation,
where the correction can be subtracted from energy kernels \emph{a posteriori}. With
variation-after-symmetry-restoration EDF calculations becoming available~\cite{stoitsov07,Rodriguez07a},
and the variational equations sometimes running into the divergences~\cite{doba05a}, setting
up a correction scheme for those variational equations becomes an important issue and will be addressed
in a forthcoming study~\cite{stoitsoVAP}.

The correction proposed in Paper~I \cite{lacroix06a} and discussed in the present one is limited to
energy functionals depending on integer powers of the density-matrices. Most functionals used in the
literature, however, have a density dependence of non-integer power, both in the functional modeling
the effective strong interaction and as an approximate Coulomb exchange term. Compared to the functionals
discussed here, such non-integer powers
of the density matrix pose two additional types of difficulties when extended to non-diagonal energy
kernels on the basis of the GWT: (i) as transition densities are complex, taking their fractional
power is ambiguous~\cite{doba05a}, and (ii) there is no well-defined basis at present to remove the
spurious branch cuts that are generated by such terms. Both points are illustrated and examined
further in Paper~III of this series~\cite{paperIII}.

In our opinion, the particular difficulties of functionals with non-integer density dependencies constitute a
strong motivation to construct energy functionals with integer powers of the densities only in view of
performing meaningful MR-EDF calculations in the future. At present, there are no such non-relativistic
functionals of high performance. Relativistic functionals  have been constructed along these lines
recently~\cite{bue02a} with a different motivation, and have already been used in PNR-EDF
calculations~\cite{niksic06b}. The construction of correctable energy functionals for multi-reference
applications becomes an urgent task for the future. A particular problem will be to find a suitable
functional for the Coulomb interaction, as using the exact exchange term is incommensurately expensive
in multidimensional MR-EDF calculations.

\begin{acknowledgments}

We thank J.~Dobaczewski, W.~Nazarewicz, P.-G.~Reinhard and M.~V.~Stoitsov for providing
us with their analysis of the PNR-HFB problem in the complex plane a very early stage
which triggered the present work. Part of the work by M.~B.\ was
performed within the framework of the Espace de Structure Nucl{\'e}aire Th{\'e}orique (ESNT) at Saclay.
This work was supported by the U.S.\ National Science Foundation under Grant No.\ PHY-0456903.
We also thank the [Department of Energy's] Institute for Nuclear Theory at the University of Washington
for its hospitality and the Department of Energy for partial support during the elaboration of this work.

\end{acknowledgments}

%
%

\begin{appendix}

\section{The energy functional}
\label{app:functional}

The energy is given as the sum of the non-interacting kinetic energy, the
Skyrme energy functional that models the strong particle-hole interaction,
a pairing functional that models the particle-particle interaction and
the Coulomb energy functional
\begin{equation}
\label{SHFbindingenergy}
\mathcal{E}
= \mathcal{E}_{\text{kin}}
  + \mathcal{E}_{\text{Skyrme}}
  + \mathcal{E}_{\text{Coulomb}}
  + \mathcal{E}_{\text{pair}}
  + \mathcal{E}_{\text{corr}}
\, .
\end{equation}
The kinetic energy is the mean value of a one-body operator; hence it does
not pose problems. From the point of view of establishing the correction to
the MR energy kernel, we identify in the following
\begin{subequations}
\begin{eqnarray}
\mathcal{E}^{\rho\rho}
& \equiv &   \mathcal{E}_{\text{Skyrme}}^{\rho\rho}
           + \mathcal{E}^{direct}_{\text{Coulomb}} \, ,
          \\
\mathcal{E}^{\rho\rho\rho}
& \equiv &  \mathcal{E}_{\text{Skyrme}}^{\rho\rho\rho} \, ,
           \\
\mathcal{E}^{\kappa\kappa}
&\equiv &   \mathcal{E}^{\kappa\kappa}_{\text{DI}}  \, ,
\end{eqnarray}
\end{subequations}
making explicit the power of the density matrices entering a given term. Let us
now specify these terms more explicitly.

\subsection{The Skyrme energy functional}
\label{appendixskyrme}
We restrict ourselves here to those terms of the Skyrme EDF depending on time-even densities and currents that
 contribute to the ground states of even-even nuclei in SR and MR-PNR calculations. Also, the functional given
below corresponds to the particular Skyrme interaction SIII used throughout this paper. For SIII, there are no
density-dependent coupling constants, but the energy functional can be divided into a bilinear
$\mathcal{E}_{\text{Skyrme}}^{\rho\rho}$ and a trilinear term $\mathcal{E}_{\text{Skyrme}}^{\rho\rho\rho}$. The
Skyrme energy functional is usually represented either in terms of isoscalar and isovector densities
\cite{perlinska04a} or in terms of the total density and the densities of the nucleon species~\cite{bonche87a}. In
the context of particle-number restoration, the most convenient representation separates contributions which are
bilinear in densities of the same isospin from those that are bilinear in densities of different isospin
\begin{widetext}
\begin{eqnarray}
\label{functional:bi:skyrme}
\mathcal{E}_{\text{Skyrme}}^{\rho\rho}
& = & \int \! d^3r \;
      \Bigg\{ \sum_{q = p,n}
      \Big[   A^{\rho\rho} \, \rho^{2}_{q}
            + A^{\rho\tau} \, \rho_{q} \tau_{q}
            + A^{\rho\Delta \rho} \, \rho_{q} \Delta \rho_{q}
            + A^{\rho \nabla J} \rho_q \nabla \cdot \vec{J}_q
      \Big]
     \nonumber \\
&  & + \sum_{q,q' = p,n \atop q \neq q'}
     \Big[ B^{\rho\rho} \, \rho_q \rho_{q'}
         + B^{\rho\tau} \, \rho_q \tau_{q'}
         + B^{\rho\Delta \rho} \, \rho_q \Delta\rho_{q'}
         + B^{\rho \nabla J} \, \rho_q \nabla \cdot \vec{J}_{q'}
     \Big]
     \Bigg\}
     ,
     \\
\label{functional:tri:skyrme}
\mathcal{E}_{\text{Skyrme}}^{\rho\rho\rho}
& = & \int \! d^3r
      \sum_{q,q' = p,n \atop q \neq q'}
      A^{\rho\rho\rho} \rho_{q}^2 \, \rho_{q'}
      \, .
\end{eqnarray}
\end{widetext}
The $A^{ff'}$ and $B^{ff'}$ denote the coupling constants,\footnote{Superscripts $ff'$ and
$fff'$ used on the r.h.s.\ of Eqs.~(\ref{functional:bi:skyrme}-\ref{functional:tri:skyrme})
refer to the local densities that appear in the functional, while the superscripts $\rho\rho$,
$\kappa\kappa$, $\rho\rho\rho$, \ldots on the l.h.s.\ of
Eqs.~(\ref{functional:bi:skyrme}), (\ref{functional:tri:skyrme}), and (\ref{pairingfunctional})
correspond to the powers in the density matrices.}
none of which is density dependent for SIII. In the canonical basis, the local densities entering the
energy functional (\ref{functional:bi:skyrme}-\ref{functional:tri:skyrme}) are given by
\begin{eqnarray}
\rho_q (\vec{r})
& = & 2 \sum_{\mu > 0} \phi_{\mu}^\dagger (\vec{r} q) \,
                       \phi_{\mu} (\vec{r} q)  \, \rho_{\mu\mu}
      \nonumber \\
\tau_q (\vec{r})
& = & 2 \sum_{\mu > 0}
      \big[ \nabla \varphi^{\dagger}_{\mu} (\vec{r} q) \big]
      \cdot
      \big[ \nabla \phi_{\nu} (\vec{r} q) \big]
      \, \rho_{\mu\mu}
      \label{densities} \\
\vec{J}_{q} (\vec{r})
& = & - i \sum_{\mu > 0}
      \left\{ \varphi^{\dagger}_{\mu}  (\vec{r} q)
              \big[ \nabla \times \hat{\sigma} \, \phi_{\mu} (\vec{r} q)
              \big]
             - \text{h.c.}
      \right\} \, \rho_{\mu\mu}
      \nonumber
\end{eqnarray}
and denote, for the isospin $q = n,p$, the matter density, the kinetic density and the spin-orbit current,
respectively. The operator $\hat{\vec\sigma}$ is the vector built out of the three cartesian Pauli matrices. The density matrix $\rho_{\mu\mu}$ is either given by Eq.~(\ref{intrdens1}) for the SR EDF, or by
Eq.~(\ref{contractph}) or (\ref{contractphcomplex}) for the PNR MR EDF. One can see from the
expressions given above that any local density $f_{q} (\vec{r})$ can be written as:
\begin{equation}
f_{q} (\vec{r})
\equiv 2 \sum_{\mu > 0} W^{f}_{\mu \mu} (\vec{r} q)  \,  \rho_{\mu\mu}
\, ,
\end{equation}
where $f \in \{\rho, \tau, \vec{J} \}$
and where the explicit form of each $W^{f}_{\mu \mu} (\vec{r} q)$
can be easily extracted from Eq.~(\ref{densities}). This will
facilitate the construction of the matrix elements needed to evaluate
the correction $\mathcal{E}^{N}_{CG}$.

%
%
\subsection{The Coulomb energy functional}
\label{appendixcoulomb}
The standard Coulomb energy functional that is used in connection with
most parameterizations of the Skyrme energy functional is given by
\begin{eqnarray}
\label{eq:coul:funct}
\mathcal{E}_{\text{Coulomb}}
& = & \frac{e^2}{2} \iint \! d^3r \; d^3r' \;
      \frac{\rho_p(\vec{r}) \rho_p (\vec{r}')}
           {|\vec{r}-\vec{r}'|}
      \nonumber \\
&   & - \frac{3}{4} e^2 \left( \frac{3}{\pi} \right)^{1/3}
        \int \! d^3r \; \rho_p^{4/3} (\vec{r})
.
\end{eqnarray}
The proton density entering Eq.~(\ref{eq:coul:funct}) is calculated as described
in the preceding section. The energy functional~(\ref{eq:coul:funct}) provides the
textbook example of an energy functional that is not self-interaction free~\cite{perdew81a}.

The Coulomb exchange term in the Slater approximation, represented by the second term on
the r.h.s.\ of Eq.~(\ref{eq:coul:funct}), resembles the density-dependent terms of modern
parameterizations of the Skyrme functional. As, at present, we do not have a correction scheme
for terms depending on non-integer powers of the density, we drop it and consider the direct
term only in the present work. Concerning absolute binding energies, the Coulomb exchange
term is the smallest of all contributions to the energy functional for nuclei and states
considered here; it does not exceed $2 \, \%$ of the total binding energy even in very heavy
nuclei with a strong Coulomb field. What is even more important for the present study is that its
value changes also by at most $2 \, \%$ when deforming a nucleus; its influence on potential
energy surfaces is smaller than what can be resolved in the plots shown in Sec.~\ref{applications}.

%
%
\subsection{Pairing energy functional}
\label{appendixpairing}

For pairing, we choose a local energy functional deduced from a simple delta interaction (DI),
often referred to as "volume pairing"
\begin{equation}
\label{pairingfunctional}
\mathcal{E}^{\kappa\kappa}_{\text{DI}}
= \sum_{q} \int \! d^3r  \, A^{\tilde{\rho} \bar{\rho}} \,
  \bar{\rho}^*_{q} (\vec{r}) \, \tilde{\rho}_{q} (\vec{r})
\, .
\end{equation}
More elaborate parameterizations of the pairing energy functional are frequently used in the literature.
When enforcing time-reversal invariance as done here, the local pair densities entering the pairing
functional are related to the pairing tensor through
\begin{eqnarray}
\label{anomalousdensity}
\tilde{\rho}_{q} (\vec{r})
& \equiv & 2 \sum_{\mu > 0} W^{\tilde{\rho}}_{\mu \bar{\mu}} (\vec{r} q) \,
           \kappa_{\bar\mu \mu}^{\varphi \varphi'} \, ,
           \\
\bar{\rho}_{q}^* (\vec{r})
& \equiv & 2 \sum_{\mu > 0} W^{\bar{\rho}}_{ \mu \bar{\mu}}{}^* (\vec{r} q) \,
           \kappa_{\bar\mu \mu}^{\varphi' \varphi \, \ast} \, ,
\end{eqnarray}
where $\kappa_{\mu \bar\mu}^{\varphi \varphi'}$ and $\kappa_{\mu \bar\mu}^{\varphi' \varphi \, \ast}$
are given by Eqns.~(\ref{intrdens2}) and~(\ref{intrdens3}) for SR-EDF calculations with
 $\varphi' = \varphi$, and by
Eqns.~(\ref{contracthh}) and~(\ref{contractpp}), or  Eqns.~(\ref{contractppcomplex})
and~(\ref{contracthhcomplex}), respectively, for PNR MR-EDF calculations.
In the case of SR EDF and PNR MR EDF calculations, $W^{\tilde{\rho}}_{\mu \bar{\mu}} (\vec{r} q)$
and $W^{\bar{\rho}}_{\mu \bar{\mu}} (\vec{r} q)$ are equal and given by
\begin{equation}
\label{anomalousdensity2}
W^{\tilde{\rho}}_{\mu \bar{\mu}} (\vec{r} q)
= W^{\bar{\rho}}_{\mu \bar{\mu}} (\vec{r} q)
= g_{\mu} \, \sum_{\sigma = \pm 1} \sigma \, \phi_{\mu} (\vec{r} \sigma q) \, \phi_{\bar{\mu}} (\vec{r} -\sigma q)
\, ,
\end{equation}
and represent the spin-singlet part of the two-body wave function. This does not hold
for other MR EDF calculations. The notation $\sigma = \pm1$ denotes the spinor component with spin
projection $\pm 1/2$. The functions
$W^{\tilde{\rho}}_{\mu\bar{\mu}} (\vec{r} q)$ and $W^{\bar{\rho}}_{\mu\bar{\mu}} (\vec{r} q)$
incorporate a cutoff $g_{\mu}$ to regularize the pairing problem, which is otherwise divergent
in a variational calculation. In the SR calculations, we use the smooth phenomenological cutoff
proposed in Ref.~\cite{bonche85a}, while in the PAV-PNR MR calculations it is set to $g_{\mu}=1$.

%
%

\section{Correction term}
\label{correctionskyrme}
%
%
\subsection{Bilinear terms}

\subsubsection{Matrix elements}

We focus here on the case where the system is time-reversal invariant, which leads to
\begin{equation}
\label{equalME}
W^{f}_{\mu\mu}
= W^{f}_{\bar{\mu}\bar{\mu}}
\end{equation}
for the time-even densities contributing to the Skyrme and Coulomb functionals. There is a minus
sign in the l.h.s.\ of Eq.~(\ref{equalME}) when considering time-odd ones that we do not have to
take into account here as the corresponding contributions from the two states
($\mu$, $\bar\mu$) cancel out both in the total energy and in the correction given by
Eq.~(\ref{spurious:bi}). For the state-dependent function entering the pair density we have
\begin{equation}
W^{\tilde{\rho}\, \ast}_{\mu \bar{\mu}} (\vec{r} q)
= W^{\tilde{\rho}}_{\mu\bar{\mu}} (\vec{r} q)
= - g_{\mu} W^{\rho}_{\mu \mu}  (\vec{r} q)
\, .
\end{equation}
For the SIII energy functional, the matrix elements that match the definition of the bilinear part as given by
Eq.~(\ref{eq:e00}) read as
\begin{equation}
\label{phmatrixelement:skyrme}
\bar v^{\rho\rho}_{\mu\nu\mu\nu}
= 2 \int \! d^3r
  \sum_{f,f'} A^{ff'} \, W^{f }_{\mu \mu} (\vec{r} q) \,
                         W^{f'}_{\nu \nu} (\vec{r} q)
\, ,
\end{equation}
where the sum over $f$, $f'$ runs over all terms appearing in Eq.~(\ref{functional:bi:skyrme}).
The quasi-local form of the Skyrme energy functional simplifies the construction of the matrix elements
$\bar{v}^{\rho\rho}_{\mu\nu\mu\nu}$ in two ways: on the one hand, they involve one triple integral only,
and on the other they contain products that are separable in $\mu$ and $\nu$. This is of great help
from the numerical point of view when coding the correction to the PNR energy as defined by
Eq.~(\ref{spurious:bi}).

The situation is different for the direct Coulomb term. Indeed, the corresponding matrix elements
(not antisymmetric as Coulomb exchange was dropped all together) are not separable
\begin{equation}
\label{phmatrixelementcoul}
\bar v^{\rho\rho}_{\mu\nu\mu\nu}
=  2 \, e^2 \iint \! d^3r \, d^3r' \,
  \frac{W^{\rho}_{\mu \mu} (\vec{r} p) \, W^{\rho}_{\nu \nu} (\vec{r}' p)}
       {|\vec{r}-\vec{r}'|}
\, .
\end{equation}
and they involve a six-fold integral. This considerably
complicates their calculation compared to the matrix elements of the Skyrme functional. Instead, the Poisson
equation for the Coulomb potential generated by the source $W^{\rho}_{\lambda \lambda} (\vec{r} p)$
\begin{equation}
\label{poisson}
U^{\rho}_{\lambda \lambda} (\vec{r})
= - 4 \pi e \, \Delta W^{\rho}_{\lambda \lambda} (\vec{r} p)
\, ,
\end{equation}
is solved first using boundary conditions constructed from the lowest-order terms in a
multipole expansion of the state-dependent field $W^{\rho}_{\lambda \lambda} (\vec{r} p)$,
and then the Coulomb energy of the other density in this field is obtained as
\begin{equation}
\bar v^{\rho\rho}_{\mu\nu\mu\nu}
= 2 \, e^2 \int \! d^3r \, W^{\rho}_{\mu \mu} (\vec{r} p) \, U_{\nu \nu} (\vec{r})
\, .
\end{equation}
For all but very light nuclei, the calculation of the correction is much more costly
than the calculation of the PNR direct Coulomb energy itself, as the correction
$U^{\rho}_{\mu \mu} (\vec{r})$ has to be determined for each single-particle state
solving Eq.~(\ref{poisson}), while for the total Coulomb energy the Coulomb potential
has to be determined only for the summed up total charge density. However,
$U^{\rho}_{\mu \mu} (\vec{r})$ entering the correction is independent of the gauge
angle, while the Coulomb potential has to be determined for each gauge angle when
calculating the total PNR Coulomb energy.

Last, but not least, the matrix elements entering the pairing functional are given by
\begin{equation}
 \label{ppmatrixelement}
\bar v^{\kappa\kappa}_{\mu\bar{\mu}\nu\bar{\nu}} =  4 \int \! d^3 r \; A^{\tilde{\rho}\tilde{\rho}} \,
                   W^{\tilde{\rho} \, \ast}_{\mu\bar{\mu}} (\vec{r} q) \,
                   W^{\tilde{\rho}}_{\nu \bar{\nu}} (\vec{r} q)
\, .
\end{equation}

\subsubsection{Correction}

Let us now write the spurious contribution $\mathcal{E}^{N}_CG$ that must be removed from the MR-PNR energy, defined by
Eq.~(\ref{spurious:bi}), for the functional introduced in Appendices~\ref{appendixskyrme}, \ref{appendixcoulomb}
and \ref{appendixpairing}.

The spurious contributions only originate from interactions between particles of the same isospin. All contributions from
the bi-linear part of the energy functional to the correction contain the same occupation factor, for which we
introduce the shorthand notation
\begin{eqnarray}
\divfac^{N}_{SG \, \mu}
& \equiv & ( u_{\mu} v_{\mu} )^{4}
           \int_{0}^{2\pi} \!  \! d\varphi \;
           \frac{e^{-i\varphi N}}{2\pi \, c^{2}_{N}} \,
           \frac{\left(e^{2 i \varphi} - 1 \right)^{2}}
                {\left(u_\mu^2 + v_{\mu}^2 \, e^{2 i \varphi}\right)^{2}}
           \nonumber \\
&        & \quad \times
           \prod_{\nu > 0 \atop q_{\nu} = q_{\mu}} (u_{\nu}^2 + v_{\nu}^2 e^{2i\varphi})
\, .
\end{eqnarray}
Hence, we obtain
\begin{widetext}
\begin{eqnarray}
\label{spurious:bi:skyrme} \mathcal{E}_{CG}^{N}
& = & 4 \int \! d^3 r \,
      \sum_{\mu >0} \divfac^{N}_{SG \, \mu}
      \Bigg[ \sum_{\{f,f'\}} A^{ff'}  \, W^{f }_{\mu \mu} (\vec{r} q) \,
                            W^{f'}_{\mu \mu} (\vec{r} q) - A^{\tilde{\rho}\tilde{\rho}} \,
                    \tilde{W}_{\mu \bar\mu} (\vec{r} q) \,
                    \tilde{W}_{\mu \bar\mu} (\vec{r} q) + e^2 \,
      W^{\rho}_{\mu \mu} (\vec{r} p) \, U_{\mu \mu} (\vec{r})
     \Bigg]
\, ,
\end{eqnarray}
%
where it is understood that the Coulomb term only contributes to the sum over proton pairs. In the MR-PNR code,
the calculation of Eq.~(\ref{spurious:bi:skyrme}) constitutes an effort similar to the evaluation of a local one-body
operator, as it can be reduced to a single sum over half of the single-particle states adding up a local function in space
that is integrated over afterwards.

%
%

\subsection{Trilinear terms}
\label{tri-linearcorrection}

\subsubsection{General expression}

We have restricted ourselves here to the special case of the Skyrme SIII functional. The zero-range
three-body force that it originates from has the particular property that it gives an energy functional
composed of terms which are bilinear in densities of one isospin times a density of the other isospin.
The absence of terms trilinear in densities of one isospin greatly simplifies the correction term
(see Paper~I), which reduces to
%
\begin{eqnarray}
\label{spurious:tri} \mathcal{E}_{CG}^{N}
& = & \frac{1}{6}\sum_{\mu > 0}
        \sum_{\lambda \gtrless 0 \atop q_\lambda \neq q_\mu}
          \,
         \Big( \bar{v}^{\rho\rho\rho}_{\mu\mu\lambda\mu\mu\lambda}
              +\bar{v}^{\rho\rho\rho}_{\bar\mu\mu\lambda\bar\mu\mu\lambda}
              +\bar{v}^{\rho\rho\rho}_{\mu\bar\mu\lambda\mu\bar\mu\lambda}
              +\bar{v}^{\rho\rho\rho}_{\bar\mu\bar\mu\lambda\bar\mu\bar\mu\lambda}
              +\bar{v}^{\rho\rho\rho}_{\mu\lambda\mu\mu\lambda\mu}
              +\bar{v}^{\rho\rho\rho}_{\bar\mu\lambda\mu\bar\mu\lambda\mu}
      \nonumber \\
&   & \phantom{\frac{1}{6}\sum_{\mu > 0}
                 \sum_{\lambda \gtrless 0 \atop q_\lambda \neq q_\mu}
                  \big(
       }
              +\bar{v}^{\rho\rho\rho}_{\mu\lambda\bar\mu\mu\lambda\bar\mu}
              +\bar{v}^{\rho\rho\rho}_{\bar\mu\lambda\bar\mu\bar\mu\lambda\bar\mu}
              +\bar{v}^{\rho\rho\rho}_{\lambda\mu\mu\lambda\mu\mu}
              +\bar{v}^{\rho\rho\rho}_{\lambda\bar\mu\mu\lambda\bar\mu\mu}
              +\bar{v}^{\rho\rho\rho}_{\lambda\mu\bar\mu\lambda\mu\bar\mu}
              +\bar{v}^{\rho\rho\rho}_{\lambda\bar\mu\bar\mu\lambda\bar\mu\bar\mu}
        \Big) \, \divfac^{N_{\mu}}_{SG \, \mu}
      \nonumber \\
&   &
     \phantom{\frac{1}{6}\sum_{\mu > 0}
        \sum_{\lambda \gtrless 0 \atop q_\lambda \neq q_\mu}
      }
      \times
      v_{\lambda}^2 \;
      \bigg[
      \int_{0}^{2 \pi} \! d\phi \;
      \frac{e^{-i\phi N_{\lambda}}}{2 \pi c^{2}_{N_{\lambda}}}
      \frac{e^{2i\phi}}{u_{\lambda}^2 + v_{\lambda}^2 e^{2i\phi}}
      \prod_{\nu > 0 \atop q_{\nu} = q_{\lambda}} (u_{\nu}^2 + v_{\nu}^2 e^{2i\phi})
      \bigg]
\end{eqnarray}
where $(N_{\lambda} = N$, $N_{\mu} = Z)$ or $(N_{\lambda} = Z$, $N_{\mu} = N)$ depending on the isospin of the
states $(\mu, \bar{\mu})$.

\subsubsection{Matrix elements}

The matrix elements of the trilinear term appearing in the SIII Skyrme functional are given by
\begin{equation}
\label{rrrmatrixelement}
\bar{v}^{\rho\rho\rho}_{\mu\nu\lambda\mu\nu\lambda}
= 6 \int \! d^3r \, A^{\rho\rho\rho} \,
                  W^{\rho}_{\mu \mu} (\vec{r} q_\mu) \,
                  W^{\rho}_{\nu \nu} (\vec{r} q_\nu) \,
                  W^{\rho}_{\nu \nu} (\vec{r} q_\lambda) \, .
\end{equation}

%
%

\subsubsection{Correction}

Finally, the spurious term to be removed from the trilinear part of the SIII Skyrme functional is
\begin{eqnarray}
\mathcal{E}_{CG}^{N}
& = & 12 \sum_{\mu > 0} \divfac^{N_{\mu}}_{SG \, \mu} \,
      \int \! d^3r \; \big[ W^\rho_{\mu \mu} (\vec{r}q_\mu) \big]^{2} \,
      \bigg[
      \int_{0}^{2 \pi} \! d\phi \;
      \frac{e^{-i\phi N_{\lambda}}}{2 \pi c^{2}_{N_{\lambda}}}
      \, \rho_{q_\lambda} (\vec{r} \phi) \,
      \prod_{\nu > 0 \atop q_{\nu} = q_{\lambda}} (u_{\nu}^2 + v_{\nu}^2 e^{2i\varphi})
      \bigg] \, ,
\end{eqnarray}
where the last term in square brackets $[ \cdots ]$ is nothing but the
particle-number projected local density of nucleons with isospin $q_{\nu}\neq q_{\mu}$.

\end{widetext}

\end{appendix}

\bibliography{papierII}

\end{document}